\documentclass[11pt]{JHEP3}
\usepackage{cite,amsmath,amssymb}

\title{Spin-half fermions with mass dimension one: theory, 
phenomenology, and dark matter}

\author{D V Ahluwalia-Khalilova$^1$ and D Grumiller$^2$\\
\rm
$^1$ ASGBG/CIU, Department of 
Mathematics, Apartado Postal C-600,\\
University of Zacatecas (UAZ), Zacatecas, Zac 98060, Mexico\\
$^2$ Institut f\"ur Theoretische Physik,
University of Leipzig, Augustusplatz 10-11,\\
D-04109 Leipzig, Germany\\
E-mail: dva-k@heritage.reduaz.mx and grumiller@itp.uni-leipzig.de\\

\vspace{11pt}
Received 8 December 2004\\
Accepted 14 June 2005\\
Published 19 July 2005\\
\textbf{ArXiv ePrint:} hep-th/0412080
}

\abstract{
We provide the first details on the unexpected theoretical discovery
of a spin-one-half matter field with mass dimension one. It is based
upon a complete set of  dual-helicity 
eigenspinors of the charge conjugation operator.
Due to its unusual properties with respect to charge conjugation and
parity, it belongs to a non-standard Wigner class. Consequently, the
theory exhibits non-locality with $(CPT)^2=-\mathbb{I}$. We briefly
discuss its relevance to the cosmological `horizon problem'.  
Because the introduced fermionic field is endowed with mass dimension
one, it can carry a quartic self-interaction.
Its dominant
interaction with known forms of matter is via Higgs, and with gravity.
This aspect leads us to contemplate the
new fermion as a
prime dark matter candidate. 
Taking this suggestion seriously we study a
supernova-like explosion of a galactic-mass dark matter cloud to set
limits on the mass of the new particle and present a calculation
on relic abundance to constrain the relevant cross-section. 
The analysis favours light mass (roughly $20\; \mbox{MeV}$)
and  relevant cross-section of about $2\;\mbox{pb}$. 
Similarities and differences with the WIMP and mirror matter proposals for
dark matter are enumerated. In a critique of the theory we
bare a hint on non-commutative aspects of spacetime, and
energy-momentum space.
}


\preprint{\textbf{Journal ref: JCAP 07 (2005) 012 }}

\dedicated{To celebrate the birthday of my wife
Dr I S Ahluwalia-Khalilova \\
and in memory of my father 
Shri B  S  Ahluwalia (1933-1977) ~\textendash~ dva-k.}

\begin{document}

\addtocounter{footnote}{2}

\def\beq{\begin{eqnarray}}
\def\eeq{\end{eqnarray}}

\def\ua{\{-,+\}}
\def\da{\{+,-\}}

\def\defn{\buildrel \rm def \over =}


\def\s{\mbox{\boldmath$\displaystyle\mathbf{\sigma}$}}
\def\J{\mbox{\boldmath$\displaystyle\mathbf{J}$}}
\def\K{\mbox{\boldmath$\displaystyle\mathbf{K}$}}
\def\A{\mbox{\boldmath$\displaystyle\mathbf{A}$}}
\def\B{\mbox{\boldmath$\displaystyle\mathbf{B}$}}

\def\ri{\mathrm{i}} 

\def\b{\mbox{\boldmath$\displaystyle\mathbf{b}$}}
\def\Jc{\mbox{\boldmath$\displaystyle\mathbf{\mathcal{J}}$}}

\def\P{\mbox{\boldmath$\displaystyle\mathbf{P}$}}
\def\p{\mbox{\boldmath$\displaystyle\mathbf{p}$}}
\def\v{\mbox{\boldmath$\displaystyle\mathbf{v}$}}
\def\r{\mbox{\boldmath$\displaystyle\mathbf{r}$}}
\def\n{\mbox{\boldmath$\displaystyle\mathbf{n}$}}
\def\l{\mbox{\boldmath$\displaystyle\mathbf{l}$}}
\def\hp{\mbox{\boldmath$\displaystyle\mathbf{\widehat{\p}}$}}

\def\hr{\mbox{\boldmath$\displaystyle\mathbf{\widehat{\r}}$}}

\def\x{\mbox{\boldmath$\displaystyle\mathbf{x}$}}
\def\y{\mbox{\boldmath$\displaystyle\mathbf{y}$}}
\def\0{\mbox{\boldmath$\displaystyle\mathbf{0}$}}
\def\bv{\mbox{\boldmath$\displaystyle\mathbf{\varphi}$}}
\def\hbv{\mbox{\boldmath$\displaystyle\mathbf{\widehat\varphi}$}}

\def\bn{\mbox{\boldmath$\displaystyle\mathbf{\nabla}$}}

\def\bl{\mbox{\boldmath$\displaystyle\mathbf{\lambda}$}}
\def\bl{\mbox{\boldmath$\displaystyle\mathbf{\lambda}$}}
\def\br{\mbox{\boldmath$\displaystyle\mathbf{\rho}$}}
\def\bfhh{\mbox{\boldmath$\displaystyle\mathbf{(1/2,0)\oplus(0,1/2)}\,\,$}}

\def\bg{\mbox{\boldmath$\displaystyle\mathbf{\gamma}$}}
\def\bbz{\mbox{\boldmath$\displaystyle\mathbf{\beta}$}}
\def\bnz{\mbox{\boldmath$\displaystyle\mathbf{\nu}$}}

\def\mn{\mbox{\boldmath$\displaystyle\mathbf{\nu}$}}
\def\amn{\mbox{\boldmath$\displaystyle\mathbf{\overline{\nu}}$}}

\def\mne{\mbox{\boldmath$\displaystyle\mathbf{\nu_e}$}}
\def\amne{\mbox{\boldmath$\displaystyle\mathbf{\overline{\nu}_e}$}}
\def\rlh{\mbox{\boldmath$\displaystyle\mathbf{\rightleftharpoons}$}}

\def\wm{\mbox{\boldmath$\displaystyle\mathbf{W^-}$}}
\def\hh{\mbox{\boldmath$\displaystyle\mathbf{(1/2,1/2)}$}}
\def\h00h{\mbox{\boldmath$\displaystyle\mathbf{(1/2,0)\oplus(0,1/2)}$}}
\def\znbb{\mbox{\boldmath$\displaystyle\mathbf{0\nu \beta\beta}$}}

\def\cO{\mbox{\boldmath$\displaystyle\mathbf{\mathcal{O}}$}} 

\def\rb{\kappa^{\left(1/2,0\right)}}
\def\lb{\kappa^{\left(0,1/2\right)}}

\def\rf{\frac{m^2+p^2-E^2}{(m-p+E)(m+p+E)\sqrt{2 m (m+E)}}}
\def\aa{m+E+p\cos(\theta)}
\def\bb{m+E-p\cos(\theta)}
\def\ab{p\, e^{-i\phi}  \sin(\theta)}
\def\ba{p \,e^{i\phi}  \sin(\theta)}

\def\f{\;\frac{1}{ \sqrt{2\,m\,E(\p)}}}
\def\fp{\;\frac{1}{ \sqrt{2\,m\,E(\p^\prime)}}}

\newcommand{\ecco}{Elko}
\def\es{{\textit{\ecco~}}}


\newcommand{\matD}{{\mathcal D}} 

\newcommand{\phase}{\Phi} 

\newcommand{\mawave}{{\mathcal G}} 

\newcommand{\locality}{{\mathcal E}} 



\section{Introduction}
\label{Sec:Introduction}

Stars, their remnants, and gas in galaxies, contribute no
more than $1\%$ of the total cosmic matter-energy content. 
Several per cent more is accounted for
by diffuse material pervading intergalactic space \cite{Nicastro:2004fb}. 
This inventory of cosmic baryons accounts for 
no more than $5\%$ of the universe.  
The problem was first brought to attention
as early as 1933 by Zwicky \cite{Zwicky:1933gu,Einasto:2004rf}.  
One now knows \cite{Rees:2004ta,Olive:2003iq}
that the deficit is accounted for by non-baryonic 
dark matter, $\sim 25 \%$, and some form 
of all pervading dark energy, $\sim 70 \%$.
That is, roughly 
$95\%$ of matter\textendash energy content of the universe is invisible
and  has no known, widely accepted, 
first-principle theoretical framework 
for its description.
Rees \cite{Rees:2004ta} has described this situation 
as `embarrassing'. The 
question we ask is: what is dark matter and why is it invisible? 
Here 
we show  that a quantum field based on dual-helicity 
eigenspinors of spin-one-half 
charge conjugation operator, i.e., the operator associated with
the particle\textendash antiparticle  symmetry, has precisely
the property called for by the dark matter.
In other words, we suggest that
whatever dark matter is, one thing that seems 
reasonably assured is that in the low-energy 
limit it behaves as one of the
representations of the Lorentz group. Since the known particles
are described
by quantum fields 
involving finite-dimensional representation spaces of usual
Wigner classes \cite{Wigner:1962bww,Lee:1966tdw,Weinberg:1995sw} 
\textemdash~
with certain questions about Higgs particles being deferred to
another place \cite{Ahluwalia:2000pj,Ruegg:2003ps} 
\textemdash~ the dark matter may belong to the 
yet unexplored unusual Wigner classes.

We do 
envisage the possibility that dark matter need not be confined 
to spin-one-half alone, even though the present paper focuses on this spin.
Furthermore, while a vast majority of the
physics community seems to be convinced of the existence of dark matter, it
is important to remain open to the possibility that in part, if not
in its entirety, 
the dark matter problem may be a reflection of
the growth of the Newtonian constant at astrophysical scales 
\cite{Reuter:2004nx} (see also  
\cite{Milgrom:1983pn,Sanders:2002pf,Ahluwalia:2002eh,Kiselev:2004vy,Bekenstein:2004ne,Merrifield:2004tr}).
Scientific caution suggests \cite{Visser:2004bf} that
existing data  be viewed with dark matter and modifications
of gravity at large scales
as complimentary  contributors to the same data.

From a formal theoretical point of view,
building on the classic works of Wigner \cite{Wigner:1939cj,Wigner:1962bww},
this paper provides an account of our attempt to
understand the particle content as implied by 
Poincar\'e spacetime symmetries. 
The literature on the subject has, so far, provided
valuable general insights \cite{Wigner:1962bww,Lee:1966tdw,Weinberg:1995sw}
but it lacks in specific constructs. Yet, a focus on specifics
can bring about important and unexpected insights which otherwise escape 
\cite{Ahluwalia:1993zt}. It is in this latter spirit also 
that this paper comes into existence. A condensed version containing some of the key results is available as \cite{Ahluwalia-Khalilova:2004sz}.

Another reason for which we venture
to make our research notes public is the following.  
The assumption of locality has
confined the physicists' focus
to only those Wigner classes for which the charge conjugation, $C$, 
and the parity, $P$, operators commute for bosons, and anticommute 
for fermions.
Yet  attempts to merge the quantum, the relativistic, and the gravitational 
realms immediately ask for an element of 
non-locality which may be realized for example in 
the framework of field theories on non-commutative spaces 
(for reviews, see for example \cite{Douglas:2001ba,Szabo:2001kg}).
Furthermore, attempts to reconcile LSND excess events  
\cite{Athanassopoulos:1996jb,Athanassopoulos:1997pv} indirectly suggest 
abandoning the locality requirement \cite{Ahluwalia:1998xb,Murayama:2000hm}. 
Such a suggestion would gain strength if MiniBooNE confirms \cite{Raaf:2004hx} 
the LSND result.
This combined circumstance should encourage us 
to take a cautious walk outside the
boundaries set by local relativistic-quantum field theories. 
Our first step in that realm constitutes preserving the Poincar\'e symmetries
but abandoning the demand of locality. This is done
by constructing and studying a 
quantum field based on the above-indicated eigenspinors of the 
spin-$\frac{1}{2}$ charge conjugation operator $C$.
We shall find that the quantum field so constructed
is rich in structure: it belongs to the Wigner class with
$[C,~P]=0$, its propagator is not that for the Dirac field,
and its mass dimension is one.

Initially, we did not set out to construct a field with the
properties outlined above, or a field which would be a candidate for
dark matter.  Instead, we were exposed to this structure when we took
an \textit{ab initio} look at the Majorana field.

\subsection{Genesis: from the Majorana field to a call for a
new dark matter field }

The spin-$\frac{1}{2}$ 
mass dimension one field came about as follows.
The Majorana field is obtained by identifying $b_h^\dagger(\p)$ with a
`phase factor $\times$ $ a_h^\dagger(\p)$', 
in the standard Dirac field 
\cite{Majorana:1937vz,Marshak:1961ecg,Weinberg:1995sw,Mohapatra:1991ng}

\beq
\Psi(x) &&= \int \frac{d^3 p}{{2\pi}^3}\frac{m}{p_0}
\sum_{h={+,-}} \left[
a_h(\p) u_h(\p) \mathrm{e}^{-\mathrm{i} p_\mu x^\mu}
+ b_h^\dagger(\p)  v_h(\p) \mathrm{e}^{+\mathrm{i} p_\mu x^\mu}\right]\,, 
\label{eq:Dirac}
\eeq
so that the charge-conjugated 
$\Psi(x)$, denoted by $\Psi^c(x)$, is \textit{physically indistinguishable}
from $\Psi(x)$ itself\footnote{We follow, 
unless stated otherwise, the notation of Ryder \cite{LHR1996}.}: 
\beq
\hspace*{-227pt}\Psi^c(x) = \mathrm{e}^{\mathrm{i} \beta} \Psi(x)\,,
\eeq
where $\beta \in \mathbb{R}$.
This single observation has inspired a whole generation of 
physicists to devote their entire academic lives to confirm experimentally
the realization of this suggestion \cite{Klapdor-Kleingrothaus:hans2001}.
After decades of pioneering work, 
the Heidelberg\textendash Moscow (HM)
collaboration  has, in the last few years, presented  
first experimental evidence \textemdash~
or, as some may prefer to say, tantalizing hints
\textemdash~
for a Majorana particle. 
The initial $3$-$\sigma$ signal now has better 
than  $4$-$\sigma$ significance 
\cite{Klapdor-Kleingrothaus:2001ke,Klapdor-Kleingrothaus:2002md,Klapdor-Kleingrothaus:2004gs,Klapdor-Kleingrothaus:2004ge}.
The field $\Psi(x)$ carries mass dimension three-halves.

Now, whether one is considering the Majorana field or the Dirac field,
both are based upon the Dirac spinors. In 1957, there was an effort
to reformulate the Majorana field \cite{Case1957,Mc1957}
in such a way that the new field was based upon  what are known as Majorana
spinors. It seems to have remained unasked as to what 
effect the choice for the 
helicity structure of the Majorana spinors has on the
physical content of the resulting field, and why
the same spinors should not be asked to satisfy an appropriate completeness
relation in the $(1/2,0)\oplus(0,1/2)$ representation space.
In the context
of generalization to higher spins, a preliminary exploration of
these issues emerged in  \cite{Ahluwalia:1993cz,Ahluwalia:1996uy}. 
The unexpected results that we present here arose 
when the present authors decided to take the research notes contained
in \cite{Ahluwalia-Khalilova:2003jt} to their logical conclusion.
It turned out, as the reader will read below, that a field based
on the dual-helicity eigenspinors of the 
spin-one-half charge conjugation
operator \textemdash~ constituting a significant extension of the original
Majorana idea   \textemdash~ did not carry the property required
for the identification with neutrinos. 

In the meantime, there has been progress on the experimental front.
While
the evidence for a \textit{Majorana particle} constitutes phenomenological 
realization of a quantum field which was never before known to
have been used by Nature,  the concurrent discovery \textemdash~awaiting
due confirmation by other groups \textemdash~
that there
exists a $6.3$ $\sigma$ DAMA-signal for \textit{dark matter} 
\cite{Bernabei:2003su} adds to the excitement.
 This also asks for a quantum field
beyond the Standard Model.  If the Majorana field was theoretically known 
since 1937, then, within the framework of known spacetime
symmetries \textemdash~ 
with parity, and
combined operation of charge conjugation and parity,
violated \textemdash~ 
there is no first-principle quantum field which fits the 
1933 Zwicky call of dark matter discovery.
Only now, some seven decades later, 
are the experiments, the observations, and the theory merging with
a call for a new (or a set of new) quantum field(s) 
which may attend to observations and experiments on dark matter.

In the context of extended spacetime symmetries,
supersymmetric partners of the Standard Model fields also
provide dark matter candidates (see, e.g.,  
\cite{Nath:1997qm,Dimopoulos:1996gy,Chattopadhyay:2001vx}); 
the most discussed
being neutralino (see, e.g., \cite{Bottino:2002ry,Feng:2000gh,Baer:1997ai}). 
But in doing that, one goes
beyond the experimentally observed spacetime symmetries.
If supersymmetry is  discovered at LHC in a few years then 
our proposal will compete for a `natural status' as 
a candidate for dark matter. 
Obviously, it is also conceivable that both supersymmetric partners 
\textit{and} the construct presented here may be the source of 
dark matter.
The remarks on mirror matter require a more detailed
discussion. These are  postponed to section \ref{Sec:mirrormatter}, while
the reader is referred to  \cite{Bertone:2004pz} for
a recent review on dark matter.

\subsection{The new spin-$\mathbf{\frac{1}{2}}$ quantum field}

In general, the charge and charge conjugation 
operators do not commute  \cite{Nigam:1956}. 
The Dirac particles are eigenstates of the charge operator.
This fact, combined with the circumstances
summarized above, 
suggests studying in detail the unexplored  Wigner classes.
The simplest of these is the spin-$\frac{1}{2}$ field 
\beq
&&
\hspace*{-75pt} 
\eta(x)=\int \frac{\mathrm{d}^3 p}{(2\pi)^3}\frac{1}
{\sqrt{2\, m\, E(\p)}} 
\sum_{\beta} \left[c_\beta(\p) \lambda^\mathrm{S}_\beta(\p) 
\mathrm{e}^{-\mathrm{i} p_\mu x^\mu} + d_\beta^\dagger(\p)  
\lambda^\mathrm{A}_\beta(\p) \mathrm{e}^{+ \mathrm{i} 
p_\mu x^\mu}\right]\,,\label{eq:4.36az} 
\eeq
where the $\lambda^\mathrm{S/A}(\p)$ are the dual-helicity
eigenspinors of the $(1/2,0)\oplus(0,1/2)$
charge conjugation 
operator (see section \ref{Sec:es}). 
We shall  abbreviate   $\lambda^\mathrm{S/A}(\p)$ as \es
for the following reason.
At the end of our path 
to obtain a meaningful and phonetically viable acronym,  we eventually 
settled for the German \es\hspace{-3pt}: {\bf E}igenspinoren des 
{\bf L}adungs{\bf k}onjugations{\bf o}perators.

As will be shown in detail below, the charged field $\eta(x)$ 
is different from that of Dirac.
On identification of 
$d_\beta^\dagger(\p)$ with $c_\beta^\dagger(\p)$ up to a possible phase, 
it yields a neutral
field which is different from that of Majorana.  
As already noted, our initial motivation was to offer 
$\eta(x)$ as a competing
candidate for the Majorana field. However, an extended and 
detailed analysis 
revealed that the new field, whether charged or neutral, carries
mass dimension one, and not three-halves. As such, it cannot be 
part of the $SU(2)_L$ doublets of the Standard Model 
which necessarily include
spin-$\frac{1}{2}$ 
particles of mass dimension three-halves. In other words, 
a description of neutrinos by $\eta(x)$, with
$d_\beta^\dagger(\p)$ and $c_\beta^\dagger(\p)$ identified with 
each other appropriately,
 results in the mixing of mass dimension
$\frac{3}{2}$ and $1$ spin-$\frac{1}{2}$ fermionic 
fields.\footnote{In this paper,
while referring to a quantum field,
we shall often take liberty of just saying `dimension' rather
than `mass dimension'. With a minor exception
in section \ref{Sec:ThirringLense}, all our considerations are confined
to the physical four-dimensional spacetime of special relativity.}  
So, we concluded that $\eta(x)$ is not a good candidate for
identification with the electroweak neutrinos\footnote{Although we 
will not consider this possibility in the current
 work it is to be noted that, in principle, a right-handed neutrino 
may be \es\hspace{-5 pt}, 
because it does not have a charge with respect to any of the 
Standard Model gauge groups and thus is a truly neutral particle. 
When coupling to the left-handed sector with Yukawa-like terms 
it is emphasized that the coupling constant in such terms will 
cease to be dimensionless; rather, it will have positive 
mass dimension 1/2. Naive power counting suggests 
that quartic \es terms may also  be  of importance. These considerations 
may be of relevance for the understanding of neutrino oscillations 
and neutrino mass generation and deserve a separate study. \label{fn:nuright}}.

Given the possible 
phenomenological and theoretical importance of the results obtained, 
a natural question may arise in the mind of our reader as to why
such a construct has not been undertaken before. 
One reason may be that any student of physics who wished 
to venture on such a journey would be immediately
discouraged by knowledgeable physicists citing an important 
1966 paper of Lee and Wick which essentially assures that any such
theory will be non-local \cite{Lee:1966tdw}. Yet, for the present
authors, that has not been a discouragement.
We simply set out
to look at an alternative to the Dirac construct for reasons we have already mentioned. 
The feature of neutrality allows one to argue naively why such fields are likely to exhibit non-locality: typically, what localizes otherwise extended field configurations like solitons is a conserved (topological) 
charge (see for example \cite{Rajaraman:1982is}). 
In the absence of such charges there is nothing that protects the 
`particle' from spreading and thus the emergence of 
non-locality it is not completely surprising\footnote{However, 
it is emphasized that in the context of solitons `non-locality' 
refers to a classical field configuration, 
while the non-locality encountered 
in the present work appears at the level of field (anti) commutators.}.
Concurrently, it may be noted
that today the  conventional wisdom 
has evolved to a position where
non-locality, and at times 
even $CPT$ violation,
is recognized as 
an expected part of a theory of quantum gravity
\cite{Ahluwalia:1998xb,Murayama:2000hm,Mavromatos:2004sz,Datta:2003dg}. 
For instance,
an argument can be made that at the interface of quantum
and gravitational realms spacetime must be non-commutative, and that
non-locality must be an integral part of any 
field-theoretic structure.
The simplest of these early 
arguments can be found, for example, in \cite{Ahluwalia:1993dd,VilelaMendes:1994zg,Doplicher:1994zv,Ahluwalia:2000iw}.
We shall be more concrete about 
these remarks in the concluding section.

\subsection{On the presentation of the paper}
 
The general plan of the paper is apparent from
the table of contents. 
Yet a few specifics may be in order.
To establish our notation, and to remind the reader
of the relation of the particle-antiparticle symmetry  with the
spacetime  symmetries, we present a brief review on the 
emergence of the  charge conjugation operator in section \ref{Sec:C}.
The next section presents  the dual-helicity
eigenspinors of the charge conjugation
operator, i.e., \textit{Elko}. 
An appropriate new dual for these spinors is introduced
in section \ref{SubSec:newdual}, while the associated orthonormality and 
completeness relations are the subject of a short section, \ref{SubSec:oc}. 
The action of the $(1/2,0)\oplus(0,1/2)$ parity operator on the 
\es is far from trivial, and we take some time to present the 
details in section \ref{Sec:P}. Apart from establishing that 
the charge conjugation and parity operators commute while 
acting on \textit{Elko}, 
we show that square of the parity operator on the \es basis
is not an identity operator; instead it is given by {\em minus} 
the identity operator.  
Similarly, section \ref{Sec:CPT} shows that while acting on \es the square
of the combined operation of charge conjugation, spatial parity,
and time reversal operators yields {\em minus} 
the identity operator.   
Section \ref{Sec:xt} is devoted
to a detailed examination of \es  at the 
representation-space level.
The dimension one aspect of the quantum field  based upon \es 
is presented in detail in section \ref{Sec:MassDim}.
The discussion of section \ref{Sec:xt} and  \ref{Sec:MassDim} 
also serves another useful purpose. It sheds additional 
light on the Dirac construct.
The statistics for the \es quantum field is the subject of
section \ref{Sec:Stat}.
Locality structure of the
theory is obtained in section \ref{Sec:ls}.
Section \ref{Sec:FurtherOnNonLocality}
outlines elements of $S$\textendash matrix theory for \es and 
briefly discusses relevance of the obtained non-locality
to the horizon problem of cosmology.
Section \ref{Sec:Identification}
is devoted to a possible identification of the \es framework
to dark matter. 
Section \ref{Sec:identification2} is focused  on  constraining the \es
mass and the relevant cross-section.
The presented construct carries some similarities, and
important differences, from the WIMP and mirror matter proposals.
This is the subject of sections \ref{Sec:Relic}--\ref{Sec:mirrormatter}.
The dual-helicity of \es states
gives rise to an important asymmetry. This is
discussed in section \ref{Sec:ThirringLense}.
The emergent \es non-locality is discussed in section \ref{Sec:cc},
which also contains a detailed critique,
and discussion pointing towards a non-commutative energy momentum
space on the one hand and a non-commutative spacetime on the 
other hand. 
Section \ref{Sec:summary0} provides a reference guide
to some of the key equations;  by following  
these equations a reader should be able to construct a 
rough and quick overview of the theoretical flow.
The unconventionally long section \ref{Sec:cc}
ends with a summary.  
A set of appendices provides
auxiliary details of calculations and some additional 
\textit{elkological} 
properties. 

In order not to allow the discussion to spread over too large
a technical landscape we have chosen to 
confine ourselves  to  the mass dimension $1$ neutral, 
rather than charged, field.
For a similar reason we shall confine ourselves to spin-$\frac{1}{2}$.
Yet, we shall phrase our arguments and presentation in such a
manner that the two-fold generalization, i.e., to higher spins and
to charged fields, will  be rendered obvious.

The subject matter at hand requires 
a somewhat pedagogic approach to the presentation. We follow this
demand without apology, even at the cost of seeming pedantic.
The reader is requested to reserve judgment until having 
read  the entire paper and is, in particular, 
asked to refrain from prematurely invoking any folklore.

\section{Emergence of the  charge conjugation 
operator: a brief review}
\label{Sec:C}

\subsection{The Dirac construct}

Both the Dirac and Majorana fields are built upon Dirac spinors.
A Dirac spinor, in Weyl representation,
is
\beq
\hspace*{-215pt}
\psi(\p) = \left(
\begin{array}{c}
\phi_\mathrm{R}(\p) \\
\phi_\mathrm{L}(\p)
\end{array}
\right)\,,
\eeq 
where the massive  Weyl  spinors $\phi_\mathrm{R}(\p)$ transform as
$(1/2,0)$ representation-space objects, and  massive  Weyl  
spinors $\phi_\mathrm{L}(\p)$
transform as  $(0,1/2)$ representation-space objects. 
The momentum\textendash space wave equation satisfied by $\psi(\p)$ thus
constructed follows uniquely \cite{LHR1996,Ahluwalia:1997lhr,Gaioli:1995ra} 
from  the interplay of 
$
\phi_\mathrm{R} (\0) = \pm \phi_\mathrm{L}(\0)  
$
and
$
\phi_\mathrm{R}(\p) = \rb \phi_\mathrm{R} (\0) $ $\;\mbox{\&}\;
\phi_\mathrm{L}(\p) = \lb \phi_\mathrm{L} (\0)
$, where
\beq
\hspace*{-60pt}
\kappa^{\left({1}/{2},0\right)} &=&\exp\left(+\, 
\frac{\s}{2}\cdot\bv\right)= \sqrt{\frac{E+m}{2\,m}}
\left(\mathbb{I}+\frac{\s\cdot\p}{E+m}\right)
\,,\label{br}\\
\hspace*{-60pt}\kappa^{\left(0,{1}/{2}\right)}&=&\exp\left(-\, 
\frac{\s}{2}\cdot\bv\right)= \sqrt{\frac{E+m}{2\,m}}
\left(\mathbb{I}-\frac{\s\cdot\p}{E+m}\right)\,.\label{bl}
\eeq
The  boost parameter, $\bv$, is  defined as
\beq
\hspace*{-105pt}
\cosh(\varphi)=\frac{E}{m},\quad
\sinh(\varphi)=\frac{\vert\p\vert}{m},\quad 
{\hbv}
=\frac{\p}{\vert \p\vert}\,;\label{bp}
\eeq
and because of the identity $\cosh^2\theta -\sinh^2\theta = 1$
encodes in it the dispersion relation 
\begin{equation}
  \label{eq:dr}
\hspace*{-248pt}  
E^2=\p^2+m^2\,.
\end{equation}
The implied wave equation is the momentum\textendash space Dirac 
equation\footnote{This result will be derived and also 
given an \textit{ab initio} and detailed attention in section \ref{Sec:xt}.}
\beq
\hspace*{-200pt}
\left(\gamma^\mu p_\mu \mp m \mathbb{I}\right)\psi(\p)=0\,.\label{zzz}
\eeq
Here, $\mathbb{I}$ are $n\times n$ identity matrices, 
their dimensionality being apparent from the context in which
they appear\footnote{So, for example, 
in equations (\ref{br}) and (\ref{bl}), the
 $\mathbb{I}$ stand for $2\times 2$ identity matrices, while
in equation (\ref{zzz})  $\mathbb{I}$  is a $4\times 4$ identity matrix.}.   
The $\gamma^\mu$ have
their standard Weyl-representation form:
\beq
\hspace*{-40pt}
\gamma^0= \left(
\begin{array}{cc}
\mathbb{O} & \mathbb{I} \\
\mathbb{I} & \mathbb{O}
\end{array}\right)
\,,\quad 
\gamma^i=
\left(
\begin{array}{cc} \mathbb{O} & - \sigma^i \\
\sigma^i & \mathbb{O}\end{array}
\right)\,,\quad
\gamma^5= \left(
\begin{array}{cc}
\mathbb{I} & \mathbb{O} \\
\mathbb{O} & -\mathbb{I}
\end{array}\right)
\,,\label{gammamatrices}
\eeq
with $\gamma^5:=-\mathrm{i}\gamma^0\gamma^1\gamma^2\gamma^3$.
For consistency of the notation, $\mathbb{O}$ 
here represents a $n\times n$ 
null matrix (in the above equation, $n=2$).
Obviously, the Dirac equation has four linearly independent solutions. 
Letting
$
p_\mu = \mathrm{i} \partial_\mu
$
\textit{and}, 
$
\psi(x) := \exp\left( \mp \mathrm{i}  p_\mu x^\mu\right)\psi(\p),
$
with upper sign for particles, and lower sign for antiparticles,
one obtains the configuration space Dirac equation:
\beq
\hspace*{-200pt}
\left(\mathrm{i} \gamma^\mu \partial_\mu -  m \mathbb{I}\right)\psi(x)=0\,.
\label{eq:pamd}
\eeq 

\subsection{Dirac's insight: not projecting out antiparticles}

\begin{quote}
\textit{\hspace*{20pt}One would thus be inclined to introduce, 
as a new assumption of the
theory, that only one of two kinds of motion occurs in practice. 
$\ldots$}\\ \vspace{36pt}
\hspace{196pt}{\tiny\sc P. A. M. Dirac, Nobel Lecture,\footnote{The quote is
from 
\cite{Dirac:1933pam}. Furthermore, it may be noted that 
Dirac's 
initial hesitation
to identify the associated particle with a new particle is well documented
by Schweber in \cite{Schweber:1994sss}. In brief:   reluctant to
introduce a new particle, Dirac initially identified the new
particle with the proton. Heisenberg,  Oppenheimer, Pauli, Tamm, 
and Weyl immediately saw that such
an identification was not tenable and the new particle must carry the 
same mass as electron, and opposite charge. By 1931 Dirac was to write
so himself: `A hole, if there were one, would be a new kind of particle,
unknown to experimental physics, having the same mass and opposite charge
 to an electron. We may call such a particle an anti-electron'.
The name `positron' was suggested to Anderson 
by Watson Davis (see \cite{Schweber:1994sss}).
In the 1933
Nobel lecture Dirac unambiguously writes: 
`There is one feature of these equations which I should
now like to discuss, a feature which led to the prediction of the 
positron'.} 1933}.
\end{quote}
Following insistence on `only two degrees of freedom for a 
spin one half\textendash particle', Dirac could have proposed
a constraint which projected out two of the four degrees of
freedom. 
The fact that he could have done so in a covariant manner
would have assured that
no one, or hardly any one, raised  an objection. 
Had Dirac taken that  path, a local
$U(1)$ gauge theory based on such a covariant framework
would have lacked physical viability.  
It would have missed Lamb shift \cite{Lamb:1947npp,Bethe1947lsp}, 
not to say antiparticles \cite{Anderson:1933mb,Blackett:1933pms}.  
The lesson is inescapable \cite{Kirchbach:2002nu,Kaloshin:2004jh}:
one should not impose mathematical constraints 
on a representation space
to obtain an interpretation which satisfies certain
empirically untested
physical intuitions, or prevalent folklore.
The physical intuition may ask for
avoiding the doubling of the degrees of freedom
or a folklore may demand a definite spin for
particles, etc. Such constraints may have a limited validity in
a classical framework. But in a quantum framework, the 
interactions will, in general, induce transitions 
between the classically allowed and the 
classically forbidden sectors unless
prohibited, by a conservation law, or a selection rule, 
for some reason. 
Here, we shall follow Dirac's insight and not project out
similar \textemdash~ i.e., anti-self-conjugate (see below) \textemdash~ 
degrees of freedom we shall encounter\footnote{This
seemingly logical position encounters an element of opposition 
when one applies 
it to a related problem of Rarita\textendash Schwinger field 
\cite{Kirchbach:2002nu} . 
In this latter context the suggestion is to consider as unphysical the
practice of projecting out the lower-spin 
components; and to, instead,
treat $\psi_\mu$ as a single physical field which carries spin-$\frac{3}{2}$
as well as spin-$\frac{1}{2}$ components. 
Apart from \cite{Kirchbach:2002nu}, 
recent work of Kaloshin and Lomov confirms our interpretation
\cite{Kaloshin:2004jh}.
}.

The derivation of the Dirac  equation as outlined here 
carries a quantum-mechanical aspect
in allowing for the fact that the two Weyl spaces may carry a relative
phase, in the sense made explicit above, and concurrently a relativistic
element via the Lorentz transformation properties of the Weyl spinors.
In turn the very existence of the latter depends on the existence of
two spacetime $SU(2)$s, with the following generators
of transformation:
\beq
\hspace*{-180pt}&& SU(2)_\mathrm{A}:\qquad \A= 
\frac{1}{2}\left(\J+ \mathrm{i}\K\right)\,,\\
\hspace*{-180pt}&& SU(2)_\mathrm{B}:\qquad \B= \frac{1}{2}
\left(\J- \mathrm{i}\K\right)\,.
\eeq
The $\J$ and $\K$ represent the generators of rotations and boosts, 
respectively, for  any of the relevant finite-dimensional 
representation spaces which may be under consideration.
For $\B=\0$, and $\J=\s/{2}$, we have the 
$(\frac{1}{2},0)$ right-handed
Weyl space, where $\K$ equals $ -\mathrm{i}(\s/{2})$.
For $\A=\0$, and $\J=\s/2$, we have the $(0,\frac{1}{2})$ 
left-handed Weyl space for which  $\K$ is $ + \mathrm{i} 
(\s/{2})$.

From the womb of this structure emerges a new symmetry, i.e., that
of charge conjugation. The operator associated with this symmetry 
is
\beq
\hspace*{-195pt}
{C} = 
\left(
\begin{array}{ccc}
\mathbb{O} & \hspace*{5pt}& \mathrm{i}\,\Theta \\
-\mathrm{i}\,\Theta & \hspace*{5pt}& \mathbb{O}
\end{array}
\right) {K}\,.\label{cc}
\eeq
Here, the operator $K$ complex conjugates any Weyl spinor that appears 
on its right, 
and  $\Theta$ is the Wigner's spin-$1/2$ time reversal 
operator. We use the representation
\beq
\hspace*{-228pt}
\Theta=
\left(
\begin{array}{cc}
0 & -1 \\
1 & 0
\end{array}
\right)\,.\label{wt}
\eeq
For an arbitrary spin it is defined 
by the property $\Theta \J \Theta^{-1}= -\,\J^\ast$.
Equation (\ref{cc}) is deliberately written 
in a slightly unfamiliar form. 
The chosen form is justified on 
the following grounds, and invites the remarks:

\begin{enumerate}
\item
Even for $j=1/2$ we refrain from  identifying 
$\Theta$ with `$-\,\mathrm{i}\, \sigma_2$,'
as is done implicitly in all considerations
on the subject \textemdash~ see, for example, \cite{PR1989} \textemdash~ 
because
such  an identification does not exist for higher-spin $(j,0)\oplus(0,j)$ 
representation spaces. The existence of a Wigner time-reversal
operator for all $j$ allows for 
the introduction of $(j,0)\oplus (0,j)$ \es representation spaces.
In this paper, however,  our attention is focused on $j=1/2$. 

\item
This form readily generalizes to 
higher spins. Furthermore,  as required
by the St\"uckelberg\textendash Feynman interpretation of antiparticles 
\cite{Stueckelberg:1941th,Feynman:1949hz}\footnote{It may be worth
noting that the St\"uckelberg\textendash Feynman 
interpretation of antiparticles 
ceases to be equivalent to the standard interpretation
in cosmological context \cite{Ahluwalia:2001md}.}, 
it makes the connection between 
particle\textendash antiparticle symmetry 
and time reversal operator manifest.

\end{enumerate}
Equation~(\ref{cc}) is readily seen to yield 
the standard form, $C=-\gamma^2 K$.
The boost
operator, $\rb\oplus \lb$, and the 
$(1/2,0)\oplus(0,1/2)$-space charge conjugation 
operator, $C$, commute:
\beq
\hspace*{-180pt}
\left[C, \;\rb\oplus \lb\right] = 0\,.
\eeq
This makes the notion of particle/antiparticle frame 
independent\footnote{\label{f:Weinberg}
However, in general, boosts do not leave the time-order of events unchanged.
This leads to interesting paradoxes, and again this necessitates
existence of antiparticles. This has been discussed elegantly 
in section 13 of chapter 2 of Weinberg's classic on gravitation and cosmology,
and since we cannot do a better job than that the reader is
referred to \cite{Weinberg:1972gcb}.}.

\textit{So, particles and antiparticles are offsprings of a 
fine interplay between
the quantum realm and the realm of spacetime symmetries.}
This brief review makes it transparent\footnote{
A more formal treatment of this result can be found in 
the classic work of Streater and Wightman \cite{Streater:1989vi}.}.

The operation of $C$ takes, up to a spinor-dependent 
global phase\footnote{The spinor dependence may be removed by 
appropriate redefinitions without changing the physical content 
of the theory.}, Dirac's particle spinors into Dirac's antiparticle 
spinors and vice versa \textemdash~ see equation (\ref{ceq}) below.
Keeping with our pedagogic style, we note:
the Dirac spinors are thus not eigenspinors of the charge conjugation
operator.

\section{Dual-helicity eigenspinors of  charge 
conjugation operator, or
Eigenspinoren des
Ladungskonjugationsoperators (\ecco) }
\label{Sec:es}

We have just summarized the origin and form of the 
charge conjugation operator. We now proceed
to obtain its eigenspinors. 
Towards this task one may take a direct and
purely mathematical
approach, or adopt a slightly indirect but physically 
insightful path. We shall follow the latter,
and will shortly argue that  
if $\phi_\mathrm{L}(\p)$ transforms as a left-handed spinor, then
$\left(\zeta_\lambda \Theta\right) \,\phi_\mathrm{L}^\ast(\p)$
transforms as a right-handed spinor \textemdash~ 
where $\zeta_\lambda$ is
an unspecified phase \textemdash~
with a similar assertion holding true for $\phi_\mathrm{R}(\p)$.
This allows us to define $(1/2,0)\oplus(0,1/2)$
spinors which are different from that of Dirac 
\textemdash~ which, of course, also belong to
the $(1/2,0)\oplus(0,1/2)$ representation space \textemdash~
and which become eigenspinors of the $C$ operator 
if $\zeta_\lambda$ is given some specific values.

\subsection{Formal structure of \ecco}
 
The details are as follows:
because the boost operators written in equations (\ref{br}), (\ref{bl}) 
are Hermitean and inverse to each other, we have
\beq
\hspace*{-30pt}
\left( \kappa^{\left(0,{1}/{2}\right)} \right)^{-1} = 
\left( \kappa^{\left({1}/{2},0\right)} \right)^\dagger\,,\qquad
\left( \kappa^{\left({1}/{2},0\right)} \right)^{-1} = 
\left( \kappa^{\left(0,{1}/{2}\right)} \right)^\dagger \, .
\label{identities}
\eeq
Further,   $\Theta$, the Wigner's spin-$1/2$ time reversal 
operator, has the property
\beq
\hspace*{-187pt}
\Theta \left[\s/2\right] \Theta^{-1} = -\, \left[\s/2\right]^\ast\,. 
\label{wigner}
\eeq
When combined, these observations imply that: 
(a)
if $\phi_\mathrm{L}(\p)$ transforms as a left-handed spinor, then
$\left(\zeta_\lambda \Theta\right) \,\phi_\mathrm{L}^\ast(\p)$
transforms as a right-handed spinor \textemdash~ where $\zeta_\lambda$ is
an unspecified phase;
(b)
if $\phi_\mathrm{R}(\p)$ transforms as a right-handed spinor, then
$\left(\zeta_\rho \Theta\right)^\ast \,\phi_\mathrm{R}^\ast(\p)$
transforms as a left-handed spinor \textemdash~  where $\zeta_\rho$ is
an unspecified phase. These results are in agreement
with Ramond's observation in \cite{PR1989}.
As a consequence, the following spinors 
belong to the $(1/2,0)\oplus(0,1/2)$
representation space:
\beq
\hspace*{-33pt}
\lambda(\p) =
\left(
\begin{array}{c}
\left(\zeta_\lambda \Theta\right) \,\phi_\mathrm{L}^\ast(\p)\\
\phi_\mathrm{L}(\p)
\end{array}
\right)\,,\qquad
\rho(\p)=
\left(
\begin{array}{c}
\phi_\mathrm{R}(\p)\\
\left(\zeta_\rho \Theta\right)^\ast \,\phi_\mathrm{R}^\ast(\p)
\end{array}
\right)\,.\label{eq:zimpok13}
\eeq
Confining ourselves to real eigenvalues (the demand of observability), 
these become  eigenspinors of the charge conjugation
operator with eigenvalues, $\pm 1$, 
if  the phases, $\zeta_\lambda$ and
$\zeta_\rho$, are restricted to the values
\beq
\hspace*{-200pt}
\zeta_\lambda= \pm\,\mathrm{i}\,,\qquad \zeta_\rho=\pm\,\mathrm{i}\,. 
\label{eq:zimpok14z}
\eeq
With this restriction imposed, we have
\beq
\hspace*{-125pt}
{C} \lambda(\p) = \pm  \lambda(\p)\,,\qquad
 {C} \rho(\p) = \pm  \rho(\p)\,. \label{eq:chconj}
\eeq 
The plus sign 
yields  \textit{self-conjugate spinors:} $\lambda^\mathrm{S}(\p)$
 and $\rho^\mathrm{S}(\p)$.
The minus sign results in the \textit{anti-self-conjugate spinors:} 
  $\lambda^\mathrm{A}(\p)$ and
$\rho^\mathrm{A}(\p)$.
To obtain explicit expressions for $\lambda (\p )$ we first write  
the rest spinors. These are
\beq
\hspace*{-50pt}
\lambda^\mathrm{S}(\0) = 
\left(
\begin{array}{c}
+\,\mathrm{i} \,\Theta \,\phi_\mathrm{L}^\ast(\0)\\
\phi_\mathrm{L}(\0)
\end{array}
\right)\,,\qquad
\lambda^\mathrm{A}(\0) = 
\left(
\begin{array}{c}
-\,\mathrm{i} \,\Theta \,\phi_\mathrm{L}^\ast(\0)\\
\phi_\mathrm{L}(\0)\, 
\end{array}
\right)\, .
\eeq
Next, we choose the $\phi_\mathrm{L}(\0)$ to be helicity eigenstates
\beq
\hspace*{-190pt}
\s\cdot{\hp} \;\phi_\mathrm{L}^\pm (\0)= \pm\;\phi_\mathrm{L}^\pm(\0)\,,
\label{x}
\eeq  
and concurrently note that
\beq
\hspace*{-147pt}
\s\cdot\hp \, \Theta \left[\phi_\mathrm{L}^\pm (\0)\right]^\ast
= \mp\, \Theta\left[\phi_\mathrm{L}^\pm(\0)\right]^\ast\,.
\label{y}
\eeq 
The derivation of equation (\ref{y})
is given in appendix \ref{app:a}, while
the explicit forms of $\phi^\pm_\mathrm{L}(\0)$ 
are given in appendix \ref{app:b}. 
The physical content of the result  (\ref{y}) is
the following: $\Theta \left[\phi_\mathrm{L}^\pm (\0)\right]^\ast$
has opposite helicity of $\phi_\mathrm{L}^\pm (\0)$.
Since $\s\cdot\hp$ commutes with the boost operator 
$\kappa^{\left(1/2,0\right)}$ this result holds for all $\p$. 

\subsection{Distinction between Elko and Majorana spinors}
\label{Sec:distinction}

So as not to obscure the physics by notational differences,
it is helpful to note \textemdash~ 
a choice we confine to this subsection only \textemdash~ 
that since $\mathrm{i} \Theta = \sigma_2$
we may write
\beq
\hspace*{-70pt}
\lambda(\p) =
\left(
\begin{array}{c}
\pm \sigma_2 \,\phi_\mathrm{L}^\ast(\p)\\
\phi_\mathrm{L}(\p)
\end{array}
\right)\,,\qquad
\rho(\p)=
\left(
\begin{array}{c}
\phi_\mathrm{R}(\p)\\
\mp\sigma_2 \,\phi_\mathrm{R}^\ast(\p)
\end{array}
\right)\,, \nonumber
\eeq
where the upper sign is for self-conjugate spinors, and the lower
sign yields the anti-self-conjugate spinors.
We now have a choice in selecting the helicity of the $(1/2,0)$ and
$(0,1/2)$ components of $\lambda(\p)$.
 We find that 
this choice has important physical consequences for reasons
which parallel Weinberg's detailed 
analysis of Dirac spinors (see section 5.5 of \cite{Weinberg:1995sw}).
In particular, as we shall confirm, that
the choice affects the parity and locality properties of
the constructed field.
For the moment it suffices to note that
if one chooses the helicity for the $(1/2,0)$ and $(0,1/2)$ components
to be \textit{same}, then the $\lambda(\p)$ are characterized by a
single-helicity and become identical to the standard Majorana spinors
(see, e.g., \cite{Peskin:1995mep,Marshak:1961ecg}).  This choice
violates the spirit of the result contained in equation (\ref{y}).  We
fully respect the spirit and the content of the result contained in
equation (\ref{y}) and therein lies our point of departure from Majorana
spinors.  That is, for our \es we start with the $(0,1/2)$ component
$\phi_\mathrm{L}(\p)$ in one or the other helicity. Then, 
when constructing the
$(1/2,0)$ component, $\pm\, \mathrm{i} \,\Theta\, \phi_\mathrm{L}
^\ast(\p)$ (or,
equivalently $\pm \sigma_2 \,\phi_\mathrm{L}^\ast(\p)$), we take the same
original $\phi_\mathrm{L}(\p)$ in the same helicity, i.e., we do not flip its
helicity by hand. This causes the $(1/2,0)$ transforming component to
carry the opposite helicity to that of  the original $\phi_\mathrm{L}(\p)$.
This is dictated by equation (\ref{y}).  For this reason \es we consider
are dual-helicity objects.

Similar remarks apply to $\rho(\p)$, which incidentally do
not constitute an independent set of \textit{Elko}\footnote{Section
\ref{Sec:distinction} was added to the manuscript 
as an answer to remarks by E. C. G. Sudarshan \cite{Sudarshan:2004ecg}. }.

\subsection{Explicit form of \ecco}

Having thus seen the formal structure of \es it is 
now useful to familiarize oneself by constructing them in
their fully explicit form.

The results of the above discussion lead to four rest spinors. 
Two of which are self-conjugate, 
\beq
\hspace*{-55pt}
\lambda_{\{-,+\}}^\mathrm{S}(\0) = 
\left(
\begin{array}{c}
+\,\mathrm{i} \,\Theta \,\left[\phi^+_\mathrm{L}(\0)\right]^\ast\\
\phi^+_\mathrm{L}(\0)
\end{array}
\right)\,,\qquad
\lambda_{\{+,-\}}^\mathrm{S}(\0) = 
\left(
\begin{array}{c}
+\,\mathrm{i} \,\Theta \,\left[\phi^-_\mathrm{L}(\0)\right]^\ast\\
\phi^-_\mathrm{L}(\0)\, 
\end{array}
\right)\, , \qquad \label{ls}
\eeq
and the other two are anti-self-conjugate,
\beq
\hspace*{-80pt}
\lambda_{\{-,+\}}^\mathrm{A}(\0) = 
\left(
\begin{array}{c}
-\,\mathrm{i} \,\Theta \,\left[\phi^+_\mathrm{L}(\0)\right]^\ast\\
\phi^+_\mathrm{L}(\0)
\end{array}
\right)\,,\qquad
\lambda_{\{+,-\}}^\mathrm{A}(\0) = 
\left(
\begin{array}{c}
-\,\mathrm{i} \,\Theta \,\left[\phi^-_\mathrm{L}(\0)\right]^\ast\\
\phi^-_\mathrm{L}(\0)\, 
\end{array}
\right)\, .\label{la}
\eeq
The first helicity entry refers to the $(1/2,0)$ transforming component of the
 $\lambda(\p)$, while the second entry encodes the helicity of
the $(0,1/2)$  component. 
The boosted spinors are now obtained via the operation
\beq
\hspace*{-94pt}
\lambda_{\{h,-h\}}(\p)=\left(
\begin{array}{cc}
\kappa^{\left({1}/{2},0\right)} & \mathbb{O} \\
\mathbb{O} & \kappa^{\left(0,{1}/{2}\right)}
\end{array}
\right)\lambda_{\{h,-h\}}(\0)\,.\label{z}
\eeq
In the boosts, we replace $\s\cdot\p$ by $p\, \s\cdot{\hp}$,
and then exploit equation (\ref{y}). After simplification,
equation (\ref{z}) yields
\beq
\hspace*{-77pt}
\lambda_{\{-,+\}}^\mathrm{S}(\p)
=
\sqrt{\frac{E+m}{2\,m}}\left(1-\frac{ p}{E+m}\right)
\lambda_{\{-,+\}}^\mathrm{S}(\0)\,,\label{lsup}
\eeq
which, in the massless limit, {\em identically vanishes,\/}
 while in the same limit
\beq
\hspace*{-81pt}
\lambda_{\{+,-\}}^\mathrm{S}(\p)
=
\sqrt{\frac{E+m}{2\,m}}\left(1+\frac{ p}{E+m}\right)
\lambda_{\{+,-\}}^\mathrm{S}(\0)
\label{lsdown}
\eeq
does not.
We hasten to warn the reader that one should not be tempted to read the
two different pre-factors to $\lambda^\mathrm{S}(\0)$
in the above expressions as the boost operator that appears in equation
(\ref{z}). For one thing, there is only one (not two) 
boost operator(s) in the
$(1/2,0)\oplus(0,1/2)$ representation space. The simplification that
appears here is due to a fine interplay between equation (\ref{y}), the boost
operator, and the structure of the $\lambda^\mathrm{S}(\0)$.
Similarly, the anti-self-conjugate set of the boosted spinors reads
\beq
\hspace*{-95pt}
&&\lambda_{\{-,+\}}^\mathrm{A}(\p)
=
\sqrt{\frac{E+m}{2\,m}}\left(1-\frac{p}{E+m}\right)
\lambda_{\{-,+\}}^\mathrm{A}(\0)\,,\label{laup}\\
\hspace*{-95pt}&&\lambda_{\{+,-\}}^\mathrm{A}(\p)
=
\sqrt{\frac{E+m}{2\,m}}\left(1+\frac{ p}{E+m}\right)
\lambda_{\{+,-\}}^\mathrm{A}(\0)\,.\label{ladown}
\eeq
In the massless limit, the first of these spinors 
{\em identically vanishes\/}, while the second does not.

\subsection{A new dual for \ecco}
\label{SubSec:newdual}

For any $(1/2,0)\oplus(0,1/2)$ spinor  $\xi(\p)$, 
the  Dirac dual spinor $\overline{\xi}(\p)$ is defined as
\beq
\hspace*{-230pt}
\overline{\xi}(\p) := \xi^\dagger(\p) \gamma^0\,. \label{eq:dico}
\eeq 
With respect to the Dirac dual, the 
\es
have an imaginary  
bi-orthogonal 
norm as was already noted in
\cite{Ahluwalia:1993cz,Ahluwalia:1996uy}.
For the sake of a ready reference, 
this is recorded explicitly in  appendix \ref{app:c}. 
The imaginary norm of \es
is a hindrance to physical interpretation and quantization.
Enormous simplification of interpretation and calculation
occurs if we define a new dual with respect to which \es
have a real norm.
The new dual must have the 
property that: (a) it yields an invariant real definite norm, and
(b) in addition, it must 
must secure a  positive-definite norm for two of
the four  \textit{Elko}'s, 
and negative-definite norm
for the remaining two. Any other choice will introduce 
an unjustified element of asymmetry. 
 Up to a relative sign, a unique 
definition of such a dual, which we call 
{\em \es dual,\/} is
\beq
\hspace*{-135pt}&& \lambda^\mathrm{S}(\p): \qquad  
\stackrel{\neg}\lambda^\mathrm{S}
_{\pm,\mp}(\p)
:= + \left[\rho^\mathrm{A}_{\mp,\pm}(\p)
\right]^\dagger \gamma^0 \,, \label{eq:madu1} \\
\hspace*{-135pt}&&\lambda^\mathrm{A}(\p):  \qquad  
\stackrel{\neg}\lambda_{\pm,\mp}^
\mathrm{A}(\p):=
- \left[\rho^\mathrm{S}_{\mp\pm}(\p)\right]^\dagger 
\gamma^0 \,, \label{eq:madu2} 
\eeq
where the $\rho(\p)$ are given in appendix \ref{app:d}. 

The \es dual can also 
be expressed in the following
equivalent, but very useful, form:
\beq
\hspace*{-115pt}
\mbox{\sc {\ecco}\hspace{6pt}Dual:}\qquad
\stackrel{\neg}\lambda_\alpha(\p):= \mathrm{i}\, \varepsilon_\alpha^\beta\,
\lambda_\beta^\dagger(\p)\,\gamma^0\,,\label{eq:md}
\eeq
with the antisymmetric symbol $\varepsilon^{\{-,+\}}_{\{+,-\}}:= -1=-\varepsilon^{\{+,-\}}_{\{-,+\}}$.
The upper and lower position of indices has been chosen only to avoid expressions like $\varepsilon_{\{+,-\}\{-,+\}}$ and not to imply the use of a metric to raise and lower indices.
Equation \eqref{eq:md} holds for self-conjugate 
as well as anti-self-conjugate $\lambda(\p)$.
The Dirac dual, for comparison, may then be re-expressed in the
following equivalent form:
\beq
\hspace*{-120pt}
\mbox{\sc Dirac Dual:}\qquad
\overline{\psi}_h(\p):=
\delta_h^{h^\prime}\, \psi^\dagger_{h^\prime}(\p)\,\gamma^0\,,
\eeq
where $\psi(\p)$ represents any of the four
Dirac spinors and $\delta_h^{h^\prime}$ is the Kronecker symbol.
Explicitly, equation (\ref{eq:md}) yields
\beq
\hspace*{-165pt}&& \stackrel{\neg}\lambda^\mathrm{S/A}_{\{-,+\}}(\p)
= + \,\mathrm{i}\,\left[
\lambda^\mathrm{S/A}_{\{+,-\}}(\p)\right]^\dagger\gamma^0\,,\\
\hspace*{-165pt}&& \stackrel{\neg}\lambda^\mathrm{S/A}_{\{+,-\}}(\p)
= -\, \mathrm{i}\,\left[
\lambda^\mathrm{S/A}_{\{-,+\}}(\p)\right]^\dagger\gamma^0\,,
\eeq
which, on use of results given in appendix \ref{app:d},
shows these to be equivalent to definitions (\ref{eq:madu1})
and  (\ref{eq:madu2}). We have belaboured this point as different
expression are useful in various contexts.

\subsection{Orthonormality and completeness relations for \ecco}
\label{SubSec:oc}

With the \es dual thus defined, we  now have (by construction)
\beq
\hspace*{-190pt}&&\stackrel{\neg}\lambda^\mathrm{S}_{\alpha}(\p)\;
\lambda^\mathrm{S}_
{\alpha^\prime}(\p) = +\; 2 m\; \delta_{\alpha\alpha^\prime}\,,
\label{zd1}\\
\hspace*{-190pt}&&\stackrel{\neg}\lambda^\mathrm{A}_{\alpha}(\p)\;
\lambda^\mathrm{A}_{\alpha^\prime}(\p) = 
-\; 2 m \;\delta_{\alpha\alpha^\prime}\,.
\label{z5}
\eeq
The subscript $\alpha$ ranges  over two
possibilities: $ \{+,-\}, \{-,+\}$. 
The completeness relation
\beq
\hspace*{-102pt}
\frac{1}{2 m}\sum_\alpha 
 \left[\lambda^\mathrm{S}_{\alpha}(\p) \stackrel{\neg}\lambda^\mathrm{S}
_{\alpha}(\p) 
      - \lambda^\mathrm{A}_{\alpha} (\p)\stackrel{\neg}\lambda^\mathrm{A}
_{\alpha}(\p)\right]
 = \mathbb{I}\,, \label{z1}
\eeq
clearly shows the necessity of the anti-self-conjugate
spinors.
Equations (\ref{zd1})\textendash (\ref{z1}) have their 
direct counterpart in Dirac's construct:
\beq
\hspace*{-200pt}&& \overline{u}_h(\p)\, u_{h^\prime}(\p) =  +\;2 m\;
 \delta_{h {h^\prime}}\,,\label{d1}\\
\hspace*{-200pt}&&\overline{v}_h(\p)\, v_{h^\prime}(\p) =  -\;2 m\;
 \delta_{h {h^\prime}}\,,\label{5}
\eeq
and
\beq
\hspace*{-104pt}
 \frac{1}{2 m} \sum_{h=\pm 1/2}
\Big[
 u_{h}(\p) \overline{u}_h(\p) -
  v_{h}(\p) \overline{v}_h(\p)\Big] = \mathbb{I}\,.\label{1}
\eeq

\section{Establishing $\mathbf{\left(C P T\right)^2 = -} \mathbf{\mathbb{I}}$
for Elko}\label{Sec:cpt}

In this section we present the detailed 
properties of \es spinors under the operation of spatial
parity. This prepares us to show that the 
square of the combined operation of charge conjugation, spatial  parity,
and time-reversal operators, when acting upon the \textit{Elko}, 
meets the expectations of Wigner.

\subsection{Commutativity of C and P, and  parity asymmetry}
\label{Sec:P}

To set the stage for this section we begin by quoting
the  unedited textbook wisdom\cite{Ryder:1986mc}:
\beq
\hspace*{-30pt}
\left\{
\begin{array}{l}
\mbox{bosons: particle and antiparticle have same parity} \\
\mbox{fermions:  particle and antiparticle have opposite parity.}
\end{array}
\right\} 
\eeq
To our knowledge the only textbook which tells a more intricate story is
that by Weinberg \cite{Weinberg:1995sw}. The only known
explicit construct of a theory which challenges the 
conventional wisdom was reported only about a decade ago in 1993 
\cite{Ahluwalia:1993zt}. In that pure spin-$1$ bosonic theory particles 
and antiparticles carry opposite, rather than same, relative 
intrinsic parity. 
It manifests itself 
through the anticommutativity, as opposed to the commutativity, 
of the $(1,0)\oplus(0,1)$-space's 
charge conjugation and parity operators. 
In a somewhat parallel fashion we shall now show that 
for the spin-$\frac{1}{2}$ \es  the charge conjugation operator
and parity operator commute, rather than anticommute (as they do
for the Dirac case). We shall have more to say about these matters
in the concluding section where we bring to our reader's
attention the classic work of Wigner \cite{Wigner:1962bww}, 
and that of Lee and Wick \cite{Lee:1966tdw}.

Given these remarks it does not come as a surprise
that the parity operation is slightly subtle for 
\textit{Elko}. 
In the $(1/2,0)\oplus(0,1/2)$ representation space it reads
\beq
\hspace*{-250pt}
P= \mathrm{e}^{\mathrm{i}\phase} \gamma^0 {\mathcal R}\,.
\eeq
With $\p:=
p \,\left(\sin(\theta)\cos(\phi), \sin(\theta)\sin(\phi), 
\cos(\theta)\right)$,
the ${\mathcal R}$ reads
\beq
\hspace*{-135pt}
{\mathcal R} \equiv
\left\{ \theta\rightarrow\pi-\theta, \;\phi\rightarrow \phi+\pi,\;
p\rightarrow p\right\}\,.
\eeq
This has the consequence that eigenvalues, $h$,  of the helicity
operator $\s\cdot\hp / 2$ 
change sign under the operation of $\mathcal R$:
\beq
\hspace*{-232pt}
{\mathcal R}: h \rightarrow h^\prime = - h\,.
\eeq
Furthermore, while acting on the Dirac spinors,
\beq
\hspace*{-50pt}
P u_h(\p) = \mathrm{e}^{\mathrm{i}\phase} \gamma^0 {\mathcal R} u_h(\p) =
 \mathrm{e}^{\mathrm{i}\phase} \gamma^0 u_{-h}(-\p) = 
- \mathrm{i} e^{\mathrm{i}\phase} u_h(\p)\,.
\eeq  
Similarly,
\beq
\hspace*{-220pt}
P v_h(\p) =\mathrm{i} \mathrm{e}^{\mathrm{i}\phase} v_h(\p)\,.
\eeq
Because for the theory  based upon Dirac spinors relative
intrinsic parity is an observable,
we must require the eigenvalues of $P$ to be real. This
fixes the phase
\beq
\hspace*{-270pt}
\mathrm{e}^{\mathrm{i}\phase} = \pm \mathrm{i}\,. \label{pf}
\eeq
The remaining ambiguity, as contained in the sign, 
still remains.  This
ambiguity does not affect the physical consequences.
It is  fixed by recourse to text-book 
convention by taking the sign on the right-hand side
of equation (\ref{pf}) to be positive.
The parity operator is therefore fixed to be
\beq
\hspace*{-263pt}
P= \mathrm{i} \gamma^0 {\mathcal R}\,.
\eeq
Thus 
\beq
\hspace*{-233pt}&& P u_h(\p) = +\, u_h(\p)\,,\label{peqa}\\
\hspace*{-233pt} && P v_h(\p) = -\, v_h(\p)\,.\label{peqb}
\eeq 
That is, Dirac spinors are  eigenspinors of the 
parity operator.
Equations (\ref{peqa}) and (\ref{peqb}) imply
\beq
\hspace*{-70pt}
\mbox{\sc Dirac Spinors}: \quad P^2= \,\mathbb{I}\,,\quad
[\mbox{cf equation (\ref{cf2})}]\,.\label{cf1}
\eeq 

To calculate the anticommutator, $\{C,P\}$, when acting on
the $u_h(\p)$ and $v_h(\p)$ we now need, in addition,
the action of $C$ on these spinors. This action
can be summarized as follows:
\beq
C :\qquad{\Bigg\{}\begin{array}{l}
  u_{+1/2}(\p) \rightarrow - v_{-1/2}(\p)\,,
 \hspace{36pt} u_{-1/2}(\p) \rightarrow   v_{+1/2}(\p)\,,\\
  v_{+1/2}(\p) \rightarrow   u_{-1/2}(\p) \,,
 \hspace{44pt} v_{-1/2}(\p) \rightarrow - u_{+1/2}(\p)\,.\label{ceq}
\end{array}
\eeq
Using equations (\ref{peqa}), (\ref{peqb}),  
and (\ref{ceq}) one can readily obtain the 
action of the anticommutator, $\{C,P\}$, on the four $u(\p)$ and 
$v(\p)$ spinors. For each case it is found to vanish: 
\beq
\hspace*{-52pt}
\mbox{\sc Dirac Spinors}: \qquad
\{C,P\}=0\,,\quad
[\mbox{cf  equation (\ref{eq:cpelko})}]\,. \label{eq:dcp}
\eeq 
\noindent
The $P$ acting on the  \es
yields the result
\beq
\hspace*{-25pt}
P\lambda^\mathrm{S}_{\ua} (\p)= +\, \mathrm{i}\, 
\lambda^\mathrm{A}_{\da}(\p)\,,\qquad
P\lambda^\mathrm{S}_{\da} (\p)= -\, \mathrm{i} 
\,\lambda^\mathrm{A}_{\ua}(\p)\,,&&\label{peqla}\\
\hspace*{-25pt}
P\lambda^\mathrm{A}_{\ua} (\p)= - \,\mathrm{i} 
\,\lambda^\mathrm{S}_{\da}(\p)\,,\qquad
P\lambda^\mathrm{A}_{\da} (\p)= + \,\mathrm{i} 
\,\lambda^\mathrm{S}_{\ua}(\p)\,.\label{peqlb}
\eeq
That is, \es are \textit{not} eigenspinors of the 
parity operator.
Following the same procedure as before, we now use  (\ref{peqla}), 
(\ref{peqlb}), and (\ref{eq:chconj}) 
\textemdash~ taking a special note of equation \eqref{cc} \textemdash~
to evaluate the action of the commutator $[C,P]$ 
on each of the four  \textit{Elkos}.
We find that it vanishes for 
each of them: 
\beq
\hspace*{-102pt}
\mbox{\sc \ecco}: \qquad
[C,P]=0\,,\quad[\mbox{cf  equation (\ref{eq:dcp})}]\,. \label{eq:cpelko}
\eeq 

The commutativity and anticommutativity of the $C$ and $P$ operators
is an important distinction between the
Dirac spinors and the \textit{Elko}. 
In this aspect,
our results coincide with the possibilities offered by 
Wigner's general analysis  \cite{Wigner:1962bww}. Despite 
similarities, our construct 
differs from the Wigner\textendash Weinberg 
analysis in a crucial aspect. We outline
this in section  \ref{Sec:Wigner-Weinberg}. Yet, this difference
does not seem to affect many of the general conclusions. 
Even though a full formal generalization of the Wigner\textendash Weinberg
analysis may be desirable, our specific construct
does not require it.

Unlike the Dirac spinors, as already noted,  
equations (\ref{peqla}) and (\ref{peqlb}) 
reveal that \es are not eigenstates of $P$.   
Furthermore, an apparently  
paradoxical asymmetry is contained in these equations.
For instance, the second equation in (\ref{peqla}) reads
\beq
\hspace*{-180pt}
P\lambda^\mathrm{S}_{\da} (\p)= -\, 
\mathrm{i} \,\lambda^\mathrm{A}_{\ua}(\p)\,.
\eeq
As a consequence of (\ref{lsdown}) and (\ref{laup}), 
in the massless/high-energy limit the $P$-reflection of 
$\lambda^\mathrm{S}_{\da} (\p)$ identically vanishes. 
The same happens to the $\lambda^\mathrm{A}_{\da} (\p)$ spinors
under $P$-reflection. This situation is in sharp
contrast to the charged-particle spinors.
The origin of the asymmetry under $P$-reflection 
resides in the fact that the \textit{Elko}, 
in being dual-helicity objects, 
combine Weyl spinors of {\em opposite\/}
helicities. However,  in the massless limit, the structures of 
$\kappa^{\left({1}/{2},0\right)}$ and 
$\kappa^{\left(0,{1}/{2}\right)}$ 
force only positive-helicity $\left({1}/{2},0\right)$-Weyl 
and negative-helicity $\left(0,{1}/{2}\right)$-Weyl spinors
to be non-vanishing.
For this reason, in 
the massless limit the \textit{Elko}, $\lambda^\mathrm{S}_{\ua}(\p)$
and  $\lambda^\mathrm{A}_{\ua}(\p)$,
carrying negative-helicity $\left({1}/{2},0\right)$-Weyl 
and positive-helicity $\left(0,{1}/{2}\right)$-Weyl spinors
identically vanish.

 Furthermore, 
the consistency of  equations 
(\ref{peqla}) and (\ref{peqlb}) requires 
$P^2 = - \mathbb{I}$ and in the process shows that the remaining two, i.e.,
the first and the third equation in that set, do not contain additional
physical content:
\begin{equation}
\hspace*{-110pt}
\mbox{\sc \ecco}: \quad P^2=-\,\mathbb{I}\,.
\qquad
[\mbox{cf  equation (\ref{cf1})}]\,. \label{cf2}
\end{equation}

The $(1/2,0)\oplus(0,1/2)$ 
is a $P$ covariant representation space.  Yet, in the \es 
formalism, it carries
$P$-reflection asymmetry. 
This result has a similar precedence in the Velo-Zwanziger observation, who
noted \cite{Velo:1969bt}
`the main lesson to be drawn from our analysis is that special relativity
is not automatically satisfied by writing equations which transform 
covariantly'.

\subsection{Agreement with Wigner: $\mathbf{\left(C P T\right)^2 = -
 \mathbb{I}}$}
\label{Sec:CPT}

The time-reversal operator $T= \mathrm{i} \gamma^5 C$
acts on \es as follows:

\begin{equation}
\label{eq:time}
\hspace*{-110pt}
T \lambda_\alpha^\mathrm{S}(\p) = - \mathrm{i} 
\lambda_\alpha^\mathrm{A}(\p),\qquad  
T \lambda_\alpha^\mathrm{A}(\p) = + \mathrm{i} 
\lambda_\alpha^\mathrm{S}(\p)\,,
\end{equation}
implying $T^2=-\mathbb{I}$. 
With the action of all three of the $C,\;P$ and $T$ on \es now known,
one can immediately deduce that, in addition to 
(\ref{eq:cpelko}), we have
\beq
\hspace*{-150pt}
\mbox{\sc \ecco}: \quad
[C,T]=0\,,\quad \{P,T\}=0\,,
\eeq 
and that at the same time,
\beq
\hspace*{-242pt}
\left(C P T \right)^2 = -\mathbb{I}\,, \label{eq:cptsq}
\eeq
thus confirming Wigner's expectation. For a discussion of
differences with Weinberg, we refer the reader to section 
\ref{Sec:Wigner-Weinberg}.

\section{Spacetime evolution}
\label{Sec:xt}

The existing techniques to specify spacetime evolution 
do not fully suffice for \textit{Elko}.
The path we take carries its inspiration from standard quantum field
theory \cite{LHR1996,Weinberg:1995sw},
but in the end we had to develop much of the formalism
ourselves.  So, what follows
constitutes in large part our \textit{ab initio} effort.  

Section~\ref{nodirac} establishes that massive 
\es do not satisfy the Dirac equation. The next subsection, 
i.e., section \ref{sub:spinsums},
briefly reflects on the connection between 
`spin sums', wave operators, and propagators. 
The remaining three subsections are devoted to establishing
a contrast between \es and Dirac spinors. This exercise 
not only gives a sharper independent existence to \es
but it also sheds  new light on the well known Dirac spinors.

\subsection{Massive {\ecco} do not satisfy the Dirac equation}

\label{nodirac}

For the task at hand it is helpful to make the following local change in 
notation:
\beq
&&\hspace{-17pt} \mbox{\sc For  Dirac Spinors}:\qquad
u_+(\p)\rightarrow d_1, \,\;\;
u_-(\p) \rightarrow d_2, \,\;\;
v_+(\p)\rightarrow d_3, \,\;\;
v_-(\p)\rightarrow d_4\,.\nonumber \\ \\
&&\hspace{-17pt} \mbox{\sc For {\ecco}}:\qquad
\lambda^\mathrm{S}_{\ua}(\p)\rightarrow e_1, \,\;\;
\lambda^\mathrm{S}_{\da}(\p) \rightarrow e_2, \,\;\;
\lambda^\mathrm{A}_{\ua}(\p) \rightarrow e_3, \,\;\;
 \lambda^\mathrm{A}_{\da}(\p) \rightarrow e_4.\nonumber\\
\eeq
Adopting the procedure introduced in  \cite{Ahluwalia:xa},
the \es can now be written as
\beq
\hspace*{-155pt}
e_i= \sum_{j=1}^4 \Omega_{ij} d_j\,,\qquad i=1,2,3,4\,,\label{md}
\eeq
where
\begin{equation}
\hspace*{-127pt}
\Omega_{ij}=
 \begin{cases}
+\left({1}/{2 m}\right) \overline{d}_j\, e_i \mathbb{I}\,, & \qquad{\rm for}
\,\, j =1,2\,,\cr
-\left({1}/{2 m}\right) \overline{d}_j \,e_i \mathbb{I}\,, & \qquad{\rm for} 
\,\,j =3,4\,.\cr
\end{cases}
\end{equation}
In matrix form,  $\Omega$ reads
\beq
\hspace*{-180pt}
\Omega =
\frac{1}{2}\left(
\begin{array}{cccc}
\mathbb{I} & - \ri \mathbb{I} & - \mathbb{I} & - \ri\mathbb{I} \\
\ri\mathbb{I} & \mathbb{I} & \ri\mathbb{I} & - \mathbb{I} \\
\mathbb{I} & \ri\mathbb{I} & - \mathbb{I} & \ri\mathbb{I} \\
- \ri\mathbb{I} & \mathbb{I} & - \ri\mathbb{I} & - \mathbb{I}
\end{array}
\right)\,.\label{omega}
\eeq
\newcommand{\matB}{{\mathcal B}}
With the definition $\matB:=(\mathbb{I}+\sigma_2)$, 
equation (\ref{omega})
can be recast into the form\footnote{In what follows the second 
entry in the direct product always refers to the spinorial part, 
while the first one refers to the `$ij$' part as implied by (\ref{md}).}
\begin{equation}
  \label{eq:omega}
\hspace*{-212pt}
  \Omega =
\frac{1}{2}\left(
\begin{array}{cc}
\matB & -\matB^\ast \\
\matB^\ast & -\matB
\end{array}\right) 
\otimes \mathbb{I}\,.
\end{equation}
Equations~(\ref{md}) and (\ref{omega}) immediately tell us that each
of the
spinors in the set defined by \es is a linear combination
of the  Dirac {\em particle and antiparticle\/} 
spinors. In momentum space, the Dirac spinors 
are annihilated by $\left(\gamma^\mu p_\mu \pm m \mathbb{I}\right)$
\beq
\hspace*{-75pt}
\begin{cases}
\mbox{For particles:}
\qquad\left(\gamma^\mu p_\mu - m \mathbb{I}\right) u(\p)=0\,,\qquad\cr
\mbox{For antiparticles:}\qquad
\left(\gamma^\mu p_\mu + m \mathbb{I}\right) v(\p)=0\,.\cr
\end{cases}\label{deqs}
\eeq
That is, Dirac's $u(\p)$ and $v(\p)$
are eigenspinors of the $\gamma^\mu p_\mu$ operator with
eigenvalues $+m$ and $-m$, respectively \textemdash~
a fact emphasized  by Weinberg 
(see, page 225 of \cite{Weinberg:1995sw})
with the observation that it is a result of how the
$(1/2,0)$ and $(0,1/2)$ representation spaces have been 
put together to carry simple properties under spatial
reflection.
 Since the mass terms carry opposite signs, and  
hence are different for the particle and antiparticle, 
the \es 
cannot be annihilated by 
$\left(\gamma^\mu p_\mu - m \mathbb{I}\right)$, nor by
$\left(\gamma^\mu p_\mu + m \mathbb{I}\right)$. 
That is, they cannot be eigenspinors of the
the  $\gamma^\mu p_\mu$ operator. We shall make this
result more precise below.
Moreover,
since the time evolution of the of $u(\p)$ occurs via
$\exp(-\mathrm{i} p_\mu x^\mu)$ while that for   
$v(\p)$ spinors occurs via $\exp(+ \mathrm{i} 
p_\mu x^\mu)$, one cannot naively go from
momentum\textendash space expression (\ref{md}) 
to its configuration space counterpart.

For formal simplification, we introduce
\beq
\hspace*{-162pt}
e :=
\left(
\begin{array}{c}
e_1\\
e_2\\
e_3\\
e_4
\end{array}
\right)\,,\qquad
d :=
\left(
\begin{array}{c}
d_1\\
d_2\\
d_3\\
d_4
\end{array}
\right)\,,
\eeq
and 
\beq
\hspace*{-240pt}
\Gamma := \mathbb{I}\otimes \gamma_\mu p^\mu
\,.
\eeq
In this language, equation (\ref{md}) becomes
$e=\Omega d$.
Applying from the left the operator $\Gamma$ and using
$\left[\Gamma,\Omega\right] =0$ yields
\beq
\hspace*{-257pt}
\Gamma e = \Omega \Gamma d\,.
\eeq
But, equations (\ref{deqs}) imply
$\Gamma d = m\,\gamma^5\otimes\mathbb{I}\,d$.
Therefore, on using $d=\Omega^{-1} e$, we obtain
\beq
\hspace*{-200pt}
\Gamma e = 
\Omega \left( m\,\gamma^5\otimes\mathbb{I}\right) \Omega^{-1} e
\,.
\eeq
An explicit evaluation of
$\mu := \Omega \left( m\,\gamma^5\otimes\mathbb{I}\right) \Omega^{-1}$
reveals 
\beq
\hspace*{-200pt}
\mu= m\, \left(
\begin{array}{cc}
\sigma_2 & \mathbb{O} \\
\mathbb{O} & - \sigma_2
\end{array}
\right)\otimes\mathbb{I}\,.
\eeq
Thus, making the direct product explicit again, 
finally we reach the result
\beq
\hspace*{-90pt}
\left(
\begin{array}{cccc}
\gamma_\mu p^\mu & \mathbb{O} & \mathbb{O} & \mathbb{O} \\
 \mathbb{O} & \gamma_\mu p^\mu & \mathbb{O} & \mathbb{O}  \\
  \mathbb{O} & \mathbb{O}& \gamma_\mu p^\mu & \mathbb{O}   \\
  \mathbb{O} & \mathbb{O} & \mathbb{O} & \gamma_\mu p^\mu   \\
\end{array}
\right)
\left(
\begin{array}{c}
\lambda^\mathrm{S}_{\ua}(\p) \\
\lambda^\mathrm{S}_{\da}(\p) \\
\lambda^\mathrm{A}_{\ua}(\p) \\
\lambda^\mathrm{A}_{\da}(\p)
\end{array}
\right)
-
\ri m  \mathbb{I} \left(
\begin{array}{c}
- \,\lambda^\mathrm{S}_{\da}(\p) \\
 \,\lambda^\mathrm{S}_{\ua}(\p) \\
 \,\lambda^\mathrm{A}_{\da}(\p) \\
- \,\lambda^\mathrm{A}_{\ua}(\p)
\end{array}
\right) = 0\,, \label{eq:coupled}
\eeq
which establishes that 
$\left(\gamma^\mu p_\mu \pm m \mathbb{I}\right)$ do not annihilate
the neutral particle spinors\footnote{The result contained in the above 
equation confirms earlier result of \cite{Dvoeglazov:1995kn}
and \cite{Kirchbach:2002mkn}.}. 
On recalling section \ref{SubSec:newdual}'s  
antisymmetric symbol 
defined as $\varepsilon^{\{-,+\}}_{\{+,-\}}:=-1$, the above equation
reduces to
\beq
\hspace*{-185pt}&& \left(\gamma_\mu p^\mu \delta_\alpha^\beta + \ri m\mathbb{I}
\varepsilon_\alpha^\beta\right)\lambda_\beta^\mathrm{S}(\p)=0\,,\label{meq:s}\\
\hspace*{-185pt}&& \left(\gamma_\mu p^\mu \delta_\alpha^\beta - \ri m\mathbb{I}
\varepsilon_\alpha^\beta\right)\lambda_\beta^\mathrm{A}(\p)=0\,.\label{meq:a}
\eeq
These are counterparts of equations (\ref{deqs}). The presence
of $\delta_\alpha^\beta$ in the 
$\gamma_\mu p^\mu$ term, and the existence of the 
$\varepsilon_\alpha^\beta$ in the mass term, now make
it impossible for the \es to be eigenspinors of the 
$\gamma_\mu p^\mu$ operator, thus making precise the 
observation made above.
To obtain the 
configuration-space evolution, we make the standard substitution
$p^\mu \rightarrow \ri \partial^\mu$, and define
\beq
\hspace*{-120pt}
\lambda^\mathrm{S/A}(x):=\lambda^\mathrm{S/A}(\p) 
\;\exp\left( \epsilon^\mathrm{S/A}\times 
\ri p_\mu x^\mu\right)\,.\label{ste}
\eeq
Consistency with equations (\ref{meq:s}) and  (\ref{meq:a}) determines
$\epsilon^\mathrm{S} = -1$ and $\epsilon^\mathrm{A} = +1$, while yielding
\beq
\hspace*{-160pt}
 \left(\ri\gamma_\mu \partial^\mu \delta_\alpha^\beta + \ri m\mathbb{I}
\varepsilon_\alpha^\beta\right)\lambda_\beta^\mathrm{S/A}(x)=0\,.\label{new}
\eeq 
Its counterpart for the Dirac case, equation (\ref{eq:pamd}), has several
formal similarities and differences:
\begin{enumerate}
\item 
The Dirac operator, 
$\ri \gamma^\mu \partial_\mu -  m \mathbb{I}$,  
annihilates each of the four $u_h(x)$ and  $v_h(x)$.
It is not so for the wave operator for \textit{Elko}.  
The wave operator in equation (\ref{new}) couples the $\{-,+\}$
degree of freedom with the  $\{+,-\}$, and vice versa. 
This is true for self-conjugate as well as anti-self-conjugate
\textit{Elko}. Equation (\ref{new}) asks for a
eight-component formalism \cite{Kirchbach:2004qn,Kirchbach:2003mk3} 
in exactly the same
manner as the coupled equations for the right-handed Weyl and 
left-handed Weyl spinors \textendash~
each of which is a two-component spinor \textendash~
yields the wave equation for the four-component Dirac 
spinor. As we proceed, we shall see that the eight-component formalism 
is not called for as it introduces eight independent degrees of freedom
for an inherently four-dimensional representation space.

\item
The off-diagonal nature of the mass term in (\ref{new}) is different
from a phenomenological off-diagonal Majorana mass term which is often
introduced in the context of Dirac equation. This is so because of
the  observation on the nature of wave operator
for \textit{Elko} just enumerated.

\item
The Dirac operator can be considered as a `square root of the 
Klein\textendash Gordon operator' 
\textendash~ often, in introductory lectures, 
it is even constructed in that way \textendash~ in the sense that 
$(\gamma_\mu p^\mu-m\mathbb{I})(\gamma_\mu p^\mu+m\mathbb{I}
)=(p_\mu p^\mu - m^2)\mathbb{I}$. This feature remains true for 
\textit{Elko}: 
$(\gamma_\mu p^\mu\delta_\alpha^\beta+\ri m\mathbb{I}
\varepsilon_\alpha^\beta)(\gamma_\mu p^\mu
\delta_\alpha^\beta-\ri m\mathbb{I}
\varepsilon_\alpha^\beta)=(p_\mu p^\mu - m^2)\mathbb{I}\,
\delta_\alpha^\beta$. Thus, both Dirac and \es particles 
have to fulfill the Klein\textendash Gordon equation. Turning this argument 
around, it is possible to provide a `quick and dirty derivation' 
of the wave equation: we would like to consider different square 
roots of the Klein\textendash Gordon operator times a two-dimensional 
Kronecker 
$\delta$ of `helicities' (or referring to (anti-)self-dual spinors 
in the neutral case), i.e., of $(p_\mu p^\mu - m^2)\mathbb{I}
\,\delta_\alpha^\beta$. It is well known how to take the 
`square-root' of $p_\mu p^\mu\mathbb{I}$: it yields 
$\gamma_\mu p^\mu$. Thus, we only have to choose which root of the 
Kronecker symbol we take. Since its roots are always invertible, 
the wave equation can always be brought into a form where the first 
term reads $\gamma_\mu p^\mu\,\delta_\alpha^\beta$, so only the mass 
term has to be considered. Taking the trivial root, 
$\pm\delta_\alpha^\beta$, yields the Dirac equation. 
Taking $(\sigma_2)_\alpha^\beta=\pm \ri\varepsilon_\alpha^\beta$ 
instead produces the wave equations derived above, (\ref{meq:s}) 
and (\ref{meq:a}). The other Pauli matrices, $\pm\sigma_1$ and 
$\pm\sigma_3$, i.e., the trace-free symmetric ones, 
are also possible roots, but they are not considered here.
\end{enumerate}
Before attending to quantum field theoretic structure of the 
theory we need to collect together some essential new details
about the spin sums, and further 
develop the formalism at the representation space level.

\subsection{When wave operators and spin sums do not coincide: a pivotal
observation}
\label{sub:spinsums}

We draw our reader's attention to the fact that, 
using the results of appendix \ref{app:i}, the spin sum over 
self-conjugate spinors reads
\begin{equation}
  \label{eq:spinsum}
\hspace*{-155pt}
  \sum_{\alpha} \lambda^\mathrm{S}_\alpha(\p)\stackrel{\neg}
\lambda_\alpha^\mathrm{S}(\p) = m \left(
\begin{array}{cc}
\mathbb{I} & {\mathcal A}^\mathrm{S}\\
{\mathcal A}^\mathrm{S} & \mathbb{I}
\end{array}
\right)\,,
\end{equation}
and the one for the anti-self-conjugate spinors is given by
\begin{equation}
  \label{eq:spinsum2}
\hspace*{-155pt}
  \sum_{\alpha} \lambda^\mathrm{A}_\alpha(\p)
\stackrel{\neg}\lambda_\alpha^\mathrm{A}(\p) = 
m \left(
\begin{array}{cc}
-\mathbb{I} & {\mathcal A}^\mathrm{S}\\
{\mathcal A}^\mathrm{S} & -\mathbb{I}
\end{array}
\right)\,.
\end{equation}
The matrix $\mathcal{A}^\mathrm{S}$ defines the phase relationship between
the $(1/2,0)$ and $(0,1/2)$ transforming components
of the $\lambda(\p)$ spinors. Its explicit form will be obtained 
in the next section.
For the moment, note may be taken that 
these spin sums reproduce the completeness relation (\ref{z1}),
and that
in the representation in which the four energy\textendash momentum 
vector is given by,
$
p^\mu:=\left(E, \,p\sin(\theta)\cos(\phi), \,p\sin(\theta)
\sin(\phi), \,p\cos(\theta)\right),$
$\mathcal{A}^\mathrm{S}$ reads
\beq
\hspace*{-237pt}
{\mathcal A}^\mathrm{S}  =
\left(
\begin{array}{cc}
0 & \lambda^\ast \\
\lambda & 0
\end{array}
\right)\,,\label{eq:as}
\eeq
where $\lambda:=\ri \mathrm{e}^{\ri\phi}$. 

\paragraph{}
We now make what is one of the pivotal observations for the theory.
It affects the entire particle interpretation,
the realized statistics, the propagator, and the locality structure:
\textit{The right-hand sides of the
spin sums are not proportional, or unitarily connected, 
to the momentum\textendash space wave operators in equations 
(\ref{meq:s}) and (\ref{meq:a}).}
This structure contrasts sharply with 
the Dirac case where
the spin sum over the particle spinors
\beq
\hspace*{-150pt} \sum_{h=\pm ({1}/{2})} 
u_h(\p) \overline{u}_h(\p) = 
\gamma_\mu p^\mu + m \mathbb{I}\,,\label{eq:zimpok1}
\eeq
and the one for the antiparticle spinors
\beq 
\hspace*{-150pt}\sum_{h=\pm ({1}/{2})} 
v_h(\p) \overline{u}_h(\p) = 
\gamma_\mu p^\mu - m \mathbb{I}\,,\label{eq:zimpok2}
\eeq
correspond to the momentum\textendash space wave
operators for the Dirac spinors. 

To realize the importance of these contrasting behaviours, 
the reader may recall that spin sums
enter at a profound  level in the locality and statistics 
structure of the theory. So, with these observation in mind,
it is important to decipher the origins of spin sums and 
their relation to wave operators. This we do next on our way
to developing the particle interpretation.

\subsection{Non-trivial connection between
the spin sums and wave operators: introducing $\cO$ }
\label{sec:spinsumconnection}


The question which is now posed is: is there an additional 
operator which annihilates the 
$\lambda(\p)$  and is different from the ones 
in (\ref{meq:s}) and (\ref{meq:a}). Such an operator is
required to have
the property that it does not couple one of the $\lambda(\p)$
with the other; and that it annihilates these $\lambda(\p)$ singly.
Furthermore, such an operator is expected to shed light
on the structure of the spin sums which appear in equations 
(\ref{eq:spinsum})
and (\ref{eq:spinsum2}).
The meaning of these statements will become more clear as we proceed.

This section is devoted to establishing the existence of such  
operators, and to reveal their origin and associated 
properties. We present a unified method which applies not only 
to the \es but equally well 
to the Dirac framework. 
The method is a generalization of the textbook 
procedure to obtain `wave operators' 
\cite{LHR1996} with corrections noted 
in \cite{Ahluwalia:1993zt,Gaioli:1995ra,Ahluwalia:2001md,Ahluwalia:1997lhr}.  
We introduce a general $(1/2,0)\oplus(0,1/2)$ spinor,
\beq
\hspace*{-200pt}
\xi(\p) = \left(
\begin{array}{c}
\chi^{\left({1}/{2},0\right)}(\p)\\
 \chi^{\left(0,{1}/{2}\right)}(\p)
\end{array}
\right)\,.\label{eq1}
\eeq
Our task is to obtain the operator(s) defined above. 
For the Dirac case, it will be found that this
operator is  nothing but $\left(\gamma_\mu p^\mu \pm m \mathbb{I}\right)$.
For the \es it becomes identical,
up to a constant multiplicative factor, to the relevant spin sums.
Our walk in search of this operator is leisurely, and we do not refrain
from stopping to look at other aspects which these operators may carry.
In a particle's rest frame, by definition 
\cite{LHR1996,Ahluwalia:1997lhr,Hladik:spb,Gaioli:1995ra,Ahluwalia:2001md},
\beq
\hspace*{-185pt}
\chi^{\left({1}/{2},0\right)}(\0)
={\mathcal  A}\;
 \chi^{\left(0,{1}/{2}\right)}(\0)\,. \label{eq2}
\eeq
Here, the complex $2\times 2$ matrix ${\mathcal A}$ 
encodes $C$, $P$, and $T$ properties of the spinor.
It is left unspecified at the moment except that we require it to
be invertible. Its most general form 
may be written, if required, as a general 
invertible $2\times 2$ matrix times $K$ \textemdash~ where $K$ 
complex conjugates a spinor to its right.
Once 
$\chi^{\left({1}/{2},0\right)}(\0)$ and $
\chi^{\left(0,{1}/{2}\right)}(\0)$ are specified
the $\chi^{\left({1}/{2},0\right)}(\p)$ and $
\chi^{\left(0,{1}/{2}\right)}(\p)$ follow from
\beq
\hspace*{-160pt}
\chi^{\left({1}/{2},0\right)}(\p) =&&\rb\;
\chi^{\left({1}/{2},0\right)}(\0) \,, \label{eq3}\\
\hspace*{-160pt}\chi^{\left(0,{1}/{2}\right)}(\p) =&&\lb\;
\chi^{\left(0,{1}/{2}\right)}(\0)\,. \label{eq4}
\eeq
Equation~(\ref{eq2}) implies
\beq
\hspace*{-175pt}
\chi^{\left(0,{1}/{2}\right)}(\0) ={\mathcal A}^{-1}
\chi^{\left({1}/{2},0\right)}(\0)\,,
\eeq
which on immediate use of equation (\ref{eq3}) yields 
\beq
\hspace*{-120pt}
\chi^{\left(0,{1}/{2}\right)}(\0) &=&
{\mathcal A}^{-1} \left(\rb\right)^{-1} 
\;\chi^{\left({1}/{2},0\right)}(\p) \,.\label{eq79}
\eeq
Similarly
\beq
\hspace*{-120pt}
\chi^{\left({1}/{2},0\right)}(\0) =
{\mathcal A}\;\left(\lb\right)^{-1} \;\chi^{\left(0,{1}/{2}\right)}(\p)
\,.\label{eq80}
\eeq
With the useful definition
\begin{equation}
  \label{eq:D}
\hspace*{-187pt}
  \matD:=\rb {\mathcal A}\; \left(\lb\right)^{-1}\,,
\end{equation}
substituting for  $\chi^{\left({1}/{2},0\right)}(\0)$
from equation (\ref{eq80}) in equation (\ref{eq3}), and re-arranging,
gives
\beq
\hspace*{-150pt}
-\;\chi^{\left({1}/{2},0\right)}(\p) \;+ \;\matD\;
\chi^{\left(0,{1}/{2}\right)}(\p) = 0\,;
\eeq
while similar use of equation (\ref{eq79}) in equation 
(\ref{eq4}) results in
\beq
\hspace*{-147pt}
\matD^{-1} \; \chi^{\left({1}/{2},0\right)}(\p)
\;-\; \chi^{\left(0,{1}/{2}\right)}(\p)  =0\,.
\eeq
The last two equations, when combined into a matrix form, result in
\beq 
\hspace*{-200pt}
\left(
\begin{array}{ccc}
- \mathbb{I} &\hspace{7pt} &\matD \\
\matD^{-1} &\hspace{7pt}  &- \mathbb{I}
\end{array}
\right)\xi(\p)=0\,.\label{meq}
\eeq
The operator
\beq
\hspace*{-208pt}
{\mathcal O}:= \left(
\begin{array}{ccc}
- \mathbb{I} &\hspace{7pt} & \matD \\
\matD^{-1} &\hspace{7pt} &  - \mathbb{I}
\end{array}
\right)\,,\label{o}
\eeq
which, as we will soon see,
is the momentum\textendash space operator we are searching for.
We now study its various properties.

\subsection{The $\cO$ for Dirac spinors} 

The Dirac representation space is specified by giving
$\mathcal A$. 
The $\mathcal A$ can be read off from the Dirac rest 
spinors
\beq
\hspace*{-147pt}
\vspace{-\abovedisplayskip}
{\mathcal A} =\begin{cases}+ \;
\mathbb{I}\,,\qquad\mbox{for}\,\,u(\p)\,\,\mbox{spinors} \cr
-\;\mathbb{I}\,,\qquad\mbox{for}\,\,v(\p)\,\,\mbox{spinors}\,.\end{cases}
\label{eq:zimpok20}
\eeq
Parenthetically, we remind the reader that the writing down of the 
Dirac rest spinors, as shown by Weinberg
and also by our independent studies,
follows from the following two requirements:
(a)
parity covariance 
\cite{Weinberg:1995sw,Ahluwalia:2001md}; and that (b) in a quantum field theoretic framework (with locality imposed), the Dirac field describes 
fermions \cite{Weinberg:1995sw}.

Using information contained in equation 
(\ref{eq:zimpok20}) in equation (\ref{o}),
along with the explicit expressions for $\rb$ and $\lb$,
yields
\beq
\hspace*{-125pt}&& \mathcal{O}_{u(\p)}=
+ \;\left(
\begin{array}{ccc}
-\;\mathbb{I} & \hspace{2pt} &\exp\left(\s\cdot\bv\right)\\
\exp\left(-\;\s\cdot\bv\right) & \hspace{2pt} &-\;\mathbb{I}
\end{array}
\right)\,,\label{ueq} \\
\hspace*{-125pt} && \mathcal{O}_{v(\p)}=- \;\left(
\begin{array}{ccc}
\mathbb{I} & \hspace{2pt} &\exp\left(\s\cdot\bv\right)\\
\exp\left(-\;\s\cdot\bv\right) &\hspace{2pt} & \mathbb{I}
\end{array}
\right)\,.\label{veq}
\eeq
Exploiting the fact that $(\s\cdot\hp)^2=\mathbb{I}$, and using the 
definition of the boost parameter $\bv$  given in equations 
(\ref{bp}), the exponentials that appear in the above equations take the form
\beq
\hspace*{-182pt}
\exp\left(\pm \;\s\cdot\bv\right)
={\frac{E\mathbb{I} \pm \s\cdot\p}{m}}\,.\label{lin}
\eeq
Using these expansions in equations (\ref{ueq}) and  
(\ref{veq}), recalling $p_\mu=\left(E,-\p\right)$, and
introducing $\gamma^\mu$ as in equations (\ref{gammamatrices}),
gives equations (\ref{ueq}) and (\ref{veq}) the form 
\beq
\hspace*{-195pt}&& \mathcal{O}_{u(\p)}=+\; 
\frac{1}{m}\left(p_\mu\gamma^\mu 
-m \mathbb{I}\right)
\,,  \label{eq:zimpok3}\\
\hspace*{-195pt}&& \mathcal{O}_{v(\p)}=-\; 
\frac{1}{m}\left(p_\mu\gamma^\mu +m \mathbb{I}\right)
\,.\label{eq:zimpok4}
\eeq
Up to a factor of $1/m$, these are the well known momentum\textendash space 
wave operators for the representation space under consideration.
The {\em linearity\/} of these operators in $p_\mu$ is due to the
form of 
$\mathcal A$, and the property of Pauli matrices, $(\s\cdot\hp)^2
=\mathbb{I}$  
\textemdash~ see equation (\ref{lin}). Comparing equations
 (\ref{eq:zimpok1}) and
(\ref{eq:zimpok2}) with  equations  (\ref{eq:zimpok3}) and
(\ref{eq:zimpok4}) 
results in the following spin sum:
\beq
\hspace*{-178pt}&& \sum_{h=\pm \frac{1}{2}} 
u_h(\p) \overline{u}_h(\p) = -\; m\;\mathcal{O}_{v(\mathbf{p})}\,,\\ 
\hspace*{-178pt}&& \sum_{h=\pm \frac{1}{2}} 
v_h(\p) \overline{v}_h(\p) = +\; m\;\mathcal{O}_{u(\mathbf{p})}\,.
\eeq
This result makes it transparent that 
$\mathcal{O}$   encodes the spin sums.
\textit{As a result, for the Dirac and Majorana fields 
$\mathcal{O}$ not only determines the wave operator but it also
determines the structure of the Feynman\textendash Dyson propagator.}
The counterpart
of this result for \es will be proved in the next section.

\subsection{The $\cO$ for \ecco} 

Similarly as above, 
the \es is specified
by giving $\mathcal{A}$. 
The requirement that the $\lambda(\p)$ be dual-helicity
eigenspinors of the charge conjugation operator completely 
determines  $\mathcal A$
to be
\beq
\hspace{-258pt}
{\mathcal A} = \zeta_\lambda\,\Theta\,\beta\,,\label{eq:A}
\eeq
where 
\beq
\hspace*{-170pt}
\beta= \left(
\begin{array}{ccc}
\exp\left(\ri\phi\right)&\hspace{2pt} & 0 \\
0&\hspace{2pt} & \exp\left(-\;\ri\phi\right)
\end{array}
\right)\,,
\eeq
and $\phi$ is the angle defined by the 4-momentum \textemdash~
cf the end of the paragraph above equation (\ref{eq:as}).
The result (\ref{eq:A}) is slightly non-trivial but can be extracted from explicit 
forms of $\lambda(\0)$ given in equations (\ref{ls}) and (\ref{la}) 
and by making use of the information in appendix \ref{app:b}.
The given representation accounts for the 
complex conjugation involved.
Explicitly, for the self-conjugate \textit{Elko}, $\mathcal{A}$ is given by  
\beq
\hspace*{-230pt}
{\mathcal A}^\mathrm{S}  =
\left(
\begin{array}{ccc}
0 &\hspace{2pt}& \lambda^\ast \\
\lambda & \hspace{2pt}& 0
\end{array}
\right)\,,
\eeq
where $\lambda:=\ri \mathrm{e}^{\ri\phi}$.
Note that ${\mathcal A}^\mathrm{S}$ is Hermitean 
and a square root of the unity matrix, i.e., ${\mathcal A}^\mathrm{S}=
({\mathcal A}^\mathrm{S})^\dagger=({\mathcal A}^\mathrm{S})^{-1}$.  For the 
anti-self-conjugate \textit{Elko}, $\mathcal{A}$ reads
\beq
\hspace{-258pt}
{\mathcal A}^\mathrm{A} = - {\mathcal A}^\mathrm{S}\,. \label{eq:zimpok5}
\eeq

In order to obtain $\mathcal{O}$ we must
obtain the explicit form of $\matD$ for \textit{Elko}. 
Use of the 
general procedure
implemented for the Dirac representation space gives
\begin{equation}
  \label{eq:matD}
\hspace*{-96pt}
  \matD = \pm \frac{E+m}{2m}\left(\mathbb{I}+\frac{\s\cdot\p}{E+m}\right)
{\mathcal A}^\mathrm{S}\left(\mathbb{I}+\frac{\s\cdot\p}{E+m}\right)\,,
\end{equation}
where the plus (minus) sign refers to the (anti-) self-conjugate 
case. Since the anticommutator 
$\{\s\cdot\p,{\mathcal A}_\mathrm{S}\}$ vanishes, we have
\begin{equation}
  \label{eq:matD2}
\hspace*{-96pt}
  \matD = \pm \frac{E+m}{2m}\left(\mathbb{I}
+\frac{\s\cdot\p}{E+m}\right)\left(\mathbb{I}
-\frac{\s\cdot\p}{E+m}\right){\mathcal A}^\mathrm{S}\,.
\end{equation}
But because $(\s\cdot\p)^2=p^2$,
\begin{equation}
  \label{eq:matD4}
\hspace*{-164pt}  
\matD=\pm\frac{(E+m)^2-p^2}{2m(E+m)}{\mathcal A}^\mathrm{S}
=\pm{\mathcal A}^\mathrm{S}\,,
\end{equation}
where the last identity is due to the dispersion relation (\ref{ds1}).
Consequently,
\beq
\hspace*{-196pt}&&\mathcal{O}_{\lambda^\mathrm{S}(\p)}= +\;\left(
\begin{array}{ccc}
-\;\mathbb{I} &\hspace{2pt}& {\mathcal A}^\mathrm{S}\\
{\mathcal A}^\mathrm{S} &\hspace{2pt} & -\;\mathbb{I}
\end{array}
\right)\,,\label{Seq} \\
\hspace*{-196pt}&&\mathcal{O}_{\lambda^\mathrm{A}(\p)}= -\;\left(
\begin{array}{ccc}
\mathbb{I} &\hspace{2pt}& {\mathcal A}^\mathrm{S} \\
{\mathcal A}^\mathrm{S}&\hspace{2pt} & \mathbb{I}
\end{array}
\right)\,.\label{Aeq}
\eeq
Keeping relation (\ref{eq:zimpok5}) in mind, the comparison of 
equations  (\ref{eq:spinsum}) and
(\ref{eq:spinsum2}) with  equations  (\ref{Seq}) and
(\ref{Aeq}) yields

\beq
\hspace*{-170pt}&&  \sum_{\alpha} \lambda^\mathrm{S}_\alpha(\p)
\stackrel{\neg}\lambda_\alpha^\mathrm{S}(\p) 
= -\; m\;  \mathcal{O}_{\lambda^\mathrm{A}(\mathbf{p})}\,,\\
\hspace*{-170pt}&&  \sum_{\alpha} \lambda^\mathrm{A}_\alpha(\p)
\stackrel{\neg}\lambda_\alpha^\mathrm{A}(\p) 
= +\; m \; \mathcal{O}_{\lambda^\mathrm{S}(\mathbf{p})}\,.
\eeq
Thus, as in the Dirac case,  it is transparent that in this case as well
$\mathcal{O}$   encodes the spin sums.
As a result, for the two new fields based upon Elko,
$\mathcal{O}$ \textit{does not} determine 
the wave operator but it \textit{still 
determines} the structure of the Feynman\textendash Dyson 
propagator. The part of the assertion which refers to 
the propagator will be proved below.

\paragraph{}
For later convenience 
we introduce the $\phi$-dependent matrix
 \begin{equation}
   \label{eq:gammaphi}
\hspace*{-225pt}   \mawave := \left(
 \begin{array}{ccc}
 \mathbb{O} &\hspace{2pt}    & {\mathcal A}^\mathrm{S} \\
 {\mathcal A}^S &\hspace{2pt} & \mathbb{O}
 \end{array}
 \right) \,.
 \end{equation}
In terms of  $\mawave$ the spin sums read
\beq
\hspace*{-170pt}&&  \sum_{\alpha} \lambda^\mathrm{S}_\alpha(\p)
\stackrel{\neg}\lambda_\alpha^\mathrm{S}(\p) 
= +\, m( \mathbb{I}+\mawave)\,, \label{spinsumS}\\
\hspace*{-170pt}&&  \sum_{\alpha} \lambda^\mathrm{A}_\alpha(\p)
\stackrel{\neg}\lambda_\alpha^\mathrm{A}(\p) 
=  - m\,( \mathbb{I}-\mawave)\,. \label{spinsumA}
\eeq
Some remarks are in order. First, it should be noted that 
the operators 
\beq
\hspace*{-228pt}\mathcal{P}^\pm:=\frac12 (\mathbb{I}\pm\mawave)\,,
\eeq 
form a complete 
set of projection operators. 
 Second, the definition of $\mawave$ implies the identity
\beq
\hspace*{-218pt}\mawave(\phi) = - \mawave(\phi+\pi)\,. \label{eq:mawave}
\eeq
This corresponds to the behaviour of $\mawave$ in going from $+\,\p$ to
$-\, \p$, i.e., to $\mawave(\p)= - \,\mawave(- \p)$.
It is curious that the spin sums for the 
\es depend only on $\phi$, and are independent
of $\theta$ and of $p$.
This matter is considered in appendix \ref{Sec:noncomm}.
Furthermore, it is observed that although the spin sums do not coincide with 
the wave operator they can be still be considered as 
`square roots of the Klein\textendash Gordon operator' in the following sense. 
\textit{If one does not employ any dispersion 
relation}, from \eqref{eq:matD2} it follows that setting 
the product $\mathcal{P}^+\mathcal{P}^-$ to zero 
{\em implies} the dispersion relation \eqref{ds1}, which is nothing 
but the Klein\textendash Gordon operator in momentum space. More explicitly,
we recall that arriving at results (\ref{spinsumS}) and (\ref{spinsumA})
has already invoked the dispersion relation. One may, going a step backward,
define 
\beq
\hspace*{-120pt}&& \mathcal{P}^\mathrm{S}  =  -\,\frac{1}{2}  
\mathcal{O}_{\lambda^\mathrm{A}(\mathbf{p})} 
= + \,\frac{1}{2 m}
\sum_{\alpha} \lambda^\mathrm{S}_\alpha(\p)
\stackrel{\neg}\lambda_\alpha^\mathrm{S}(\p) \,,\\
\hspace*{-120pt}&& \mathcal{P}^\mathrm{A}  =  -\,\frac{1}{2}  
\mathcal{O}_{\lambda^\mathrm{S}(\mathbf{p})} = 
 -\frac{1}{2 m}\,
\sum_{\alpha} \lambda^\mathrm{A}_\alpha(\p)
\stackrel{\neg}\lambda_\alpha^\mathrm{A}(\p)\,,
\eeq
where the  $\mathcal{O}_{\lambda^\mathrm{S,A}(\p)}$ 
are now written without invoking
the second equality in equation
(\ref{eq:matD4}), but instead are defined using equation (\ref{eq:matD2}) 
\textemdash~
the only difference being 
that now no dispersion relation is invoked explicitly. 
Then, $\mathcal{P}^\mathrm{S}\mathcal{P}^\mathrm{A} = 0$ 
implies the dispersion 
relation.

Invoking the dispersion relation, as in equation (\ref{eq:matD4}),
makes  $\mathcal{P}^\mathrm{S}$ and $\mathcal{P}^\mathrm{A}$ identical to
$\mathcal{P}^+$ and $\mathcal{P}^-$, respectively.
This exercise may, in addition,  also be viewed as a simple
consistency check.

\section{Particle interpretation and mass dimensionality}
\label{Sec:MassDim}

It will be established in the next section that 
the quantum field associated with \es is
\beq
\hspace{-60pt} \eta(x)=\int \frac{\mathrm{d}^3 p}{(2\pi)^3}\frac{1}
{\sqrt{2\, m\, E(\p)}} 
\sum_{\beta} \left[c_\beta(\p) \lambda^\mathrm{S}_\beta(\p) 
\mathrm{e}^{-\ri p_\mu x^\mu} + c_\beta^\dagger(\p)  
\lambda^\mathrm{A}_\beta(\p) \mathrm{e}^{+\ri p_\mu x^\mu}
\right]\,,\label{eq:4.36ap} 
\eeq
with the corresponding \es dual given by
\beq
\hspace{-42pt}
\stackrel{\neg}\eta(x)
=\int \frac{\mathrm{d}^3 p}{(2\pi)^3}\frac{1}{\sqrt{2\,m\, E(\p)}} 
\sum_{\beta} \left[c^\dagger_\beta(\p) 
\stackrel{\neg} \lambda^S_\beta(\p)
\mathrm{e}^{+ \ri p_\mu x^\mu} + c_\beta(\p)  
\stackrel{\neg}\lambda^A_\beta(\p) 
\mathrm{e}^{-\ri p_\mu x^\mu}\right]\,.\label{eq:4.36bp}
\eeq
Here
\beq
\hspace*{-135pt}&& \left\{  
c_\beta(\p),\; c^\dagger_{\beta^\prime}(\p^\prime)\right\}
=  \left(2\pi\right)^3  \delta^3\left(\p-\p^\prime\right)
\delta_{\beta\beta^\prime}\,,\label{eq:anticomm} \\
\hspace*{-135pt}&&
\left\{  
c^\dagger_\beta(\p),\; c^\dagger_{\beta^\prime}(\p^\prime)\right\}
=  
\left\{c_\beta(\p),\; c_{\beta^\prime}(\p^\prime)\right\}
=0\,.
\label{eq:anticommz}
\eeq

 \subsection{The Elko propagator}

Once these fields and anticommutators are given,
the amplitude for a positive-energy self-conjugate particle 
to propagate from $x$ to $x^\prime$ is\footnote{This argument follows 
closely  Hatfield's discussion in \cite{Hatfield:bri1992}
for the Dirac case.}
\beq
\hspace*{-255pt}\langle s(x^\prime)\vert s(x)\rangle\,. \label{eq:4.44}
\eeq
The state $\vert s(x)\rangle$ contains $1$ positive-energy
self-conjugate particle of mass $m$. From equation (\ref{eq:4.36bp})
we decipher that the 
state which contains  $1$ positive-energy
self-conjugate particle at $x$ is $\stackrel{\neg}{\eta}(x)\vert\hspace{8pt}\rangle$,
where $\vert\hspace{8pt}\rangle$ represents the physical vacuum.
Therefore, the covariant amplitude (\ref{eq:4.44}) is\footnote{We
call 
$
\langle
\hspace{8pt}\vert \eta(x^\prime)\stackrel{\neg}\eta(x)
\vert\hspace{8pt}\rangle
$
the covariant amplitude, to distinguish it from
$
\langle
\hspace{8pt}\vert \eta(x^\prime)\eta^\dagger(x)\vert\hspace{8pt}\rangle
$,
which is often referred to as an amplitude.}
\beq
\hspace*{-177pt}\mathcal{Q}_{x\rightarrow x^\prime}=
\varpi
\langle
\hspace{8pt}\vert \eta(x^\prime)\stackrel{\neg}\eta(x)
\vert\hspace{8pt}\rangle\,.
\label{eq:4.45}
\eeq
The pre-factor,  $\varpi \in \mathbb{C}$, 
shall be fixed by the requirement that 
$\mathcal{Q}_{x\rightarrow x^\prime}$, when integrated over all spacetime,
yields \textit{unity} (or, to be more precise, $\exp(\ri \gamma)$, with
$\gamma\in \mathbb{R}$). This requirement is imposed by the
quantum-mechanical 
interpretational structure for such amplitudes:
the $\mathcal{Q}_{x\rightarrow x^\prime}$ integrated over all
spacetime, call it $\mathcal{A}$,
  yields the amplitude for the particle to be found anywhere
in the universe; consequently, $\mathcal{A}^\ast \,
\mathcal{A}$ is the corresponding probability.
For a free particle this interpretation is free from interpretational
problems. 
For the interacting case, the relation of 
$\mathcal{Q}_{x\rightarrow x^\prime}$ with an 
appropriate Green function provides the additional
interpretational structure.
Moreover, the integration over the entire spacetime is
required, rather than integration over the part of spacetime which carries the 
lightcone of $x$ as its (open) boundary, as the
classically forbidden region remains accessible
quantum mechanically\footnote{We are following this
first-principle derivation so as to avoid using full 
quantum field theoretic formalism which implicitly contains the
assumption of locality. The latter property, as we shall see, 
is not fully respected by the \es quantum field.}.
The Feynman\textendash Dyson 
propagator is then nothing but  
a numerical constant times $\mathcal{Q}_{x\rightarrow x^\prime}$, where the 
constant is determined by the requirement that it coincides with the 
appropriate Green  function.  
The quantum-mechanical propagation of the self-conjugate particle, like
the one in case of the Dirac particle, is the process where a positive-energy
self-conjugate particle is created out of the vacuum at $x$, 
propagates to $x^\prime$,
where it is reabsorbed into the vacuum. 
In a given inertial frame\footnote{It is important to make
this reference to an inertial frame 
because quantum mechanically a particle has a finite probability
of `tunneling' beyond the light cone of $x$. Such a tunneling 
destroys the time ordering of events. See also footnote \ref{f:Weinberg}.} 
one cannot destroy a particle
before its creation; therefore,  $t^\prime > t$.  

As in the Dirac case, there is a distinct but physically equivalent process 
that we must also take into account. If we consider positive-energy 
self-conjugate particles to carry negative  charge
$- \ell$, ($\ell > 0$)\footnote{This charge, which 
is pseudo-scalar under the operation of parity, is distinct from
Dirac charge \textemdash~ irrespective of the fact whether 
it is zero, or non-vanishing.
The pseudo-scalar nature of $\ell$ follows from the fact that
the \es lose their self-/anti-self-conjugacy under the
transformation $\lambda(x)\rightarrow \exp\left[\ri 
\alpha(x)\right] \lambda(x)$;
$\alpha\in\mathbb{R}$.
It is preserved under the local gauge transformation 
$\lambda(x)\rightarrow \exp\left[\ri \alpha(x) \gamma^5\right]\lambda(x)$.}
then the process above lowers this charge
by one unit at $x$ and subsequently raises it by one unit at
$x^\prime$. Negative-energy self-conjugate particles
propagating backward in time, that is, positive-energy 
anti-self-conjugate particles, carry the opposite charge, $\ell$.
If we create an anti-self-conjugate 
particle of mass $m$ at $x^\prime$, transport it to $x$,
where we destroy it, then we are also raising the new charge at
$x^\prime$ by one unit and lowering it by the same amount at $x$.
From equation (\ref{eq:4.36ap}) and the particle interpretation,
we see that $\eta(x)$ creates anti-self-conjugate particles; so the 
covariant amplitude
for this process is
\beq
\hspace*{-221pt}
\varpi
\langle
\hspace{8pt}\vert \stackrel{\neg}\eta(x)
\eta(x^\prime)\vert\hspace{8pt}\rangle\,.
\label{eq:4.46}
\eeq
Once again, we cannot destroy a particle before we create it; hence
this process is physically meaningful only for $t > t^\prime$.
Since fermionic amplitudes are antisymmetric under the 
exchange $x\leftrightarrow x^\prime$, the total covariant 
amplitude for the
process under consideration is
\beq
\hspace*{-42pt}
\varpi
\langle
\hspace{8pt}\vert \eta(x^\prime)\stackrel{\neg}\eta(x)\vert\hspace{8pt}\rangle
\theta(t^\prime -t) - \varpi \langle
\hspace{8pt}\vert\stackrel{\neg}\eta(x)\eta(x^\prime)\vert\hspace{8pt}\rangle
\theta(t-t^\prime)\,.
\eeq
Invoking the fermionic time-ordering operator $\mathcal{T}$ 
this may be recast as
\beq
\hspace*{-203pt}
\varpi
\langle \hspace{8pt}\vert \mathcal{T} (\eta(x^\prime)\stackrel{\neg}
\eta(x))\vert\hspace{8pt}\rangle \,.
\eeq
We now evaluate $\langle \hspace{8pt}\vert \mathcal{T} 
(\eta(x^\prime)\stackrel{\neg}
\eta(x))\vert\hspace{8pt}\rangle$ term by term. The first term
may be written as
\beq
\hspace*{-70pt}\Big\langle\hspace{11pt} \Big\vert
\int \frac{\mathrm{d}^3p}{(2\pi)^3} &&
\frac{1}{\sqrt{2\,m\,E(\p)}}
\int \frac{\mathrm{d}^3p^\prime}{(2\pi)^3}
\frac{1}{\sqrt{2\,m\,E(\p^\prime)}} \nonumber \\
\hspace*{-70pt}&&\hspace*{-35pt} \times \sum_{\beta^\prime}\sum_\beta 
\Big[c_{\beta^\prime}(\p^\prime)  
\lambda_{\beta^\prime}^\mathrm{S}(\p^\prime) 
\mathrm{e}^{- \ri p^\prime_\mu x^{\prime\mu}}
+
c^\dagger_{\beta^\prime}(\p^\prime)  
\lambda_{\beta^\prime}^\mathrm{A}(\p^\prime) 
\mathrm{e}^{+ \ri p^\prime_\mu
x^{\prime\mu}} \Big]\nonumber\\
\hspace*{-70pt}&&\hspace*{-35pt} \times \Big[c^\dagger_{\beta}(\p) 
\stackrel{\neg}\lambda_{\beta}^\mathrm{S}(\p) 
\mathrm{e}^{+ \ri p_\mu x^{\mu}}
+
c_{\beta} (\p) \stackrel{\neg}\lambda_{\beta}^\mathrm{A}(\p)) 
\mathrm{e}^{+ \ri p_\mu x^{\mu}} \Big]
\Big\vert\hspace{11pt}\Big\rangle \theta(t^\prime-t)\,.
\eeq
The only non-vanishing contribution comes from the terms in the 
expansion of the form 
$c_{\beta^\prime}(\p^\prime)c^\dagger_{\beta}(\p)$.
Under the integral, 
$\langle \hspace{8pt} \vert
c_{\beta^\prime}
(\p^\prime)c^\dagger_{\beta}(\p)\vert\hspace{8pt}\rangle $,
using  equation (\ref{eq:anticomm}),  
may be replaced by 
$\langle \hspace{8pt} \vert
- c^\dagger_{\beta}(\p) c_{\beta^\prime}
(\p^\prime)\vert\hspace{8pt}\rangle 
+ (2\pi)^3 \delta^3(\p^\prime-\p)
\delta_{\beta^\prime\beta}$ (with the already 
made implicit assumption
that the vacuum state is normalized to unity).
Taking further note of the fact that
$\langle \hspace{8pt} \vert
c^\dagger_{\beta}(\p) c_{\beta^\prime}
(\p^\prime)\vert\hspace{8pt}\rangle $
identically vanishes, this then yields the first term in  
 $\langle \hspace{8pt}\vert \mathcal{T} (\eta(x^\prime)\stackrel{\neg}
\eta(x))\vert\hspace{8pt}\rangle$ to be
\beq
\hspace*{-50pt}\int \frac{\mathrm{d}^3p}{(2\pi)^3}\;
\frac{1}{2 m E(\p)}
\sum_\beta 
\lambda_\beta^\mathrm{S}(\p) \stackrel{\neg}\lambda^\mathrm{S}_\beta(\p)
\mathrm{e}^{- \ri p_\mu(x^{\prime\mu} - x^\mu)}\theta(t^\prime-t)\,.
\eeq
A similar evaluation of the second term gives
\beq
\hspace*{-35pt}- \;\int \frac{\mathrm{d}^3p}{(2\pi)^3}\;\frac{1}{2 m E(\p)}
\sum_\beta
 \lambda_\beta^\mathrm{A}(\p) \stackrel{\neg}\lambda^\mathrm{A}_\beta(\p)
\mathrm{e}^{+ \ri p_\mu(x^{\prime\mu} - x^\mu)}\theta(t-t^\prime)\,.
\eeq
Combining both of these evaluations leads to the result
\beq
 \hspace*{-100pt} \mathcal{Q}_{x\rightarrow x^\prime} = \varpi
 \int \frac{\mathrm{d}^3p}{(2\pi)^3}&&
\frac{1}{2 m E(\p)} 
 \sum_\beta \Big[ \theta(t^\prime-t)
 \lambda_\beta^\mathrm{S}(\p) \stackrel{\neg}\lambda^\mathrm{S}_\beta(\p)
\mathrm{e}^{- \ri p_\mu(x^{\prime\mu} - x^\mu)} \nonumber\\
\hspace*{-30pt} &&\hspace*{-55pt}  -\;
 \theta(t-t^\prime) \lambda_\beta^\mathrm{A}(\p) 
\stackrel{\neg}\lambda^\mathrm{A}_\beta(\p)
\mathrm{e}^{+ \ri p_\mu(x^{\prime\mu} - x^\mu)}
\Big]\,.\label{eq:6.a}
\eeq
For further simplification we invoke the spin sums for $\lambda^\mathrm{S}(\p)$
and $\lambda^\mathrm{A}(\p)$ given in equations 
(\ref{spinsumS}) and (\ref{spinsumA}):
\beq
 \hspace*{-183pt} 
\mathcal{Q}_{x\rightarrow x^\prime} = \varpi
 \int && \hspace{-3pt}\frac{\mathrm{d}^3p}{(2\pi)^3}
\; \frac{1}{2 E(\p)}
 \Big[ \theta(t^\prime-t)
 (\mathbb{I}+\mawave(\phi))\,
\mathrm{e}^{- \ri p_\mu(x^{\prime\mu} - x^\mu)} \nonumber\\
 \hspace*{-70pt} &&\hspace*{-25pt} +\;
 \theta(t-t^\prime)  (\mathbb{I}-\mawave(\phi))\,
\mathrm{e}^{+ \ri p_\mu(x^{\prime\mu} - x^\mu)}
\Big]\,.\label{eq:6.b}
\eeq
Next focus on the second term. 
Letting $\p\rightarrow -\p$, and using (\ref{eq:mawave}),
we get
\beq
 \hspace*{-125pt} 
\mathcal{Q}_{x\rightarrow x^\prime} = \varpi
 \int &&\hspace{-3pt}\frac{\mathrm{d}^3p}{(2\pi)^3} \frac{1}{2 E(\p)}
 \Big[ \theta(t^\prime-t)
 (\mathbb{I}+\mawave(\phi))\,
\mathrm{e}^{-\ri E(\mathbf{p})(t^\prime-t)\, +\, 
\ri\,\mathbf{p} \cdot(\mathbf{x}^\prime-\mathbf{x})} \nonumber\\
  \hspace*{-30pt} &&\hspace*{-20pt} +\;
 \theta(t-t^\prime)  (\mathbb{I}+\mawave(\phi))\,
\mathrm{e}^{+ \ri E(\mathbf{p}) (t^\prime -t) + \, \ri\,\mathbf{p} 
\cdot(\mathbf{x}^\prime-\mathbf{x})}
\Big]\,.\label{eq:propc}
\eeq
The above equation can be cast into a covariant form by the use of an integral
representation of the Heaviside step function
\beq
\hspace*{-55pt}&& \mbox{For the first term:}\qquad \theta(t^\prime-t) = 
\mathop{\mathrm{lim}}\limits_{\epsilon\rightarrow 0^+}
\int \frac{\mathrm{d} \omega}{2\pi \ri} 
\frac{\mathrm{e}^{\ri \omega (t^\prime-t)}}
{\omega-\ri\epsilon}\,, \\
\hspace*{-55pt}&& \mbox{For the second term:}\qquad \theta(t-t^\prime) = 
\mathop{\mathrm{lim}}\limits_{\epsilon\rightarrow 0^+}
\int \frac{\mathrm{d} 
\omega}{2\pi \ri} \frac{\mathrm{e}^{\ri \omega (t-t^\prime)}}
{\omega-\ri\epsilon}\,.
\eeq
Inserting these integrals into (\ref{eq:propc}) yields
\beq
&& \hspace*{-55pt} \mathcal{Q}_{x\rightarrow x^\prime}
= - \ri\,\varpi \mathop{\mathrm{lim}}\limits_{\epsilon\rightarrow 0^+}
 \int \frac{\mathrm{d}^3p}{(2\pi)^3} \frac{1}{2 E(\mathbf{p})}
\int\frac{\mathrm{d}\omega}{2\pi}\nonumber \\
\hspace*{20pt}&& \times 
\bigg[ \frac{ (\mathbb{I}+\mawave(\phi))} {\omega - \ri \epsilon}
\left( \mathrm{e}^{\ri(\omega-E(\mathbf{p}))(t^\prime -t)} \,\mathrm{e}^{\ri 
\mathbf{p}.(\mathbf{x}^\prime
-\mathbf{x})}
+
\mathrm{e}^{- \ri (\omega-E(\mathbf{p}))(t^\prime -t)} \,\mathrm{e}^{\ri
\mathbf{p}.(\mathbf{x}^\prime
-\mathbf{x})}\right)\bigg]\,.\nonumber\\ \label{eq:propd}
\eeq
Now 
we change variables: in the first integral 
$\omega\rightarrow p_0=- \left(\omega -E (\p)\right)$,
while in the second integral  $\omega\rightarrow p_0=\omega -E (\p)$.
This substitution alters (\ref{eq:propd}) to
\beq
\hspace{-23pt} 
\mathcal{Q}_{x\rightarrow x^\prime}
= - \ri\,\varpi \mathop{\mathrm{lim}}
\limits_{\epsilon\rightarrow 0^+}
\int \frac{\mathrm{d}^4 p}{(2\pi)^4} \frac{1}{2 E(\p)} 
\mathrm{e}^{-\ri p_\mu(x^{\prime\mu} - x^\mu)}
\bigg[ \frac{ \mathbb{I}+\mawave(\phi)} {E(\p) - p_0 - \ri\epsilon}
+
\frac{ \mathbb{I}+\mawave(\phi)} {  E(\p) + p_0 - \ri\epsilon}\bigg]\,.
\eeq
Because of the indicated limit on the above integrals, we can drop
the terms of the order $\epsilon^2$ and write the result
as
\beq
\hspace*{-50pt}
\mathcal{Q}_{x\rightarrow x^\prime}
=  
\int \frac{\mathrm{d}^4 p}
{(2\pi)^4}\;\mathrm{e}^{-\ri p_\mu(x^{\prime\mu} - x^\mu)}\;  
\left[ \ri\,\varpi \frac{ \left(\mathbb{I}+\mawave(\phi)\right) }  
{p_\mu p^\mu - m^2 + \ri\epsilon}\right]
\,,\label{eq:noteworthy}
\eeq
where the limit $\epsilon \rightarrow 0^+$ is now understood.
The covariant amplitude is directly related to  
the Feynman\textendash Dyson propagator:
 \beq
\hspace*{-215pt} \mathcal{S}_{\mathrm{FD}}(x^\prime,x)
\propto\mathcal{Q}_{x\rightarrow x^\prime} \,,
 \label{eq:FD}
 \eeq
where the proportionality constant is determined by the requirement that 
$ \mathcal{S}_{\mathrm{FD}}(x^\prime,x)$ 
coincides with the appropriate Green 
function (see equations (\ref{eq:zimpok6})\textendash 
(\ref{eq:zimpok7}) below).

\subsection{Mass dimension one: the Elko propagator in the 
absence of a preferred direction}

If there is no preferred direction, and since we are 
integrating over all momenta, 
we are free to choose a coordinate system in which 
$\x^\prime-\x$ lies along the $\widehat{z}$ direction. 
In this special case, the $\p\cdot(\x^\prime-\x)$ depends 
only on $p$ and $\theta$, but not on $\phi$. Thus, the
only $\phi$-dependence in the whole integrand comes 
from $\mathcal{G}$ which depends on $\phi$ in
such a manner that an integral over one period vanishes. 
With this result at hand, the 
covariant amplitude reduces to

\beq
\hspace*{-60pt} 
\mathcal{Q}_{x\rightarrow x^\prime}
=  
\int \frac{\mathrm{d}^4 p}{(2\pi)^4}\;
\mathrm{e}^{-\ri p_\mu(x^{\prime\mu} - x^\mu)}\;  
\left[ \ri\,\varpi \frac{ \mathbb{I} }  
{p_\mu p^\mu - m^2 + \ri\epsilon}\right]
\,.\label{eq:noteworthy2}
\eeq

The $ \mathcal{Q}_{x\rightarrow x^\prime}$ depends not on $x$ but
on $x-x^\prime$. 
This is consistent with the observation
that we have no preferred spacetime point. For this
reason we integrate $ \mathcal{Q}_{x\rightarrow x^\prime}$ over all
possible $x-x^\prime$ and set the result to 
unity\footnote{\label{28}To be more precise,  
one should demand the amplitude to be $\exp(\ri \gamma)$, with
$\gamma\in \mathbb{R}$.}:
\beq
\hspace*{-97pt}
(2\pi)^4 \int \frac{\mathrm{d}^4 p}{(2\pi)^4} \delta^4(p^\mu) 
\left[ \ri\,\varpi \frac{ \mathbb{I} }  
{p_\mu p^\mu - m^2 + \ri\epsilon}\right] =1
\eeq
That is,
\beq
\hspace*{-230pt}
 \ri\,\varpi \frac{ \mathbb{I} }  
{- m^2 + \ri\epsilon} = \mathbb{I}\,.
\eeq
Taking the limit $\epsilon\rightarrow 0$ yields
\beq
\hspace*{-270pt}
\varpi = \ri m^2\,.
\eeq
With $\varpi$ now fixed we have

\beq
\hspace*{-56pt} \mathcal{Q}_{x\rightarrow x^\prime}
=  
- m \int \frac{\mathrm{d}^4 p}{(2\pi)^4}\;\mathrm{e}^{-\ri p_\mu(x^{\prime\mu} 
- x^\mu)}\;  
\left[  \frac{ m\,\mathbb{I} }  
{p_\mu p^\mu - m^2 + \ri\epsilon}\right]
\,.\label{eq:noteworthy3}
\eeq
Therefore, the propagator for the new theory 
is
\beq
 \hspace*{-65pt}\mathcal{S}^{\mathrm{Elko}}_{\mathrm{FD}}(x^\prime,x)
&& := -\,\frac{1}{m^2} \mathcal{Q}_{x\rightarrow x^\prime}
\label{eq:zimpok6}\\
&& \hspace*{10pt}=
 \int \frac{\mathrm{d}^4 p}{(2\pi)^4}\;\mathrm{e}^{-\ri p_\mu(x^{\prime\mu} 
- x^\mu)}\;  
\left[  \frac{ \mathbb{I} }  
{p_\mu p^\mu - m^2 + \ri\epsilon}\right]
\,,\label{eq:noteworthy4}
\eeq
and it satisfies
\beq
 \hspace*{-100pt}\left(\partial_{\mu^\prime} \partial^{\mu^\prime} \mathbb{I}
+ m^2 \mathbb{I}\right)
\mathcal{S}^{{\mathrm {Elko}}}_{\mathrm{FD}}(x^\prime,x) = - \,
\delta^4\left(x^\prime-x\right)\,.\label{eq:zimpok7}
\eeq
Thus, it is again clear that the new theory is different from
that based upon Dirac spinors. For the latter, the counterpart
of equation (\ref{eq:noteworthy4}) reads
\beq
 \hspace*{-40pt}\mathcal{S}^{{\mathrm{Dirac}}}_{\mathrm{FD}}(x^\prime,x)
=  
\int \frac{\mathrm{d}^4 p}{(2\pi)^4}\;\mathrm{e}^{-\ri 
p_\mu(x^{\prime\mu} - x^\mu)}\;  
\left[  \frac{ \gamma_\mu p^\mu + m\mathbb{I} }  
{p_\mu p^\mu - m^2 + \ri\epsilon}\right]
\,.\label{eq:dm}
\eeq
This is valid for the Majorana as well as the Dirac field.

In the absence of a preferred direction, the Elko propagator
is identical to that of a scalar Klein\textendash Gordon field. It is
this circumstance that endows the $\eta(x)$ with mass dimension
one. General discussion  on mass dimensionality of
fields and their relation with propagators can be found, for example,
 in section  12.1 of \cite{Weinberg:1995sw}. 
For the circumstance at hand it is to be noted that Weinberg's
purely dimensional arguments \textemdash~
which appear after equation (12.1.9)  of  \cite{Weinberg:1995sw}
 \textemdash~ are more robust than the arguments that precede
it. The reason lies in the assumption of locality.
For local massive fields, combining his dimensional
arguments with those following his monograph's equation (12.1.2),
one immediately sees that for a massive field
of Lorentz transformation type $(A,B)$, the mass dimension
of the field is $1 + A +B$. So for the field of the types
$(1/2,0)$ and $(0,1/2)$, the expected mass dimensionality
follows to be $3/2$. The reason that for the \es field,
which also transforms as $(1/2,0)\oplus(0,1/2)$, the mass
dimension is not $3/2$, but $1$, lies in the non-locality of the
field $\eta(x)$, a property we establish in section \ref{Sec:ls}.

A heuristic understanding of this crucial result may be gained
as follows. Equation~(\ref{eq:coupled}), or its equivalent
equation (\ref{new}), constitutes a set of 
coupled equations. The set for the self-conjugate \es reads
\beq
 \hspace*{-157pt}&& \gamma_\mu p^\mu \lambda^\mathrm{S}_{\{-,+\}}(\p)  
+ \ri m  \lambda^\mathrm{S}_{\{+,-\}}(\p) =0\,,  \label{eq:eq1}\\
 \hspace*{-157pt}&&   \gamma_\mu p^\mu \lambda^\mathrm{S}_{\{+,-\}}(\p)  
- \ri m  \lambda^\mathrm{S}_{\{-,+\}}(\p) =0 \,.\label{eq:eq2}
\eeq
The second of these equations may be re-written as
\beq
 \hspace*{-180pt}\lambda^\mathrm{S}_{\{-,+\}}(\p) 
= \frac{\gamma_\mu p^\mu }{\ri m} 
\lambda^\mathrm{S}_{\{+,-\}}(\p)\,. 
\eeq
Substitution of this in equation (\ref{eq:eq1}) gives
$
\left(\gamma_\mu  \gamma_\nu p^\mu p^\nu - m^2 \right) 
\lambda^\mathrm{S}_{\{+,-\}}(\p) 
=0$, which on
exploiting the commutator $[p^\mu, p^\nu]=0$, reads
\beq
 \hspace*{-117pt}\left(\frac{\gamma_\mu  \gamma_\nu 
+ \gamma_\nu \gamma_\mu}{2} p^\mu p^\nu 
- m^2 \mathbb{I}\right) \lambda^\mathrm{S}_{\{+,-\}}(\p) 
=0\,. \label{eq:eq3}
\eeq
Using the anticommutator
  $\{\gamma^\mu,\gamma^\nu\}= 2 \eta_{\mu\nu}\mathbb{I}$, 
equation (\ref{eq:eq3}) becomes
\beq
 \hspace*{-159pt}\left(\eta_{\mu\nu}p^\mu p^\nu \mathbb{I} 
- m^2 \mathbb{I}\right)
\lambda^\mathrm{S}_{\{+,-\}}(\p) =0\,.
\eeq
Similarly, it is seen that 
$\left(\eta_{\mu\nu}p^\mu p^\nu \mathbb{I} - m^2 \mathbb{I}\right)$
annihilates
the remaining
three $\lambda(\p)$. Thus, we have
\beq
 \hspace*{-164pt}\left(\eta_{\mu\nu}p^\mu p^\nu \mathbb{I} 
- m^2 \mathbb{I}\right)
\lambda^\mathrm{S/A}(\p) =0\,, \label{eq:kg}
\eeq
which is nothing but  the Klein\textendash Gordon 
equation in momentum space.

For the \es field the crucial difference as compared to the Dirac case 
is that the `square root' of the Klein\textendash Gordon operator is not 
a valid wave operator for a \textit{single} 
$\lambda^\mathrm{S/A}_\alpha (\p)$, 
but  rather, unless $m=0$, it
\textit{couples} the two four-component degrees of freedom, 
$\{+,-\}$ and  $\{-,+\}$ \textemdash~
see equations (\ref{eq:eq1}) and (\ref{eq:eq2}).
 This has led some authors to suggest 
an eight-component formalism \cite{Kirchbach:2004qn,Kirchbach:2003mk3}. 
However, the spacetime evolution considered above makes it
abundantly clear that the
Klein\textendash Gordon propagation is an \textit{intrinsic}
property of the \es field, and Klein\textendash Gordon operator is the
kinetic operator for this field.   
In other words, a conflict arises between the canonical
quantization and the path integral approach unless 
equation (\ref{eq:kg}) is taken to define the Lagrangian density
for the {\ecco} framework.
This circumstance makes the {\ecco} field to carry mass dimension one, rather
than three halves.

The $\mathcal{G}(\p)$ independence 
of the theory must be re-examined if a preferred direction exists.
This may come about because of a cosmic preferred direction
\cite{Ralston:2003pf},    a fixed background, a thermal bath,
a reference fluid, or some other external field such as   
a magnetic/Thirring\textendash Lense field of a neutron star.
In addition, 
the following interpretational structure is worth taking note of, explicitly.

\begin{enumerate}
\item[$\bullet$] The $c^\dagger_{\{\mp,\pm\}}$ creates a positive-energy
self-conjugate particle with dual helicity  ${\{\mp,\pm\}}$.
\item[$\bullet$] The $c_{\{\mp,\pm\}}$ destroys a negative-energy
self-conjugate particle with dual helicity  ${\{\mp,\pm\}}$. That is,
$c_{\{\mp,\pm\}}$ creates a positive-energy hole with 
the reversed dual helicity  ${\{\pm,\mp\}}$.
The holes carry the interpretation of anti-self-conjugate particles. Thus
 $c_{\{\mp,\pm\}}$ is also the creator of positive-energy 
anti-self-conjugate particles with the reversed 
helicity ${\{\pm,\mp\}}$.
\end{enumerate}

\section{Energy of vacuum and establishing the fermionic statistics}
\label{Sec:Stat}

The spacetime evolution as contained in equations (\ref{ste}) and (\ref{new}),
or equivalently in (\ref{eq:kg}),
allows us to introduce the field operator:
\beq
\hspace*{-94pt}&& 
\eta(x)=\int \frac{\mathrm{d}^3 p}{(2\pi)^3}\frac{a(p^\mu)}{2\, E(\p)} 
\sum_{\beta} \left[c_\beta(\p) \lambda^\mathrm{S}_\beta(\p) 
\mathrm{e}^{-\ri p_\mu x^\mu} + c_\beta^\dagger(\p)  
\lambda^\mathrm{A}_\beta(\p) \mathrm{e}^{+\ri p_\mu x^\mu}\right]
\,,\label{eq:4.36a} 
\eeq
where $a(p^\mu) \in \mathbb{R}$. It will  be fixed 
below.
Our aim now is to settle the statistics, i.e., commutative/anticommutative
properties of the $c_\beta(\p)$ and $c^\dagger_\beta(\p)$, and to
establish the result with which the previous section opened.
Towards this task we proceed as follows.
The field operator $\eta(x)$ satisfies
\beq
\hspace*{-180pt}\left(\mathbb{I} 
\eta_{\mu\nu} \partial^\mu \partial^\nu + m^2 \mathbb{I}\right) 
\eta(x)=0\,. \label{new2}
\eeq
To obtain the momenta conjugate to $\eta(x)$, we note that
equation (\ref{new2}) follows from the action
\beq
\hspace*{-35pt} S\left[\stackrel{\neg}\eta(x),\eta(x)\right]
&&=\int \mathrm{d}^4 x\; \mathcal{L}\left(\stackrel{\neg}\eta(x),\eta(x)\right)
\label{eq:zimpok16A}\\
&&= \int \mathrm{d}^4 x\; \left(\partial^\mu 
{\stackrel{\neg}\eta}(x) \,
                        \partial_\mu\eta(x) - m^2
\stackrel{\neg}\eta(x)\,\eta(x)\right)\,.\label{eq:zimpok16B}
\eeq
The field momentum conjugate to $\eta(x)$ is
\beq
\hspace*{-200pt}
\pi(x) =  \frac{\partial  \mathcal{L}}
{\partial\stackrel{\cdot}\eta}
= \, \frac{\partial}{\partial t}\stackrel{\neg}{\eta}(x)\,,\label{eq:zimpok16C}
\eeq
while the Hamiltonian density reads 
\beq
\hspace*{-232pt}\mathcal{H} = \pi\stackrel{\cdot}\eta(x) - \mathcal{L}\,.
\eeq
On-shell it becomes
\beq
\hspace*{-210pt}\mathcal{H} = 
\frac{\partial}{\partial t} \stackrel{\neg}{\eta}(x) \,
\frac{\partial}{\partial t}\eta(x)\,.
\eeq
The commutative/anticommutative
properties of the $c_\beta(\p)$ and $c^\dagger_\beta(\p)$ follow
from considering the field energy:
\beq
\hspace*{-30pt}H = &&\int  \mathrm{d}^3 x\;\mathcal{H} \\
\hspace*{-30pt}   =&&   \int   \mathrm{d}^3 x  \sum_{\beta,\beta^\prime}
\int\int 
\frac{\mathrm{d}^3 p^\prime}
{(2\pi)^3}\frac{a(p^{\prime\mu})}{2\, E(\p^\prime)} 
\;\;\frac{\mathrm{d}^3 p}{(2\pi)^3}\frac{a(p^\mu)}{2\, E(\p)}
\nonumber\\
\hspace*{-30pt}&& \hspace{36pt}\times \left[c^\dagger_\beta(\p) 
\stackrel{\neg}\lambda^\mathrm{S}_\beta(\p) (+ \ri E(\p))
\mathrm{e}^{+ \ri p_\mu x^\mu} + 
c_\beta(\p)  \stackrel{\neg}\lambda^\mathrm{A}_\beta(\p)
(- \ri E(\p)) \mathrm{e}^{-\ri p_\mu x^\mu}\right]\nonumber\\
\hspace*{-30pt}&&\hspace{36pt} \times \left[c_{\beta^\prime}(\p^\prime) 
\lambda^\mathrm{S}_{\beta^\prime}(\p^\prime) 
(- \ri E(\p^\prime))\mathrm{e}^{-\ri p^\prime_\mu x^\mu} 
+ c_{\beta^\prime}^\dagger(\p^\prime)  
\lambda^\mathrm{A}_{\beta^\prime}(\p^\prime) 
(+ \ri E(\p^\prime))\mathrm{e}^{+\ri p^\prime_\mu x^\mu}\right]\,.
\eeq
The spatial integration gives $(2\pi)^3 \delta^3(\p-\p^\prime)$, 
resulting in
\beq
\hspace*{-46pt}H = \int &&\!\!\! \frac{\mathrm{d}^3 p}{(2 \pi)^3}
\frac{a^2(p^\mu)}{4} 
\sum_\beta\left(c^\dagger_\beta(\p) c_\beta(\p) 
\stackrel{\neg}\lambda^\mathrm{S}_\beta(\p) \lambda^S_\beta(\p)
+  c_\beta(\p)    c^\dagger_\beta(\p)
\stackrel{\neg}\lambda^\mathrm{A}_\beta(\p) \lambda^A_\beta(\p)
\right)\,.
\eeq
Using the results (\ref{zd1}) and (\ref{z5}),
we get
\beq
\hspace*{-60pt}H = 
 \int \!\!\! \frac{\mathrm{d}^3 p}{(2\pi)^3} \frac{a^2(p^\mu)\,m}{2} 
     \sum_\beta \left(  c^\dagger_\beta(\p) c_\beta(\p) 
-   c_\beta(\p)    c^\dagger_\beta(\p)
\right)\,.\label{eq:7.11}
\eeq
While we have been careful about ordering of the various 
operators, no specific commutators/anticommutators have been
assumed. It is clear that commutative relations between
$ c_\beta(\p)$ and     $c^\dagger_\beta(\p)$ will  yield
a field energy which vanishes for
a general configuration after the usual 
normal ordering. For this reason the associated statistics 
must be fermionic:
\beq
\hspace*{-55pt}\left\{  
c_\beta(\p),\; c^\dagger_{\beta^\prime}(\p^\prime)\right\}
= b(p^\mu)\;\left(2\pi\right)^3 2 E(\p) \delta^3\left(\p-\p^\prime\right)
\delta_{\beta\beta^\prime}\,
\eeq
where $b(p^\mu) \in \mathbb{R}$ will be dictated by the
interpretation of $H$.
Implementing this anticommutator on $H$ results in
\beq
\hspace{-11pt}H=
-  \delta^3\left(\0\right)
 \int {\mathrm{d}^3 p}\; \frac{a^2 (p^\mu)\,m \, b(p^\mu)}{2} 
\sum_\beta 2 E(\p)\, \delta_{\beta\beta} 
 + 
\int  \frac{\mathrm{d}^3 p}{(2\pi)^3} \frac{a^2(p^\mu)\, m}{2} 
     \sum_\beta  2 c^\dagger_\beta(\p) c_\beta(\p)\,. \nonumber\\
\eeq
The factors of $2$ in each of the $\sum_\beta$ occur because
both the self-conjugate as well as 
the anti-self-conjugate parts 
of the field contribute. With this observation in mind, 
and in order to obtain the zero-point energy which is consistent
with fermionic fields, 
we fix $a(p^\mu)$ and 
$b(p^\mu)$ by the requirement
\beq
\hspace*{-70pt}&&\hspace{-166pt} \frac{a^2(p^\mu)\, m}{2} 
= E(\p)\,,\label{eq:az} \\
\hspace*{-70pt}&& \hspace{-166pt} \frac{a^2(p^\mu)\,m\,b(p^\mu)}{2} 
= \frac{1}{2}  \,.
\eeq
That is,
\beq
\hspace*{-136pt}
a(p^\mu)=\sqrt{\frac{2 E(\p)}{m}}\,,\qquad b(p^\mu) = \frac{1}{2 E(\p)}\,.
\eeq
Thus, we have
\beq
\hspace*{-10pt}H=
-  \delta^3\left(\0\right)  \frac{1}{2} 
 \int \!\! {\mathrm{d}^3 p}\;  \sum_\beta 2 E(\p) + 
\int \!\! \frac{\mathrm{d}^3 p}{(2\pi)^3} E(\p) 
     \sum_\beta  2 c^\dagger_\beta(\p) c_\beta(\p)\,. \label{eq:elko}
\eeq
Since $\delta^3(\p) = \left[1/(2\pi)^3\right] \int \mathrm{d}^3 x
\exp(\ri \mathbf{p}\cdot\mathbf{x})$, formally  
\beq
\hspace{-210pt}\delta^3(\0)=
\frac{1}{(2\pi)^3}\int \mathrm{d}^3 x,
\eeq
showing the first term in $H$ to be
\beq
\hspace*{-70pt}\hspace{-36pt} H_0 = - \frac{1}{(2\pi)^3}\int \mathrm{d}^3 x
 \int \!\! {\mathrm{d}^3 p}\;  \sum_\beta 2\left( \frac{1}{2} E(\p)\right)\,.
\label{eq:H}
\eeq
Since in natural units $\hbar=1$ implies $2\pi = h$, $H_0$ represents
an energy  assignment of $-\frac{1}{2} E(\p)$,
for each helicity of the self-conjugate and 
anti-self-conjugate degrees of freedom (hence the factor of 2), 
to each unit-size phase cell $(1/h^3) \mathrm{d}^3 x \mathrm{d}^3 p$ 
in the sense of statistical mechanics.
The new zero-point energy, just as in the Dirac case, comes
with a minus sign, and is infinite\footnote{Our 
discussion
of the new  zero-point energy 
is similar to Zee's in \cite{Zee:ant2003} for the Dirac case.}.
The second term in $H$ of equation (\ref{eq:elko}) shows that each of the 
four degrees of freedom \textemdash~ self-conjugate: $\{-,+\}, \{+,-\}$,
and   anti-self-conjugate: $\{-,+\}, \{+,-\}$ 
\textemdash~contributes exactly the same energy $E(\p)$ to the field
for a given momentum $\p$.

\paragraph{}
\noindent
We thus have the result

\beq
\hspace*{-115pt}\left\{  
c_\beta(\p),\; c^\dagger_{\beta^\prime}(\p^\prime)\right\}
=  \left(2\pi\right)^3  \delta^3\left(\p-\p^\prime\right)
\delta_{\beta\beta^\prime}\,.\label{eq:anticommn}
\eeq
These require imposing\footnote{Actually, as we will see below, 
the anticommutator of $\eta(x)$ with itself turns out to be 
non-vanishing. Thus, it is conceivable to modify (\ref{eq:anticommzn}) 
with the idea of introducing extra terms which do not touch the 
locality result (\ref{eq:locality}) below, but which cancel the 
terms in $\{\eta(x),\eta(x^\prime)\}$. Note that in the derivation of 
positivity of the Hamiltonian and in the evaluation of the 
propagator below only (\ref{eq:anticomm}) is employed. However, 
this idea does not seem to work. Thus, it appears to be reasonable 
to keep (\ref{eq:anticommzn}) in order to obtain the standard 
fermionic harmonic oscillator algebra for 
$c_\beta(\p),c_\beta^\dagger(\p)$.}
\beq
\hspace{-121pt}
\left\{  
c^\dagger_\beta(\p),\; c^\dagger_{\beta^\prime}(\p^\prime)\right\}
=  
\left\{c_\beta(\p),\; c_{\beta^\prime}(\p^\prime)\right\}
=0
\label{eq:anticommzn}
\eeq
to obtain correct particle interpretation. Using equation (\ref{eq:az}),
the field operator and its Elko dual become
\beq
&&\hspace{-60pt} \eta(x)=\int \frac{\mathrm{d}^3 p}{(2\pi)^3}\frac{1}
{\sqrt{2\, m\, E(\p)}} 
\sum_{\beta} \left[c_\beta(\p) \lambda^\mathrm{S}_\beta(\p) 
\mathrm{e}^{-\ri p_\mu x^\mu} + c_\beta^\dagger(\p)  
\lambda^\mathrm{A}_\beta(\p) \mathrm{e}^{+\ri 
p_\mu x^\mu}\right]\,,\label{eq:4.36apn} \\
&&\hspace{-60pt}
\stackrel{\neg}\eta(x)
=\int \frac{\mathrm{d}^3 p}{(2\pi)^3}\frac{1}{\sqrt{2\,m\, E(\p)}} 
\sum_{\beta} \left[c^\dagger_\beta(\p) 
\stackrel{\neg} \lambda^\mathrm{S}_\beta(\p)
\mathrm{e}^{+ \ri p_\mu x^\mu} + c_\beta(\p)  
\stackrel{\neg}\lambda^\mathrm{A}_\beta(\p) 
\mathrm{e}^{-\ri p_\mu x^\mu}\right]\,.\label{eq:4.36bpn}
\eeq
We thus establish the results with which the previous section
opened.

\section{Locality structure}
\label{Sec:ls}

In this section we present the three fundamental anticommutators
required to state the locality  structure of 
the theory under consideration and remark on the massless limit.

\subsection{Fundamental anticommutators for the Elko quantum field}

The emergent non-locality is at the second order. That is, while
the field\textendash momentum anticommutator exhibits the usual
form expected of a local quantum field theory, the 
field\textendash field and momentum\textendash momentum anticommutators
do not vanish.

\subsubsection{Field\textendash momentum anticommutator}

We begin with the equal-time anticommutator

\beq
\hspace*{-108pt}
\left\{
\eta(\x,t),\,\pi(\x^\prime,t)
\right\} 
=  \left\{
\eta(\x,t),\,\frac{\partial}{\partial t}\stackrel{\neg}\eta(\x^\prime,t)
\right\}\,. 
\eeq
The anticommutator on the right-hand side of the above equation
expands to
\beq
 \hspace*{2pt}\int && \frac{\mathrm{d}^3 p}{(2\pi)^3}\f 
  \int \frac{\mathrm{d}^3 p^\prime}{(2\pi)^3}\fp \nonumber\\ 
&&\hspace*{30pt} \times \sum_{\beta \beta^\prime} 
 \Big[
 \left[c_\beta(\p) \lambda^\mathrm{S}_\beta(\p) 
\mathrm{e}^{-\ri p_\mu x^\mu} + c_\beta^\dagger(\p)  
\lambda^\mathrm{A}_\beta(\p) \mathrm{e}^{+\ri p_\mu x^\mu}\right] \nonumber\\
&& \hspace*{30pt}\times \left[c^\dagger_{\beta^\prime}(\p^\prime) 
\stackrel{\neg}\lambda^\mathrm{S}_{\beta^\prime}(\p^\prime)
\,(+\ri E(\p^\prime)) 
\mathrm{e}^{+ \ri p^{\prime}_\mu x^{\prime\mu}}
 + c_{\beta^\prime}(\p^\prime)  
\stackrel{\neg}\lambda^\mathrm{A}_{\beta^\prime}(\p^\prime) 
\,(-\ri E(\p^\prime)) 
\mathrm{e}^{-\ri p^\prime_\mu x^\mu} \right]\nonumber\\
&& \hspace*{30pt}+ 
 \left[c^\dagger_{\beta^\prime}(\p^\prime) 
\left(\stackrel{\neg}\lambda^\mathrm{S}_{\beta^\prime}
(\p^\prime)\right)^\dagger
\,(+ \ri E(\p^\prime))  
\mathrm{e}^{+ i p^{\prime}_\mu x^{\prime\mu}}
+ c_{\beta^\prime}(\p^\prime)  
\left(\stackrel{\neg}\lambda^\mathrm{A}_{\beta^\prime}
(\p^\prime)\right)^\dagger 
\,(-\ri E(\p^\prime)) 
\mathrm{e}^{-ip^\prime_\mu x^\mu}\right]\nonumber\\
&& \hspace*{30pt}\times \left[c_\beta(\p) \left(\lambda^\mathrm{S}_\beta(\p)\right)^\dagger
\mathrm{e}^{-\ri p_\mu x^\mu} + c_\beta^\dagger(\p)  
\left(\lambda^\mathrm{A}_\beta(\p)\right)^\dagger 
\mathrm{e}^{+\ri p_\mu x^\mu}\right]
\Big]\,.
\eeq
Due to the anticommutators imposed by the particle
interpretation, i.e., equations (\ref{eq:anticomm})
and (\ref{eq:anticommz}), the only terms which contribute
are
\beq
\hspace*{-120pt} && \int  \frac{\mathrm{d}^3 p}{(2\pi)^3}\f 
 \int \frac{\mathrm{d}^3 p^\prime}{(2\pi)^3}\fp 
\sum_{\beta \beta^\prime}  \ri E(\p^\prime)\hfill \nonumber \\
\hspace*{-120pt}&& \hspace*{75pt}\times \Big[ 
\lambda^\mathrm{S}_\beta(\p) \stackrel{\neg}
\lambda^\mathrm{S}_{\beta^\prime}(\p^\prime) 
\left\{c_\beta(\p),c^\dagger_{\beta^\prime}(\p^\prime) \right\}
\mathrm{e}^{- \ri p_\mu x^\mu + \ri p^{\prime}_\mu x^{\prime\mu}}\hfill
\nonumber\\
\hspace*{-120pt}&&  \hspace*{75pt} - \;
\lambda^\mathrm{A}_\beta(\p)  \stackrel{\neg}
\lambda^\mathrm{A}_{\beta^\prime}(\p^\prime) 
\left\{ c_\beta^\dagger(\p) ,  c_{\beta^\prime}(\p^\prime)\right\}
\mathrm{e}^{+\ri p_\mu x^\mu - \ri p^{\prime}_\mu x^{\prime\mu}}
\Big]\,.\label{eq:locA}\hfill
\eeq
Here, we have used the fact that 
\beq
\hspace*{20pt}
\left(\stackrel{\neg}\lambda_\beta^\mathrm{S/A}(\p)\right)^\dagger 
\left(\lambda_\beta^\mathrm{S/A}(\p) \right)^\dagger
=
\left( \lambda_\beta^\mathrm{S/A}(\p) \stackrel{\neg}
\lambda_\beta^\mathrm{S/A}(\p) 
\right)^\dagger
=  \lambda_\beta^\mathrm{S/A}(\p) \stackrel{\neg}
\lambda_\beta^\mathrm{S/A}(\p)\,.
\eeq
Using  (\ref{eq:anticomm})
and (\ref{eq:anticommz}) and performing the $\p^\prime$ integration
reduces (\ref{eq:locA})
to
\beq
\hspace*{-50pt}
\int \frac{\mathrm{d}^3 p}{(2\pi)^3}\;\frac{\ri}{2\, m }
\sum_{\beta } \Big[
\lambda^\mathrm{S}_\beta(\p)  \stackrel{\neg}\lambda^\mathrm{S}_{\beta}(\p) 
\mathrm{e}^{-\ri p_\mu x^\mu + \ri p_\mu x^{\prime\mu}} -
\lambda^\mathrm{A}_\beta(\p)  \stackrel{\neg}\lambda^\mathrm{A}_{\beta}(\p) 
\mathrm{e}^{+\ri p_\mu x^\mu - \ri p_\mu x^{\prime\mu}}
\Big]\,.\label{eq:locB}
\eeq
Now first note that we are calculating an equal-time anticommutator.
So, set $t^\prime=t$. Next, in the second terms change integration
variable from $\p$ to $-\p$. This makes 
(\ref{eq:locB}) to be
\beq
\hspace*{-10pt}
\frac{\ri}{2\, m}\int \frac{\mathrm{d}^3 p}{(2\pi)^3}\,
\,\mathrm{e}^{ \ri \mathbf{p}\cdot(\mathbf{x}-\mathbf{x}^\prime)}
 \sum_\beta\left[
\lambda^\mathrm{S}_\beta(\p) \stackrel{\neg}\lambda^\mathrm{S}_{\beta}(\p) 
-
\lambda^\mathrm{A}_\beta(-\p) \stackrel{\neg}\lambda^\mathrm{A}
_{\beta}(-\p) \right]
\,.\label{eq:locC}
\eeq
The indicated spin sum equals $2\,m\left(\mathbb{I}+\mathcal{G}(\phi)\right)$.
This simplifies (\ref{eq:locC}) to 
\beq
\hspace{-82pt}
\ri \int \frac{\mathrm{d}^3 p}{(2\pi)^3}
\,\mathrm{e}^{ \ri \mathbf{p}\cdot(\mathbf{x}-\mathbf{x}^\prime)}
+
\ri \int \frac{\mathrm{d}^3 p}{(2\pi)^3}
\,\mathrm{e}^{ \ri \mathbf{p}\cdot(\mathbf{x}-\mathbf{x}^\prime)}\mathcal{G}
({\phi})\,.
\eeq
The second integral vanishes because in the absence of a preferred 
direction we are free to choose $\x-\x^\prime$ to be aligned 
with the $z$-axis, 
and thus the inner product with $\p$ becomes independent of $\phi$, 
while an integration of $\mawave(\phi)$ over one period vanishes. Therefore,
we obtain
\beq
\hspace*{-158pt}
\left\{
\eta(\x,t),\,\pi(\x^\prime,t)
\right\} 
= \ri \,\delta^3\left(\x-\x^\prime\right)\,. \label{eq:locality}
\eeq
At this point it is emphasized that the `standard result'  
\eqref{eq:locality} ceases to be valid in the presence 
of a preferred direction, in general.

\subsubsection{Field\textendash field, and momentum\textendash momentum, 
 anticommutators}
\label{Sec:ffmm}

By inspection of the above calculation one readily obtains
\beq
\hspace{-40pt}&&\left\{
\eta(\x,t),\,\stackrel{\neg}\eta(\x^\prime,t)
\right\}\,\hfill\nonumber\\ 
\hspace{-40pt}&&\hspace{56pt}=
\frac{1}{2\, m}\int \frac{\mathrm{d}^3 p}{(2\pi)^3}\,\frac{1}{E(\p)}
\,\mathrm{e}^{ \ri \mathbf{p}\cdot(\mathbf{x}-\mathbf{x}^\prime)}
 \sum_\beta\left[
\lambda^\mathrm{S}_\beta(\p) \stackrel{\neg}\lambda^\mathrm{S}_{\beta}
(\p) 
+
\lambda^\mathrm{A}_\beta(-\p) \stackrel{\neg}\lambda^\mathrm{A}_{\beta}
(-\p) \right]
\,.
\eeq
The spin sum on the right-hand side of the above equation
vanishes identically, giving
\beq
\hspace{-202pt}\left\{
\eta(\x,t),\,\stackrel{\neg}\eta(\x^\prime,t)
\right\}=0\,.
\eeq
However, as we shall now show, 
\beq
\hspace{-210pt}\left\{\eta(\x,t),\eta(\x^\prime,t)\right\}\ne 0\,.
\eeq
It turns out that not only is 
the result  non-vanishing, but it depends on
anticommutators of creation and annihilation operators, 
and also on commutators. We, therefore,  
evaluate its vacuum expectation value:
\beq
\hspace{-36pt}\left\langle\hspace{11pt}\Big\vert
\left\{\eta(\x,t), \eta(\x^\prime,t)\right\}
 \right\vert\hspace{11pt}\Big\rangle
&& = 
\int \frac{\mathrm{d}^3 p}{(2\pi)^3}\f 
 \int \frac{\mathrm{d}^3 p^\prime}{(2\pi)^3}\fp \nonumber\\ 
\hspace{-36pt} && \times \sum_{\beta \beta^\prime} 
 \Big[
\left\{c_\beta(\p),\; c^\dagger_{\beta^\prime}(\p^\prime)\right\}
\lambda^\mathrm{S}_{\beta}(\p) 
\left(\lambda^\mathrm{A}_{\beta^\prime}(\p^\prime) \right)^\mathrm{T}
\mathrm{e}^{-\ri p_\mu x^\mu + \ri p^\prime_\mu x^{\prime\mu}}\nonumber\\
\hspace{-36pt}&& +
\left\{
c_{\beta^\prime}(\p^\prime),\; c^\dagger_{\beta}(\p)
\right\}
\lambda^S_{\beta^\prime}(\p^\prime) 
\left(\lambda^\mathrm{A}_{\beta}(\p) \right)^\mathrm{T}
\mathrm{e}^{-\ri p_\mu^\prime x^{\prime\mu} + \ri p_\mu x^{\mu}}
\Big]\,,
\eeq
which exhibits only anticommutators and thus yields
\beq
\hspace{-70pt}&& \left\langle\hspace{11pt}\Big\vert
\left\{\eta(\x,t),\eta(\x^\prime,t)\right\}
 \right\vert\hspace{11pt}\Big\rangle\nonumber \\ 
\hspace{-70pt}  && = \frac{1}{2\,m}
\int \frac{\mathrm{d}^3 p}{(2\pi)^3}\frac{1}{E(\p)} 
\mathrm{e}^{\ri \mathbf{p}\cdot(\mathbf{x}-\mathbf{x}^\prime)} 
\sum_\beta\left[
\lambda^\mathrm{S}_{\beta}(\p) 
\left(\lambda^\mathrm{A}_{\beta}(\p) \right)^\mathrm{T}
+
\lambda^\mathrm{S}_{\beta}(-\p) 
\left(\lambda^\mathrm{A}_{\beta}(-\p) \right)^T\right] 
\,.\label{eq:needslabel}
\eeq
Inserting (\ref{eq:ss17}) and
performing the $\phi$ integration yields the expression
\beq
\hspace{-30pt}\frac{1}{4\,m\, \pi^2}
\int
_{0}^\infty \mathrm{d}p\, \frac{p^2}{\sqrt{p^2+m^2}}  
\int
_0^\pi\mathrm{d}\theta\,\sin(\theta)\,
\, \mathrm{e}^{\ri p r \cos(\theta)} 
\left(p\sin{\theta}\gamma^0\gamma^1 + \sqrt{p^2+m^2} \gamma^2\gamma^5\right)\,.
\eeq
For the $\theta$ integration the following two integrals are needed:
\beq
\hspace*{-150pt}&& \int
_0^\pi\mathrm{d}\theta\,\sin^2(\theta)\,
\, e^{\ri\, p\, r\, \cos(\theta)} 
=\frac{\pi J_1(p\,r)}{p\,r}\,,\label{eq:int1}\\
\hspace*{-150pt}&& \int
_0^\pi\mathrm{d}\theta\,\sin(\theta)\,
\, \mathrm{e}^{\ri\, p\, r\, \cos(\theta)} 
=\frac{2 \,\sin(p\,r)}{p\,r}\,,\label{eq:int2}
\eeq
where $J_1$ is the Bessel function of the first kind. This implies
\beq
\hspace{-130pt} && \Big\langle\hspace{11pt}\Big\vert
\left\{\eta(\x,t),\eta(\x^\prime,t)\right\}
 \Big\vert\hspace{11pt}\Big\rangle
  = 
\frac{1}{4\,m\,\pi\,r^3} \nonumber \\
\hspace{-130pt}&&\times \left[
\int
_{0}^\infty \mathrm{d}(p\,r)
\left\{
\frac{(p\,r)^2 \,J_1(p\,r)}{\sqrt{(p\,r)^2+(m\,r)^2}}\,\gamma^0\gamma^1
+
\frac{2}{\pi}
p\,r  \,\sin(p\,r) \, \gamma^2\gamma^5
\right\}
\right]\,.\label{eq:etaeta}
\eeq
For the final $p$ integration the following two integrals have 
to be evaluated:
\begin{equation}
  \label{eq:int1b}
\hspace*{-132pt}  \int
_0^\infty 
\frac{x^2J_1(x)}{\sqrt{x^2+(mr)^2}}\mathrm{d}x = e^{-mr}(1+mr) \,,
\end{equation}
and
\begin{equation}
  \label{eq:int2p}
\hspace*{-190pt}  
\int
_0^\infty p\sin{(p\,r)}\mathrm{d}p = \pi\delta(r^2)\,,
\end{equation}
where the last result is only true in a distributional sense, 
cf appendix \ref{app:int}. 

Therefore, the final result is established:

\begin{equation}
\hspace*{35pt}
\left\langle\hspace{11pt}\left\vert
\left\{\eta(\x,t),\eta(\x^\prime,t)\right\}
 \right\vert\hspace{11pt}\right\rangle = \frac{1+mr}{4\,m\,\pi r^3}e^{-mr} 
\gamma^0\gamma^1 + \frac{1}{2\,m\,\pi r^2}\delta(r^2)\gamma^2\gamma^5\,. 
\label{eq:nonlocres}
\end{equation}
The first term has a Yukawa-like behavior while the second one is localized. 
Note that  $\gamma^2\gamma^5$ actually is related to 
Wigner's time reversal operator:
\beq
\hspace*{-210pt}\frac{1}{\ri} \gamma^2\gamma^5 = \left(
                                \begin{array}{ccc}
                                 \mathbb{O} &\hspace{5pt}& \Theta\\
                                        \Theta &\hspace{5pt}& \mathbb{O}
                                \end{array}
                                \right)\,,
\eeq 
where $\Theta$ is given in equation \eqref{wt} 
\textemdash~ see \cite{Nachtmann:1989onz}. 
In order to interpret the result it is useful to consider non-locality 
integrated over all separations $\x-\x^\prime$. The angular part just gives 
the usual $4\pi r^2$ in the measure. The relevant radial integrals are
\begin{eqnarray}
\hspace*{-96pt} && \int
_0^\infty \mathrm{d}r \; e^{-mr}=\frac{1}{m}\,, \\
\hspace*{-96pt} && \int
_{r_0}^\infty \mathrm{d}r\; 
\frac{e^{-mr}}{r}=-\gamma-\ln{(mr_0)}+{\mathcal O}(mr_0)\,, \\
\hspace*{-96pt} && \int
_0^\infty \mathrm{d}r \;\delta(r^2) =\frac12
\int
_{-\infty}^\infty \mathrm{d}r\;\delta(r^2)
=\frac14\int
_{-\infty}^\infty\frac{\delta(r)}{|r|}\,\mathrm{d}r\,,
\end{eqnarray}
where $\gamma$ is the Euler\textendash Mascheroni constant.
Obviously, the main (singular) contribution comes from the coincidence 
limit, but the exponential tail does contribute to the finite part. This 
singular behavior is to be contrasted with the regularity of ordinary 
locality (\ref{eq:locality}). Note that we had to regularize one of the 
integrals. Thus, in order to obtain a convergent result which is independent 
from the cutoff $r_0$, we consider the quantity
\begin{equation}
  \label{eq:nonlocren}
\hspace*{-70pt}
  \frac{\mathrm{d}}{\mathrm{d}m} \left[m \int_{\x-\x^\prime}
\Big\langle\hspace{11pt}\Big\vert\left\{\eta(\x,t),\eta(\x^\prime,t)\right\}
 \Big\vert\hspace{11pt}\Big\rangle\right] = \frac{1}{m}\gamma^1\gamma^0 \,.
\end{equation}
It is non-trivial that both the divergent contribution and the 
regulator, $r_0$, drop out after multiplication with $m$ and 
taking the derivative with respect to $m$. The quantity 
\eqref{eq:nonlocren} may be used to estimate the sensitivity 
of non-locality to mass. In the large-$m$ limit, non-locality 
becomes negligible.

The evaluation of
\beq
\hspace*{-01pt}\Big\langle\hspace{11pt}\Big\vert\left\{
\pi(\x,t),\,\pi(\x^\prime,t)
\right\} \Big\vert\hspace{11pt}\Big\rangle 
=  \Big\langle\hspace{11pt}\Big\vert \left\{
\frac{\partial}{\partial t}\stackrel{\neg}
\eta(\x,t),\,\frac{\partial}{\partial t}\stackrel{\neg}\eta(\x^\prime,t)
\right\}\Big\vert\hspace{11pt}\Big\rangle \label{eq:pipi} 
\eeq
can be performed in full analogy to the previous case. Again it is found that the vacuum expectation value is non-trivial, exhibiting non-locality and containing a distributional part.

\subsection{Massless limit and non-locality}
\label{Sec:massless}

It is clear from the above discussion that the massless limit
of the quantum field theory based on \es is singular. Specifically,
it can be seen from the anticommutators studied in
section \ref{Sec:ffmm}. Therefore, the massless-limit 
decoupling of the
the right and left transforming components of \es at the representation
space level may be misleading to some extent. This is due to the fact that
the \es field $\eta(\x,\,t)$, and the associated momentum $\pi(\x,\,t)$, 
together carry additional information to that  
contained in the underlying representation space alone.

It is also our duty to bring to our reader's attention
that according to a pioneering work of Weinberg  \cite{Weinberg:1964ev}
all $(j,0)\oplus(0,j)$ quantum fields, independent of spin,
enjoy a smooth non-singular massless limit. That our result
disagrees with this expected wisdom is quite simple: Weinberg's
analysis explicitly assumes locality\footnote{Section
\ref{Sec:massless} was added to the manuscript 
as an answer to a question by Yong Liu \cite{Liu:2004yong}.}.

\subsection{Signatures of Elko non-locality in the physical amplitudes and
cross-sections}

\label{Sec:FurtherOnNonLocality}

It is worthwhile recalling that the field anticommutator 
between $\eta(x)$ and the conjugate 
momentum $\pi(x)$ as obtained in 
equation \eqref{eq:locality} displays the behavior expected 
from a {\em local} quantum field theory. It is only in the 
anticommutators of $\eta(x)$ (or $\pi(x)$) with itself that 
non-locality emerges. Technically, non-locality is a direct 
consequence of the non-triviality of \es spin sums.

The question now arises if the \es non-locality
carries its signatures in the physical amplitudes and
cross-sections. That the answer to the question
is non-trivial is already hinted at by the fact that \es spin sums
which  underlie non-locality end up modifying the fermionic
propagator (of \textit{Elko}). In the absence of a preferred
direction, this modification to the amplitude for propagation
from one spacetime point to another exhibits itself not as a 
non-local propagation but as a change in the mass dimensionality
of the field. Yet, the non-local anticommutators suggest that
there may be non-vanishing physical correlations for 
\es events carrying spacelike separation. 
In the context of the physics of 
very early universe this, if born out by
relevant S-matrix framework, may have significant bearing on the
horizon problem. The latter asks for causal thermal contact between
spacelike separated regions of the very early universe.

To address this possibility we here undertake
a preliminary investigation of the effect of interactions on physical
amplitudes and cross-sections. We find that non-locality affects
these, and hence we establish that it is endowed with 
observable physical consequences in a non-trivial manner.
In particular, it will be shown that non-locality plays a decisive role 
for higher-order correlation functions.

To this end we split the total Hamiltonian $H$ into free and 
interaction parts, $H=H_0+H_{\rm int}$, where $H_0$ is given by 
\eqref{eq:7.11}. 
For $H_{\rm int}$ one may take for example the last term in 
\eqref{eq:quartic} or \eqref{eq:Higgs} (integrated over 
space and with a relative minus sign, because we need the 
Hamiltonian rather than the Lagrangian density). One may now employ 
the free Hamiltonian to construct the free Heisenberg fields,
\begin{equation}
\hspace*{-120pt}\eta^H_0(t,\x)=\mathrm{e}^{\ri 
H_0(t-t_0)/2}\eta(t_0,\x)\mathrm{e}^{-\ri H_0(t-t_0)/2}\,,
\label{eq:8.3.1}
\end{equation}
where $\eta(t_0,\x)$ is given by equation 
\eqref{eq:4.36apn} with $t$ replaced by $t_0$.
The unusual factors of $1/2$ in the exponent should be noted; they are 
needed, as will be shown below.
From equation  \eqref{eq:7.11} it is evident 
that $H_0^\dagger=H_0$; unitarity 
holds also for the interaction Hamiltonian in \eqref{eq:quartic} or 
\eqref{eq:Higgs} because from equation 
\eqref{zd1} and equation \eqref{z5} it may be 
deduced that
\begin{equation}
\hspace{-196pt}\left(\stackrel{\neg}{\eta}(x)\eta(x)\right)^\dagger = 
\; \stackrel{\neg}{\eta}(x)\eta(x)\,. 
\label{eq:unitarity}
\end{equation} 
In order to verify that $\eta^H_0$ coincides with the free-field 
expression  \eqref{eq:4.36apn} the relations ($n\in\mathbb{N}$)
\begin{equation}
\hspace{-1pt}
(H_0)^n c_\alpha(\p) = c_\alpha(\p)\left(H_0-2E(\p)\right)^n\,,\qquad 
(H_0)^n c_\alpha^\dagger(\p) = c_\alpha^\dagger(\p)\left(H_0+2E(\p)
\right)^n\hfill
\label{eq:8.3.2}
\end{equation}
are very helpful. Again something unusual happens with factors: 
for an ordinary Klein\textendash Gordon 
field one obtains instead $H_0\pm E(\p)$ 
in the expressions on the right-hand side. These factors of $2$ are a 
consequence of the presence of two \ecco-modes: self-dual and 
anti-self-dual 
ones\footnote{Of course it is possible to change the definition of $a$ in \eqref{eq:az} such that the factors $1/2$ and $2$ disappear. But this will change the normalization of $\eta$ and $\pi$ and thus the canonical anticommutator \eqref{eq:locality} would acquire an unwanted factor of $1/2$.}. 


The full Heisenberg field may now be defined as

\begin{equation}
\hspace{-156pt}\eta^H(t,\x)=U^\dagger(t,t_0)\eta^H_0(t,\x)U(t,t_0)
\label{eq:8.3.3}
\end{equation}
with the time-evolution operator

\begin{equation}
\hspace{-160pt}
U(t,t_0)=\mathrm{e}^{\ri H_0(t-t_0)/2}\mathrm{e}^{-\ri H(t-t_0)/2}\,.
\label{eq:8.3.4}
\end{equation}

Standard methods lead to the well known result (again with a 
somewhat unusual factor of $1/2$):
\begin{equation}
\hspace{-140pt}
U(t,t^\prime)=\mathcal{T}\,\exp{\left(-\ri\int_{t^\prime}^td\tilde{t}\;
H^H_{\rm int}(\tilde{t})/2\right)}\,,
\label{eq:8.3.5}
\end{equation}
where $H^H_{\rm int}(\tilde{t})$ is the interaction Hamiltonian written 
as Heisenberg operator with respect to the free fields 
(`interaction picture') and $\mathcal{T}$ denotes time ordering. 
The time-evolution operator fulfills the Schr\"odinger equation

\begin{equation}
\hspace{-184pt}
\ri \frac{\partial}{\partial t}U(t,t^\prime)=
\frac{H^H_{\mathrm{int}}(t)}{2}U(t,t^\prime)\,.
\label{eq:8.3.6}
\end{equation}
With the initial condition $U(t,t)=1$, its solution is given by

\begin{equation}
\hspace{-105pt}
U(t,t^\prime)=e^{\ri H_0(t-t_0)/2}\mathrm{e}^{- \ri 
H(t-t^\prime)/2}\mathrm{e}^{-\ri H_0(t^\prime-t_0)/2}\,.
\label{eq:8.3.7}
\end{equation}
Note that $U(t,t_0)U(t_0,t^\prime)=U(t,t^\prime)$ and $U(t,t_0)(U(t^\prime,t_0))^\dagger=U(t,t^\prime)$, as expected.

Thus, as in standard perturbative quantum field theory one may express correlation functions (with respect to the ground state of the interacting theory) between interacting fields in terms of (a Taylor series of) correlation functions (with respect to the ground state of the free theory) of free Heisenberg fields. Therefore it is sufficient to consider correlators of the form
\begin{equation}
  \label{eq:8.3.8}
\hspace{+54pt} 
 \Big\langle\hspace{11pt}\Big\vert
\mathcal{T} \left\{\stackrel{\neg}{\eta}_0^H(x_1)\eta_0^H(x_1)\stackrel{\neg}{\eta}_0^H(x_2)\eta_0^H(x_2)
\dots\stackrel{\neg}{\eta}_0^H(x_n)\eta_0^H(x_n)\right\}
 \Big\vert\hspace{11pt}\Big\rangle\,.
\end{equation}

The next step is, by analogy to Wick's theorem, to study the relation 
between the time-ordered product of correlators and the normal-ordered 
product. We will not attempt to do this in full generality but rather 
restrict ourselves to the bilinear case and address below essential 
features of the quartic case. The notation
\beq
\hspace{-89pt}
& \hspace*{-5pt}\eta^+:=\mathcal{I}_\beta c_\beta 
\lambda_\beta^\mathrm{S}(\p)\mathrm{e}^{-\ri
p x}\,,
\qquad & \eta^-:=\mathcal{I}_\beta c_\beta^\dagger 
\lambda_\beta^\mathrm{A}(\p)\mathrm{e}^{\ri p x}\,,\\
\hspace{-89pt}&\stackrel{\neg}{\eta}^+:=\mathcal{I}_\beta c_\beta 
\stackrel{\neg}{\lambda}_\beta^\mathrm{A}(\p)\mathrm{e}^{-\ri p x}\,,\qquad & 
\stackrel{\neg}{\eta}^-:=\mathcal{I}_\beta c_\beta^\dagger 
\stackrel{\neg}{\lambda}_\beta^\mathrm{S}(\p)\mathrm{e}^{\ri p x}\,,
\eeq
with the abbreviation

\begin{equation}
  \label{eq:8.3.8b}
\hspace{-179pt}  \mathcal{I}_\beta:=
\int\frac{\mathrm{d}^3p}{(2\pi)^3}\frac{1}{\sqrt{2mE(\p)}}\sum_\beta\,,
\end{equation}
allows the decomposition
\begin{equation}
  \label{eq:8.3.9}
 \hspace{-180pt} 
 \eta=\eta^++\eta^-\,,\qquad\stackrel{\neg}{\eta}
\,=\,\stackrel{\neg}{\eta}^++\stackrel{\neg}{\eta}^-\,.
\end{equation}
For $x^0>y^0$ one obtains 
\begin{equation}
  \label{eq:8.3.10}
\hspace*{-70pt}   
\mathcal{T}\left(\eta(x)\stackrel{\neg}{\eta}(y)\right)=\mathcal{N}\left(\eta(x)\stackrel{\neg}{\eta}(y)\right)+\{\eta^+(x),\stackrel{\neg}{\eta}^-(y)\}\,,
\end{equation}
where $\mathcal{N}$ denotes (fermionic) normal ordering, 
i.e., all $c_\beta$ operators are on the right-hand side and all 
$c_\beta^\dagger$ operators on the left-hand side in each expression. 
By virtue of \eqref{spinsumS}, the anticommutator 
evaluates to\footnote{As usual all anticommutators are to be 
understood as being matrix-valued in spinor-space, rather than 
scalar-valued. That is why instead of the orthonormality relations 
\eqref{zd1}, \eqref{z5} the spin sums \eqref{spinsumS}, 
\eqref{spinsumA} appear subsequently.}
\begin{equation}
  \label{eq:8.3.11}
\hspace{-115pt}
  \{\eta^+(x),\stackrel{\neg}{\eta}^-(y)\} = 
\int\frac{\mathrm{d}^3p}{(2\pi)^3}\frac{\mathbb{I}+\mawave}{2E(\p)}
\mathrm{e}^{-\ri p(x-y)}\,.
\end{equation}
For $x^0<y^0$ analogous steps yield
\begin{equation}
  \label{eq:8.3.12}
 \hspace{-75pt} \mathcal{T}\left(\eta(x)\stackrel{\neg}{\eta}(y)\right)=
\mathcal{N}\left(\eta(x)\stackrel{\neg}{\eta}(y)\right)
+\{\eta^-(x),\stackrel{\neg}{\eta}^+(y)\}\,.
\end{equation}
The spin sum \eqref{spinsumA} yields
\begin{equation}
  \label{eq:8.3.13}
\hspace{-111pt} 
 \{\eta^-(x),\stackrel{\neg}{\eta}^+(y)\} = 
-\int\frac{\mathrm{d}^3p}{(2\pi)^3}\frac{\mathbb{I}-\mawave}{2E(\p)}
\mathrm{e}^{\ri p (x-y)}\,.
\end{equation}
Comparison with \eqref{eq:6.b} establishes a standard result: time-ordering 
decomposes into normal-ordering and contraction, where the latter procedure 
leads to $i$ times the \ecco-propagator $\mathcal{S}_\mathrm{FD}
^\mathrm{Elko}(x,y)$, 
which will be recalled for convenience:
\begin{equation}
  \label{eq:elkopropagator}
\hspace*{-35pt}
  \mathcal{S}_\mathrm{FD}^\mathrm{Elko}(x,y) 
= \lim_{\epsilon\to 0^+}\int \frac{\mathrm{d}^4 p}{(2\pi)^4}\;
\mathrm{e}^{-\ri p_\mu(x^\mu - y^\mu)}\;  
\left[\frac{ \left(\mathbb{I}+\mawave(\phi)\right) }  
{p_\mu p^\mu - m^2 + \ri\epsilon}\right]\,.
\end{equation}
It equals the Klein\textendash Gordon propagator 
only in the absence of a preferred 
direction. Thus, in the presence of some non-trivial background (which 
may be provided also by {\ecco} itself in higher-order perturbation theory) 
the terms proportional to $\mawave$ no longer will be negligible.

In order to unravel the appearance of non-locality one has to 
study higher-order correlation functions. 
Therefore we now address the quartic expression

\begin{equation}
\label{eq:8.3.14}\hspace*{-100pt}
\mathcal{T}\left(\eta(x)\stackrel{\neg}{\eta}(y)\eta(z)\stackrel{\neg}
{\eta}(w)\right)=\mathcal{N}\left(\eta(x)\stackrel{\neg}{\eta}(y)\eta(z)
\stackrel{\neg}{\eta}(w)\right)+\,{\rm contractions}\,.
\end{equation}
The decomposition into $\pm$ components yields 16 terms which we 
denote by $[\pm\pm\pm\pm]$, 5 of which are already normal-ordered 
($[++++], [-+++], [--++], [---+], [----]$). In two 
terms a single contraction of the type 
\eqref{eq:8.3.11} (or \eqref{eq:8.3.13}) 
is sufficient ($[+-++]$, $[--+-]$) and in two terms two contractions 
of that type appear ($[+-+-]$, $[-+-+]$). But in the remaining seven 
terms an additional feature arises which is unprecedented, namely, 
is absent for Dirac fermions. We will discuss in full detail one of these 
seven terms, $[++-+]$, in order to make this peculiarity 
explicit and to establish the connection with non-locality:
\beq
\eta^+(x)\stackrel{\neg}{\eta}^+(y)&& \eta^-(z)\stackrel{\neg}{\eta}^+(w)  =  
 - \eta^+(x)\eta^-(z)\stackrel{\neg}{\eta}^+(y)\stackrel{\neg}{\eta}^+(w) 
+ \eta^+(x)\{\eta^-(z),\stackrel{\neg}{\eta}^+(y)\}\stackrel{\neg}
{\eta}^+(w) \nonumber\\
&&\hspace{-67pt} = 
 \eta^-(z)\eta^+(x)\stackrel{\neg}{\eta}^+(y)\stackrel{\neg}{\eta}^+(w) 
- \{\eta^+(x),\eta^-(z)\}\stackrel{\neg}{\eta}^+(y)\stackrel{\neg}{\eta}^+(w) 
 + \eta^+(x) ({\rm prop.~part}) \stackrel{\neg}{\eta}^+(w) \nonumber\\
&&\hspace{-67pt}= 
\mathcal{N}\left(\eta^+(x)\stackrel{\neg}{\eta}^+(y)\eta^-(z)\stackrel
{\neg}{\eta}^+(w)\right)
- \{\eta^+(x),\eta^-(z)\}\stackrel{\neg}{\eta}^+(y)\stackrel{\neg}
{\eta}^+(w) + {\mbox{ordinary contraction.}} \nonumber
\eeq
The crucial point here is that the anticommutator in the middle term in the last line is {\em non-vanishing}, in stark contrast to Dirac fermions. 
A simple calculation yields
\begin{equation}
\label{eq:8.3.15}
\hspace*{30pt}
\{\eta^+(x),\eta^-(z)\} = \int\frac{\mathrm{d}^3p}
{(2\pi)^3}\frac{1}{\sqrt{2mE(\p)}}
\sum_\beta \lambda_\beta^\mathrm{S}(\p)
\left(\lambda_\beta^\mathrm{A}(\p)\right)^T 
\mathrm{e}^{-\ri p (x-z)}\,,
\end{equation}
which may be evaluated following \eqref{eq:ss17}. Incidentally, 
this is the {\em same} spin sum, 
the non-vanishing of which had been responsible for the emergence of non-locality in \eqref{eq:needslabel}. 

Thus, Wick's theorem in abbreviated form reads
\beq
\label{eq:8.3.16}
\hspace*{-11pt}
\mathcal{T}\left(\eta(x^1)\stackrel{\neg}{\eta}(x^1)
\dots\eta(x^n)\stackrel{\neg}{\eta}(x^n)\right)=
\mathcal{N}\left(\eta(x^1)\stackrel{\neg}{\eta}(x^1)\dots\eta(x^n)
\stackrel{\neg}{\eta}(x^n)\right)+\,{\rm contractions}
\,,\nonumber\\
\eeq
where `contractions' decompose into `ordinary contractions' 
between $\eta$ and $\stackrel{\neg}{\eta}$, 
yielding the propagator $\ri \mathcal{S}_{\mathrm{FD}}^\mathrm{Elko}$, 
and into `non-local contractions' between $\eta$ and itself or 
between $\stackrel{\neg}{\eta}$ and itself. That such contractions 
are non-vanishing is a novel feature of {\ecco} as compared to the Dirac case,
 and directly linked to the non-locality anticommutator \eqref{eq:needslabel}.

Equipped with the tools of Wick's theorem (which may be derived 
in a standard fashion by complete induction) one may now derive 
the Feynman rules for {\ecco}; no unusual features are found in 
this way, besides the remarkable behavior encoded in the 
$\mawave$-part of the \ecco-propagator and the non-triviality 
of contractions of {\ecco} with itself.

With these results the  primary
task of this subsection has been established,
i.e., we have shown that for an interacting 
theory the physical amplitudes and cross-sections 
carry non-trivial signatures of non-locality.
To examine the quantitative impact on the cosmological horizon problem 
now requires calculation of correlations between 
\es scattering amplitudes for spacelike separated events.
Such a calculation is far from trivial, but the analysis of this
section makes it, in principle, well defined\footnote{This
section was added on suggestion from a \textit{JCAP} referee.}.

\section{Identification of Elko  with dark matter}
\label{Sec:Identification}

Having established the kinematic structure of an \textit{Elko}-based framework
via the introduction
of $\eta(x)$ and having studied its various properties, our first task 
is simply to name the particles the field $\eta(x)$ describes.
We suggest the symbols $\varsigma$ and 
$\stackrel{\neg} {\varsigma}$ for these particles 
\textemdash~ with obvious symbolic distinction from the Dirac framework.

The next question which arises is what interactions
can  $\varsigma$ and 
$\stackrel{\neg} {\varsigma}$ carry with the Standard Model
fields. 
Towards the goal of answering this question we 
note that one of the reasons 
for the success of the Standard Model stems from the fact that only a very 
limited number of terms appear in the action. Simple arguments of 
power counting renormalizability prohibit for example 4-Fermi interactions or 
non-polynomial potentials for the Higgs. While these non-renormalizable 
terms are not strictly forbidden, in the low-energy regime that we are 
able to explore experimentally they play essentially no role because they 
are suppressed by factors $(k/M_\mathrm{U})^d$ for 
$k\ll M_\mathrm{U}$ ($k$ characterizes
the low-energy scale),   
where $4+d$ is the mass dimension of the
interaction, and $M_\mathrm{U}$ is the GUT or the Planck scale 
(cf e.g. section 12.3 in \cite{Weinberg:1995sw}).

Having addressed
the relevance of non-local contributions for \textit{higher-order} 
perturbation theory in the previous
section, and to proceed further, we explicitly state an
 observation and the \textit{working assumption} we must make 
\begin{quote}

Non-locality of \es appears to prohibit a naive application of power-counting 
arguments. Nevertheless, we take them as good starting point
in the hope that it could well be that non-locality is 
a higher-order effect because, after all, the equal-time 
anticommutator between $\eta(x)$ 
and  the associated canonical momentum $\pi(x)$ is local. 
So, non-locality manifests itself when at least 
two \es fields or two momenta appear together in the same expectation 
value.
\end{quote}

Therefore,
the interactions of 
$\varsigma$ and 
$\stackrel{\neg} {\varsigma}$ with the Standard Model fields
are entirely governed by mass dimension one of $\eta(x)$ and power-counting 
arguments. As a result the following 
additional  structure 
for the Standard Model  Lagrangian density comes to exist:

\begin{align}
  \label{eq:elkorenSM}
  & \mathcal{L}^{\rm new}  (x)   = \mathcal{L}^\mathrm{\ecco} (x) + 
\mathcal{L}^{\rm int}(x)\,,\\
  & \mathcal{L}^\mathrm{\ecco} (x)  = \partial^\mu \stackrel{\neg}{\eta}(x) 
\partial_\mu\eta(x) - m^2 \stackrel{\neg}{\eta}(x) \eta (x) + 
\alpha_\mathrm{E}\; 
\left[\stackrel{\neg}{\eta}(x)\eta(x)\right]^2\,, \label{eq:quartic}\\
  & \mathcal{L}^{\rm int}_{\phi\eta}(x) = 
\lambda_\mathrm{E} \;\phi^\dagger(x) \phi(x)\,
\stackrel{\neg}{\eta}(x) 
\eta(x)\,, 
\label{eq:Higgs}
\end{align}
where $\phi(x)$ is the Higgs doublet, 
$m$ the Elko mass, and $\lambda_\mathrm{E},\alpha_\mathrm{E}$
 are \textit{dimensionless} coupling constants. 
Obviously, if more than one \es field is present 
mixings between them are possible. However, as there seems to be no way to 
distinguish different \es particles \textemdash~ 
as opposed to the Standard Model 
fermions which are distinguished by their charges \textemdash~ 
it is sensible to 
introduce only one \es field. Of course, if more than one scalar field 
should be present in Nature additional interactions of the form 
\eqref{eq:Higgs} are possible. The fact that \eqref{eq:elkorenSM} 
contains no interactions with gauge fields or other Standard Model (SM) 
fermions explains why \es has not been detected yet. However, since 
it does interact with the Higgs there is a possibility that it might 
be discovered at LHC. Clearly, a thorough analysis of this issue would
 be desirable.

While 
$\varsigma$ and 
$\stackrel{\neg} {\varsigma}$
 may carry a coupling to an Abelian gauge field 
with associated field strength $F_{\mu\nu}(x)$, for example of the form 
\beq
\hspace*{-80pt}
\mathcal{L}^{\rm int}_{\eta F}(x) = \epsilon_\mathrm{E}
\stackrel{\neg}{\eta}(x)\left[\gamma^\mu,\gamma^\nu\right] 
\eta(x)\, F_{\mu\nu}(x)\,, \label{eq:zimpok17}
\eeq 
the coupling constant has to be very small. Such terms 
affect photon propagation because they
lead to an effective-mass term for the photon and the latter has been
severely constrained \cite{Accioly:2004bm,Luo:2003rz}.
The indicated smallness is not unexpected as $\eta(x)$ is
neutral with respect to local $U(1)$ gauge transformations.
Thus, the dominant interaction between 
\es and particles of the Standard Model is expected to be via 
(\ref{eq:Higgs})\footnote{We are grateful to Dima Vassilevich 
for raising a question in this regard.}.

It is also worth noting that the SM-counterpart
of the quartic self-interaction contained in (\ref{eq:quartic}) is suppressed
as $(k/M_\mathrm{U})^2$ in sharp contrast to \textit{Elko}. 
This \es self-interaction could be of significant 
physical consequence, as we shall remark below.

We end this section with a conclusion that dawns on us 
unexpectedly: the fact that 
$\varsigma$ and 
$\stackrel{\neg} {\varsigma}$
almost do not interact with the matter content 
of the Standard Model makes  them 
prime dark matter candidates.

\section{Constraining the
Elko mass and the relevant cross-section}
\label{Sec:identification2}

In the next two subsections
we strive to constrain the mass and the relevant cross-section
of these candidate particles. While the first of these two subsections
involves a conventional exercise on relic density, the second subsection
studies the gravitational collapse of a primordial Elko cloud.
It is encouraging that the latter calculation finds
an element of consistency with the relic density analysis.
The   collapse, with eventual rebound,  
should be considered as a process
that precedes
virialization of the dark matter cloud.
We thus consider both of these
approaches complementary to an extent. The collapse analysis also predicts
an explosive event in the early history of each galaxy
containing dark matter. It should be considered a candidate for
cosmic gamma-ray bursts \cite{Meszaros:2001ig}.

\subsection{Relic density of \ecco: constraining the relevant 
cross-section, mass, and a comparison with WIMP}
\label{Sec:Relic}

In a series of publication the case for a MeV-range dark matter
particle has been gaining strength
\cite{Boehm:2003bt,Boehm:2004gt,Beacom:2004pe,Casse:2004gw,Fayet:2004kz}. 
At the same time, as will be seen in section \ref{Sec:cem}, same indication 
arises from an entirely different consideration.
For these reasons we  
concentrate on the MeV-range \es mass.

In its essentials the analysis 
that follows is an adaptation
of the one  presented by Bergstr\"om and Goobar
\cite{Bergstroem:2004Goobar}, Dodelson \cite{Dodelson:2003sco}, and 
Kolb and Turner  \cite{Kolb:1994Tur} for
weakly interacting massive particles (WIMPs). 
Since a comparison with a generic WIMP scenario (of which the 
supersymmetry-implied candidates are a subset) may be useful,
we shall present the analysis in such a manner as to make
this apparent\footnote{For convenience,
we here use units where $c$, $\hbar$, and 
Boltzmann constant $k_B$ are set to 
unity.}.

A generic WIMP scenario for dark matter considers two
heavy particles $X$ which annihilate into two light particles.
The light particles are considered to be tightly coupled
to the cosmic plasma, so that their number density in the
cosmic comoving frame 
equals $n^\mathrm{EQ}$, the equilibrium value. 
This leaves one unknown density, i.e.,
that associated with the WIMP dark matter particle. Its
evolution is determined by
the Boltzmann equation.
For \es the thermodynamic situation is as follows.
The thermal contact of the cosmic plasma made of the Standard Model
particles with 
\es is provided by the Higgs: 
$\varsigma + \stackrel{\neg}\varsigma \rightleftharpoons \phi_H + 
\phi_H$\footnote{Here, $\phi_H$ represents a Higgs particle
and shall be assumed to have a mass of roughly $150\;\mbox{GeV}$.
We do not use the symbol $H$ for Higgs particle as we wish to
reserve that symbol for the Hubble rate.}.
The coupling constant $\lambda_\mathrm{E}$ 
of equation  (\ref{eq:Higgs}) then determines
the \es annihilation rate through
the
 thermally averaged product of the cross-section 
$\sigma_{\varsigma\,\stackrel{\neg}\varsigma \rightarrow\phi_H \phi_H}$
and the M\o ller velocity (which in the cosmic comoving frame can be
identified with the relative velocity of the two annihilating
\textit{Elko}):  $\langle
\sigma_{\varsigma\,\stackrel{\neg} \varsigma \rightarrow\phi_H \phi_H}
\vert \v_{M{\o} l} \vert\rangle$.

In the considered scenario, $m_\varsigma \ll m_{\phi_H}$,  whereas for
the WIMP scenario $m_X \sim m_{\phi_H}$.
The quartic self-interaction of \es given in equation (\ref{eq:quartic})
does \textit{not} change the
comoving density of \textit{Elko},  $n_\varsigma$, at the tree level.
Instead, in conjunction with gravitational interactions a 
non-zero $\alpha_\mathrm{E}$
of equation  (\ref{eq:quartic}) will contribute to fluctuations in
$n_\varsigma$ which seed the large-scale structure 
formation in the Cosmos.

With these remarks in mind, and under the
assumption that Fermi statistics can be ignored for a zeroth-order 
understanding of the relic abundance of $\varsigma$, 
the \es number density  $n_\varsigma$ is governed by\footnote{See, e.g., 
section 
9.2 of \cite{Bergstroem:2004Goobar} for the physical meaning of each
of the terms in equation (\ref{Eq:BoltzmanEq}). 
More detailed textbook derivations
with a view towards various involved assumptions 
can be found, for instance, in 
\cite{Dodelson:2003sco,Kolb:1994Tur}.
}
\beq
\hspace*{-65pt}
a^{-3} \frac{\mathrm{d} \left(n_\varsigma a^3\right)}{\mathrm{d} t}
= \left\langle\sigma_{\varsigma\, \stackrel{\neg} \varsigma \rightarrow\phi_H \phi_H}
\vert \v_{M\o l}\vert\right\rangle \left\{ \left(
n_\varsigma^\mathrm{EQ}\right)^2 
- n_\varsigma^2\right\}.\label{Eq:BoltzmanEq}
\eeq
Here, $a$ is the cosmic scale factor. Since $CPT$ is respected 
by the \es framework the $n_\varsigma$ divides equally among $\varsigma$ and
$\stackrel{\neg}\varsigma$.\footnote{We qualify the statement regarding $CPT$
with the following observations: for the Standard Model fields
$(CPT)^2 = +\mathbb{I}$, while for the unusual Wigner class under
consideration, i.e., \textit{Elko}, we have $(CPT)^2 = - \mathbb{I}$, 
as proven in section \ref{Sec:cpt}.
This makes any statement on $CPT$ a bit subtle.  If one
is considering purely the Standard Model physics $CPT$ symmetry is fine. The same remains true if one is considering purely
unusual Wigner classes. But as soon as one has both the standard and unusual
Wigner fields present, the differing actions of $CPT$ can induce a relative
sign between the amplitude terms which involve only the Standard Model fields,
and those involving an unusual Wigner class. However, as long as an amplitude 
does not contain the contraction of \es with a Dirac field, no such
relative sign comes into existence (for the quadratic appearances
of Dirac fields, as well as \textit{Elko} fields,
the effect of $(CPT)^2$ is identical). So, at least for the process
under consideration , i.e., $\varsigma + \stackrel{\neg}\varsigma 
\rightleftharpoons \phi_H + \phi_H$, $CPT$ is respected.
The general situation is likely to be more subtle and
 is expected to violate the $CPT$ symmetry.}
 The symbol $n_\varsigma$ is
used as an abbreviation for 
$n_{\varsigma\stackrel{\neg}\varsigma}$, 
the combined comoving number density of $\varsigma$
and $\stackrel{\neg}\varsigma$.

Before $T\approx m_{\phi_H}$, Standard Model particles
and the \es slosh back and forth thermally. Metaphorically
speaking, Higgs serves
as a two\textendash headed fountain: on its one end it sucks and 
sprays the \es into the cosmic plasma of the early
universe and on the other it does the same 
for the Standard Model particles. Once the temperature
falls below the Higgs mass, the high-energy tails
of the \es thermal distribution continues this
process for a while.  For a representative 
$m_\varsigma = 20\;\mbox{MeV}$ for $\varsigma$ mass,
and with an example value of  
$\left\langle\sigma_{\varsigma\,\stackrel{\neg}\varsigma 
\rightarrow\phi_H \phi_H}\vert \v_{M\o l}\vert\right\rangle 
= 1 \;\mbox{pb}$,
freeze out occurs at a temperature of about $2.1\;\mbox{MeV}$ (see
below).
The  viability of \es as a serious
dark matter candidate is established 
by showing that,
for this representative example,
the  \es energy density $\rho_{\varsigma}$ today is of the same  
order as the critical energy density at the present epoch.

The
analysis of the  angular power spectrum of the cosmic microwave background
 implies a 
spatially flat Friedmann\textendash Robertson\textendash Walker
universe \cite{Sievers:2002tq}. 
For such a scenario, the scale factor $a$ and 
the Hubble rate $H$ are connected via the usual definition 
$H(t):=
\dot{a}/a = \sqrt{8 \pi G \rho(T)/3}
=:H(T)$,
where $\rho(T)$ is the energy density of all the matter
at a given temperature, and the overdot indicates a derivative
with respect to time. Since non-relativistic matter components 
contribute negligibly, it is given by
\begin{equation}
\hspace*{-40pt}\rho(T) = \frac{\pi^2}{30} T^4
\left[\mathop{{\sum}'}_{i=bosons} 
g_i + \frac{7}{8} \mathop{{\sum}'}_{i=fermions} g_i\right] := g_\ast(T)\,  \frac{\pi^2 \,T^4}{30} 
\,,
\end{equation}
where $g_i$ represents helicity degrees of freedom associated with
a specific component in the cosmic plasma, 
and the primed summation sign means that
the summation is confined to those particles which
may be considered relativistic at the relevant temperature, $T$. 
At temperatures of the order of $1-100$ $\mbox{MeV}$, the
contributions arise from photons ($g_\gamma=2$),
three flavors of neutrinos and associated antineutrinos
($g_\nu=6$), and electrons and positrons ($g_{e^\pm}
=4$). This yields $g_\ast(1-100\;\mbox{MeV}) = 10.75$. If there
is a relativistic \es component in the indicated point in the  
temperature range then  $g_\ast(1-100\;\mbox{MeV}) = 14.25$.

Now, 
multiplying and dividing the factor of $\left( n_\varsigma a^3\right)$ on
the left-hand side of equation (\ref{Eq:BoltzmanEq}) by $T^3$, and taking
note of the fact that
$T$ roughly scales as $a^{-1}$, allows us
to move $\left(a T\right)^3$ outside the derivative. This yields
\beq
\hspace*{-74pt}
T^3  \frac{\mathrm{d} }{\mathrm{d} t}
\left(\frac{n_\varsigma}{T^3}\right)
= \left\langle\sigma_{\varsigma\,\stackrel{\neg}
\varsigma \rightarrow\phi_H \phi_H}
\vert \v_{M\o l}\vert\right\rangle \left\{ 
\left(n_\varsigma^\mathrm{EQ}\right)^2 
- n_\varsigma^2\right\}.\label{Eq:BoltzmanEqZ}
\eeq
Now define two new dimensionless variables,
$Y:={n_\varsigma}/{T^3}$, and $x 
:= {m_\varsigma}/{T}$,
as well as
\[\hspace*{-05pt}
H(m_\varsigma):= 
\sqrt{\frac{ 4 \pi^3 G \,g_\ast(m_\varsigma) \,m_\varsigma^4}{45}}\,,\quad
\lambda_\varsigma:=  \frac{m_\varsigma^3}{H(m_\varsigma)} 
\left\langle\sigma_{\varsigma\,\stackrel{\neg}
\varsigma \rightarrow\phi_H \phi_H}
\vert \v_{M\o l}\vert\right\rangle\,.
\]
That done, the evolution equation becomes
\beq
\hspace*{-205pt}
  \frac{\mathrm{d} Y}{\mathrm{d} x}
= \frac{\lambda_\varsigma} {x^2}
\left\{ Y_\mathrm{EQ}^2 
- Y^2\right\}\,.\label{Eq:BoltzmanEqB}
\eeq
For a representative 
$m_\varsigma = 20\;\mbox{MeV}$ \es
particles, and taking 
$\langle\sigma_{\varsigma\,\stackrel{\neg}
\varsigma \rightarrow\phi_H \phi_H}
\vert \v_{M\o l}\vert\rangle = 1 \;\mbox{pb}$ 
we find $H(m_\varsigma) =  1.8 \times 10^{-22}
\;\mbox{GeV}$. Thus\footnote{$1\;\mbox{pb} = 2.5681 \times 10^{-9}
\;\mbox{GeV}^{-2}$.}
\beq
\hspace*{05pt}
\lambda_\varsigma = 1.2 \times 10^{8}\quad \left[\mbox{for}\;
\left\langle\sigma_{\varsigma\,\stackrel{\neg}
\varsigma \rightarrow\phi_H \phi_H}
\vert \v_{M\o l}\vert\right\rangle = 1 \;\mbox{pb},\;
m_\varsigma= 20\;\mbox{MeV}\right]\,.\label{eq:lambda_varsigma}
\eeq
Given $\lambda_\varsigma \gg 1$, 
the $Y(T)$ tracks $Y_\mathrm{EQ}$ 
at early times, i.e., for $1 < x \ll x_\mathrm{f}$ (where
superscript/subscript  $\mathrm{f}$  generically 
represent \textit{freeze out}).
For late times, i.e., $x \gg x_\mathrm{f}$, 
$Y(T)$ 
tracks $Y_\mathrm{EQ}(T)$ very poorly, and
$Y(T) \gg Y_\mathrm{EQ}(T)$. Under this circumstance, the evolution of
$Y(T)$ is determined by
\beq
\hspace*{-180pt}
  \frac{\mathrm{d} Y}{\mathrm{d} x}
\approx -\frac{\lambda_\varsigma} {x^2}\, 
 Y^2, \quad (x \gg x_\mathrm{f})\,. \label{Eq:BoltzmanEqC}
\eeq
Integrating this equation from the freeze-out 
epoch $x = x_\mathrm{f}$ to 
very late times $x=\infty$ (i.e., when all annihilations 
into Higgs have become extremely rare), and using the fact that
on physical grounds 
$Y_\mathrm{f} \gg Y_\infty$ 
(i.e., \es annihilations in evolution from
$x_\mathrm{f}$ to 
$x_\infty$ deplete $\varsigma$ and $\stackrel{\neg}\varsigma$), we get
\beq
\hspace*{-267pt}
Y_\infty \approx \frac {x_\mathrm{f}}{\lambda_\varsigma}\,.\label{eq:yinfty}
\eeq 

A good analytical estimate for $x_\mathrm{f}$ can be obtained as follows.
Define the departure from equilibrium by
$
\Delta := Y - Y_\mathrm{EQ}
$,
and use equation (\ref{Eq:BoltzmanEqB})
with the substitution
$Y^2 - Y_\mathrm{EQ}^2 = \Delta 
\left(2  Y_\mathrm{EQ} + \Delta\right)$.
This results in
\beq
\hspace{-144pt}
\frac{\mathrm{d} \Delta}{\mathrm{d} x} =
-\frac{\lambda_\varsigma}{x^2}\,
 \Delta 
\left(2  Y_\mathrm{EQ} + \Delta\right) - 
\frac{d Y_\mathrm{EQ}}{d x}\,.\label{eq:BoltzmanD}
\eeq
Now at freeze out $x = x_\mathrm{f}$ corresponds
to $Y$ ceasing to track  $Y_\mathrm{EQ}$; that is,
at freeze out  
$\Delta(x_\mathrm{f}) \simeq Y_\mathrm{EQ}(x_\mathrm{f})$.
Substituting this into equation 
(\ref{eq:BoltzmanD}), and re-arranging
yields
\beq
\hspace{-124pt}
\frac{2} {Y_\mathrm{EQ}(x)} 
\frac{\mathrm{d} Y_\mathrm{EQ}(x)}{\mathrm{d} x} 
\Bigg\vert_{x =  x_\mathrm{f}} = - 
\frac{\lambda_\varsigma}
{x_\mathrm{f}^2}\, 3 Y_\mathrm{EQ}(x_\mathrm{f})\,.\label{eq:BoltzmanE}
\eeq
Using $Y_\mathrm{EQ}(x) = A\, x^{3/2}\, e^{-x}$ (for $x \geq 3$), 
with\footnote{Here $g$ represents the 
number of helicity degrees of freedom of \textit{Elko},
while $g_\ast(m_\varsigma)$ is roughly $10.75$.}
$A:=({45}/{2 \pi^4})\sqrt{{\pi}/{8}} \,
{g}/{g_\ast(m_\varsigma)}$,
the left-hand side of equation (\ref{eq:BoltzmanE}) reduces to
$-2 + (3/x_\mathrm{f})$. For $x_\mathrm{f} \gg 1$ (to be verified below), this
factor takes the value of $\approx -2$, resulting in
\beq
\hspace*{-210pt}
Y_\mathrm{EQ}(x_\mathrm{f}) = 
2 \left(\frac{x_\mathrm{f}^2}
{3\, \lambda_\varsigma}\right)\,.
\eeq
This gives  (cf equation (5.43) of \cite{Kolb:1994Tur})
\beq
\hspace*{-115pt}
x_\mathrm{f} \approx \frac{1}{\sqrt{2}}
\Big[\ln \left[3\, \lambda_\varsigma A \right] 
- 0.5 \ln\left\{
\ln \left[3 \,\lambda_\varsigma\, A\right]\right\}\Big]
\,.\label{eq:xf}
\eeq 
Using the representative value of $\lambda_\varsigma$ given  in
equation (\ref{eq:lambda_varsigma}), equation (\ref{eq:xf}) 
yields $x_\mathrm{f} \approx 9.7$. In conjunction with 
equation (\ref{eq:yinfty}) this implies that a rough
estimate of relic abundance of \es is $Y_\infty \approx 8.5\times 
10^{-8}$.
For  a $20\;\mbox{MeV}$ \es particle, at freeze out the temperature  is
$T_\mathrm{f} \approx 2.1\;\mbox{MeV}$. Given this result we envisage that 
the asymptotic value $Y_\infty$ is reached in an epoch when \textit{no}
significant
reheating of the cosmic plasma occurs. That is,
by $T_\mathrm{f}\approx 2.1\;\mbox{MeV}$ all the quarks and leptons, except
$e^\pm$, have already annihilated and heated the photons in the
cosmic plasma. Since $T_\mathrm{f}\approx 2.1\;\mbox{MeV}$ is very close
to the boundary of where a slight reheating by $e^\pm$ may occur,
and since $T_\infty \ll T_\mathrm{f}$,
this justifies the working assumption of no significant reheating of
the cosmic plasma after $T_\infty$ is reached.
 For this reason
\beq
\hspace*{-235pt}
\left(\frac{a_\infty T_\infty}{ a_0 T_0}\right)^3_\varsigma \approx 1\,,
\label{eq:noreheating}
\eeq
where the subscript $\infty$ indicates the respective values in the epoch
when $Y$ reaches its asymptotic value, and the subscript $0$ represents
the present epoch.
This contrasts dramatically with a heavy WIMP scenario, 
where significant reheating takes place, and\\
$\left({a_\infty T_\infty}/{ a_0 T_0}\right)^3_\mathrm{WIMP} 
\approx \frac{1}{30}$.
Therefore, the \es energy density \textit{today} is
\beq
\hspace*{-132pt}
\rho_\varsigma = \left(\frac{a_\infty T_\infty}{ a_0 T_0}\right)^3
m_\varsigma Y_\infty T_0^3 \approx m_\varsigma Y_\infty T_0^3 
\eeq
For \, the representative \, case considered above,\, for the present epoch
characterized by 
$T_0=2.725\;\mbox{K}$ ($1\mbox{K} = 8.617 \times 10^{-14}\;\mbox{GeV}$)
we find
\beq
\hspace*{-65pt}
\rho_\varsigma = 2.2 \times 10^{-47}\;\mbox{GeV}^4 \quad \left[\mbox{for}\;
\left\langle\sigma_{\varsigma\,\stackrel{\neg}
\varsigma \rightarrow\phi_H \phi_H}
\vert \v_{M\o l}\vert\right\rangle = 1 \;\mbox{pb},\;
m_\varsigma= 20\;\mbox{MeV}\right]\,.\label{eq:lambda_varsigmaB}
\eeq
Since the critical density today, assuming $h=0.72$,
is $\rho_{cr} = 4.20\times 10^{-47}\;\mbox{GeV}^4$, the obtained
result for \es contribution to dark matter energy density today is
very encouraging. This means that by choosing $m_\varsigma$
around $20\;\mbox{MeV}$ and 
$\left\langle\sigma_{\varsigma\,\stackrel{\neg}
\varsigma \rightarrow\phi_H \phi_H}
\vert \v_{M\o l}\vert\right\rangle$ around  $1 \;\mbox{pb}$
one can readily obtain $\Omega_{\varsigma}\defn \rho_\varsigma/\rho_{cr}
\approx 0.3$. To see this explicitly, the above formalism immediately
yields
\beq
\hspace*{-20pt}
\Omega_\varsigma = 
\frac{1.89 \times 10^{-2}}{\langle \sigma v\rangle_{\mbox{\tiny  pb}}} 
\Big[ 2 \ln\left(9.31 \times 10^5\,
 \langle \sigma v\rangle_{\mbox{\tiny  pb}}\, m_\varsigma^{\mbox{\tiny  MeV}}
\right)
 -\,\ln\big[\ln\left(9.31 \times 10^5\,
\langle \sigma v\rangle_{\mbox{\tiny  pb}}\, m_\varsigma^{\mbox{\tiny MeV}} 
\right)\big]\Big]\,,\nonumber\\\label{eq:omega_varsigmaD}
\eeq
where $m_\varsigma^{\mbox{\tiny  MeV}}$ is the \es mass in \textit{MeV}, while
$\langle \sigma v\rangle_{\mbox{\tiny  pb}}$ is the 
$\left\langle\sigma_{\varsigma\,\stackrel{\neg}
\varsigma \rightarrow\phi_H \phi_H}
\vert \v_{M\o l}\vert\right\rangle$ in \textit{picobarns}.
$\Omega_\varsigma$ carries a strong dependence on
$\langle \sigma v\rangle_{\mbox{\tiny  pb}}$, 
and only a much weaker dependence on
 $m_\varsigma^{\mbox{\tiny  MeV}}$.
For $\langle \sigma v\rangle_{\mbox{\tiny  pb}} = 2 $,
equation (\ref{eq:omega_varsigmaD}) gives $0.25 \le \Omega_\varsigma\le 0.33 $
as the \es mass is varied in the range 
$1 \le  m_\varsigma^{\mbox{\tiny  MeV}} \le 100$.
The value $\Omega_\varsigma = 0.30$ is obtained for
$ m_\varsigma^{\mbox{\tiny  MeV}} = 20$. The latter value for 
$m_\varsigma$ not only agrees with the considerations presented
in section \ref{Sec:cem}, but it is also  
supported by observations on the $0.511\;\mbox{MeV}$
gamma-ray line made by the European Space
Agency's INTEGRAL gamma-ray satellite
\cite{Boehm:2003bt,Boehm:2004gt,Beacom:2004pe,Casse:2004gw,Fayet:2004kz}. 
The emergent relevant cross-section is also
similar to the one suggested by  Fayet's analysis for light dark
matter  \cite{Fayet:2004kz}. As shown in section \ref{Sec:cem}, 
for the discussed mass range,
the collapse of a primordial \es clouds also leaves a significant
dark matter core at the center of galaxies.

The sensitivity of $\Omega_\varsigma$ on 
 $\langle \sigma v\rangle_{\mbox{\tiny  pb}}$ can be further
gauged from the observation that for
a $\langle \sigma v\rangle_{\mbox{\tiny  pb}}=1$,
equation (\ref{eq:omega_varsigmaD}) gives $0.47\le \Omega_\varsigma\le 0.64 $
for the same range of \es mass as above, i.e., 
$1 \le  m_\varsigma^{\mbox{\tiny  MeV}} \le 100$.

The  $\sigma_{\varsigma\,\stackrel{\neg}
\varsigma \rightarrow\phi_H \phi_H}$ that appears in
the above estimate of $\rho_\varsigma$ 
should be evaluated using the perturbative procedure
outlined in section \ref{Sec:FurtherOnNonLocality}. Such an 
evaluation may provide additional insights and novel
features which are not apparent from the presented calculations
of this section. That $\varsigma\,\stackrel{\neg}
\varsigma \rightarrow\phi_H \phi_H$ cross-section
will depend quadratically on $\lambda_E^2$ (see equation (\ref{eq:Higgs}))
is evident. What is not clear is what precise contribution the
\es non-locality will make. This is perhaps the most important
distinction from the WIMP scenario, and it remains to be studied
in detail.

In order that $\Omega_\mathrm{WIMP}$ is roughly  $0.3$, 
in a generic WIMP scenario the characteristic temperature for production
and annihilations of WIMPs is put in by hand to be around a few
hundred GeV.  
The thermal contact of \es with the Standard Model
cosmic plasma is severely restricted by the mass 
dimension one aspect.
This  introduces a  characteristic temperature below
which  the production and annihilations
of \textit{Elkos} are severely suppressed. This temperature is determined
by the Higgs mass. Therefore, \es particles constitute not only a  
first-principle candidate for dark matter, 
but also carry with them the
property which dictates the above-mentioned characteristic 
temperature\footnote{This
section was added on suggestion from a \textit{JCAP} referee.}.

\subsection{Collapse of a primordial Elko cloud: independent 
constraint on the Elko
mass, $10^6\,M_\odot$ dark matter central cores for galaxies,
and cosmic gamma-ray bursts}
\label{Sec:cem}

Here we consider a possible scenario which gives rise to the 
virialized dark matter clouds which overlap with luminous,
standard-model, galaxies. The overlap is assumed purely on 
observational grounds, and we provide no \textit{a priori} 
justification for this circumstance. 
We first give  a brief run-through,
and then proceed with the details.

\subsubsection{A brief run-through} 

Schematically, we study a 
galactic-mass primordial $\varsigma$ and 
$\stackrel{\neg} {\varsigma}$ cloud which  
undergoes the following set of qualitative transformations:
\beq
\hspace*{-225pt}
P_{\varsigma\stackrel{\neg} {\varsigma}} \rightarrow
R_{\varsigma\stackrel{\neg} {\varsigma}} \rightarrow
V_{\varsigma\stackrel{\neg}  {\varsigma}}\,.
\eeq
Here, 
\begin{enumerate}

\item[$\bullet$]
$P_{\varsigma\stackrel{\neg} {\varsigma}}$ represents $\mbox{a primordial}\;
{\varsigma\mbox{-}\stackrel{\neg}{\varsigma}}\;\mbox{cloud}$. Its spatial 
extent is assumed to be a few times that of a typical galaxy.

\item[$\bullet$]
$R_{\varsigma\stackrel{\neg} {\varsigma}}$ is a rebound caused
(a) either by \textit{Elko's} quartic self-interaction
(requiring cloud temperature of about $T_\varsigma \sim
m c^2/k_\mathrm{B}$), or (b)
by the \textit{Elko}-\textit{Elko} interaction
producing pair of Higgs (requiring cloud temperature 
of about $T_\mathrm{H} \sim m_\mathrm{H} 
c^2/k_\mathrm{B}$, with $m_\mathrm{H}$ as the Higgs mass). 
The possibility 
that rebounds occurs due to some quantum gravity effect is also
considered.

\item[$\bullet$]
$V_{\varsigma\stackrel{\neg}  {\varsigma}}$ is the virialized 
\es cloud which emerges after the above-indicated rebound
and thermalization-inducing
process.
\end{enumerate}
The scenario in which the rebound 
of a primordial \es cloud is induced by the process
$\varsigma + \stackrel{\neg}\varsigma \rightleftharpoons \phi_H + 
\phi_H$, carries three basic results: (a) it sets a lower bound 
of about $1$ MeV for $m_\varsigma$ (called simply  $m$ in 
section \ref{Sec:Relic}), (b) it suggests   
$10^6\,M_\odot$ dark matter central cores for typical galaxies,
and (c) it predicts an explosive event in the early life history of 
galaxy formation.

\textit{The physical criterion that
provides the above-enumerated results is the requirement that 
$T_\varsigma$,  or $T_\mathrm{H}$ (or $T_\mathrm{QG}$, associated with the
Planck scale) 
 is reached before the cloud radius crosses
the Chandrasekhar limit.} If this requirement is not met, then 
one ends up with a degenerate \es core, or a black hole. 
These may, or may not, have
association with luminous galaxies.

Some elements of our exercise are textbook like. Yet, 
the requirement just enunciated 
yields a rich set of results. As a parenthetical remark,
it is emphasized that we have not encountered 
these considerations in literature on dark 
matter\footnote{After this work was submitted to 
\textit{arXiv} \textemdash~ see
\textit{v1} of the present work 
\cite{Ahluwalia-Khalilova:2004ab}  \textemdash~
the approach taken here has been independently 
adopted by Vanderveld and Wasserman \cite{Vanderveld:2005tq}.}.
It is probably due to the fact that the
lower mass bound derived below is often superseded by stricter bounds;
for example, in the context of SUSY dark matter the lower mass 
bounds on the lightest SUSY particle are well within the $10$\textendash$100$ 
GeV range.  These,
however, do not apply to MeV-range Elko mass.

\subsubsection{Details}
 
For simplicity we
assume a  spherical distribution, characterized by
mass $M$ (of the order of a typical galactic mass), 
and initial radius $R$,
 of
$\varsigma$  particles undergoing a gravitationally induced 
collapse. 
The cloud
continues to collapse, and its temperature soars until
it reaches a temperature 
 $T_\ast \sim m_{\ast}\,c^2/k_\mathrm{B}$, where `$\ast$'
characterizes either of the two mass scales which 
appear in the additional \textit{Elko}\textendash induced 
structure of the Standard Model
Lagrangian density (see equations
(\ref{eq:elkorenSM})\textendash(\ref{eq:Higgs}))\footnote{In this section 
we exhibit $\hbar,\;c,\;k_\mathrm{B},\;G$ explicitly.}. 
At that stage the 
\es cloud has the possibility to radiate electromagnetically and/or 
to emit neutrinos,
and to cool down
\textit{provided $T_\ast$ is reached when the spatial extent of the
collapsing $\varsigma$ cloud is greater than the Chandrasekhar
limit, $R_{Ch}$} \textemdash a condition which will be examined below
(and roughly signals black hole formation). 
Under these circumstances,
the newly available radiation pressure may cause
part of the cloud to explode, while leaving an imploding 
remanent fated to
either become a black hole or a degenerate \es core, the 
\es analogue of a neutron star.
If such a scenario is to explain
the dark matter problem the exploding \es  envelope must,
at present,
carry dimensions of the order the galactic size.
This is not a prediction, but what detailed calculations should yield
if the presented scenario is to be viable as a dark matter candidate.
In essence for this to happen a substantial fraction of the 
 released explosive energy must be re-deposited to the expanding 
$\varsigma$ envelope.

To determine the mass and size of  such a remanent dark matter configuration
we make the working assumption that non-locality plays 
insignificant
role. Then, the standard arguments of balancing the Fermi pressure against the
gravitational potential energy  
yield the following critical Chandrasekhar values:
\beq
\hspace*{-130pt} 
M_\mathrm{Ch} \approx 
\left( \frac{m_\mathrm{P}}{m}\right)^3 m\,, \quad
R_\mathrm{Ch} \approx 
\left( \frac{m_\mathrm{P}}{m}\right) 
\lambda_\mathrm{C}\,,
\eeq
where $m$ is $\varsigma$ mass,  $\lambda_\mathrm{C}$ is the associated
Compton length, $\hbar/ m c$, and $m_\mathrm{P} :=\sqrt{\hbar c/G}$ is the
Planck mass.
The set $\{M_\mathrm{Ch},\;R_\mathrm{Ch}\}$ sets 
the boundary between stability and instability and also marks
the complete onset of general relativistic effects.
To proceed further it is essential to gain some knowledge of plausible
values of $M_\mathrm{Ch}$ and $R_\mathrm{Ch}$. For this 
we recast $M_\mathrm{Ch}$ and $R_\mathrm{Ch}$ in the following form:
\begin{equation}
\hspace*{-70pt}
M_\mathrm{Ch}
\approx \frac{1}{x_\varsigma^2}\; 1.6 \times 10^{12} M_\odot\,,\quad
R_\mathrm{Ch} 
\approx  \frac{1}{x_\varsigma^2}\; 6.3 \times 10^{-2} \,\,\mbox{pc}\,,
\label{eq:Chandra}
\end{equation}
where  $M_\odot$ is the solar mass, and $x_\varsigma$
 represents the \es mass $m$ expressed
in $\mbox{keV}$.

With these observations in mind,
we conjecture that the collapse physics of such a 
cloud is similar to that of a supernova explosion. It leaves behind
a  degenerate structure 
of $\varsigma$ particles, or a 
black hole with mass $M_\mathrm{Ch}$, while releasing $(M-M_\mathrm{Ch})$ 
as a sum total for the mass of rebounding 
$\varsigma$ envelope, $M_\varsigma$,
and an energy burst \textemdash
~ predominantly made of gamma rays, and neutrinos \textemdash ~
 carrying $M_{\gamma,\nu}\,c^2$. We shall assume that 
the coupling constants $\alpha_E$, $\lambda_E$, and their
interplay with gravity, are such that the \es cloud does not
develop significant density fluctuations to seed \textemdash~
over the timescale of its collapse  \textemdash~ the formation of
smaller structures.

We now examine the condition for which $T_\ast$ is reached 
when the spatial extent of the
collapsing $\varsigma$ cloud is greater than the Chandrasekhar
limit, $R_\mathrm{Ch}$.  Let $R_\ast$ represent the radius which characterizes
the spatial extent of the  collapsing $\varsigma$ cloud when it reaches the
temperature $T_\ast$.
Under the assumption that the initial $R \gg R_\ast$, $R_\ast$ is
characterized by the \es configuration when the average kinetic
energy gained
per $\varsigma$ equals the energy associated with a Higgs or $\varsigma$
(the only two mass scales which \es\hspace{-5 pt}-induced new 
Lagrangian density carries)\footnote{We do not include a factor of $2$
on the rhs of equation  (\ref{eq:th}) as all our calculations in
this section carry order-of-magnitude estimates.}:
\beq
\hspace*{-245pt}
\frac{ G M^2}{N R_\ast} =  m_\ast c^2 \,, \label{eq:th}
\eeq
where $N = M/m$ is the number of $\varsigma$ particles in the cloud, while
$m_\ast$ is either the $\varsigma$ mass $m$, 
or it represents the Higgs mass $m_H$. 
Taking note of the fact
that $m_\mathrm{P}^3\, m_\mathrm{p}^{-2}$ is a typical stellar mass 
$(= 3.77 \times 10^{33}\;\mbox{g} \approx 1.9 M_\odot)$, we
write $M \approx M_\mathrm{G}\approx \alpha_\mathrm{star} \,
m_\mathrm{P}^3\, m_\mathrm{p}^{-2}$, where $M_\mathrm{G}$ 
represents the luminous
mass  of a typical galaxy, $\alpha_{star}$ 
is approximately the number of stars in the same, 
and $m_p$ is the proton mass, and  
inserting for $R_\ast$ the absolute lower bound of 
$R_\mathrm{Ch}$, equation (\ref{eq:th}) yields
\beq
\hspace*{-240pt}
\alpha_\mathrm{star} = \frac{ m_\mathrm{p}^2\, m_\ast }
{m^3} \,.\label{eq:astar}
\eeq
This remarkable equation  
can be read in two ways: with the 
$m_\ast/m^3$  as input a rough estimate 
for the number of stars in a typical galaxy may be derived; 
on the other hand,
with $\alpha_{star}\approx 10^{11}$ as input,
one may obtain the ratio $m_\ast/m^3$.  Since observationally
$\alpha_{star}\approx 10^{11}$ is known, 
this immediately implies the following results for the mass of the
$\varsigma$ particles:
\beq
\hspace*{-100pt}
{m} =\begin{cases} \frac{m_\mathrm{p}}{\sqrt{\alpha_\mathrm{star}}}
\;\quad\qquad\qquad\; \mbox{for}\; m_\ast = m \,,\cr
\left(\frac{m_\mathrm{p}^2 \,m_\mathrm{H}}{\alpha_\mathrm{star}}
\right)^{1/3}
\qquad\quad \mbox{for}\; m_\ast = m_\mathrm{H}\,.\label{eq:zimpok15}
\end{cases}
\eeq
Taking a representative value of $\alpha_\mathrm{star} = 10^{11}$, 
and
writing the Higgs mass $m_\mathrm{H} = x_\mathrm{H} \,100\; 
\mbox{GeV}/c^2$, with
$1\le x_\mathrm{H} \le 2$, the above equation
yields
\beq
\hspace*{-85pt}
{m} =\begin{cases} 3 \;\mbox{keV}/c^2\qquad\qquad\qquad
 \mbox{for}\; m_\ast = m \,,\cr
1.0 - 1.2 \;\mbox{MeV}/c^2
\;\qquad \mbox{for}\; m_\ast = m_\mathrm{H}\,.
\end{cases}
\eeq
These values of $m$ are to be considered as 
lower bounds because the requirement one has 
to impose is $R_\ast \geq R_\mathrm{Ch}$, 
while \eqref{eq:astar} was obtained by saturation  
of this inequality.

To decide between these two values of the \es mass, we now 
make the observation that 
\beq
\hspace*{-50pt}
M_\varsigma + M_{\gamma,\nu} := M - M_\mathrm{Ch} = 
\left(1 - \frac{m_\mathrm{p}^2}{\alpha_\mathrm{star}\, m^2}\right)   
\frac{\alpha_\mathrm{star}\,m_\mathrm{P}^3}{m_\mathrm{p}^2}\,.\label{eq:zimpok10}
\eeq
On the extreme rhs of the above equation,
identifying the factor outside the bracket as $M$ and 
using equation  (\ref{eq:astar}) inside the bracket, we get
\beq
\hspace*{-175pt}
M_\varsigma + M_{\gamma,\nu} = \left(1 - 
\frac{m}{m_\ast}\right) M\,. \label{eq:zimpok11}
\eeq
Since without a rebound of the \es cloud the viability criterion
cannot be met\footnote{That is, observationally,
the mass of the dark matter cloud should
be of the order of a typical galactic mass and its spatial extent must extend
beyond luminous extent of galaxies.
This we call the minimal viability criteria for \es to be a dark matter
candidate.},
 the sum $M_\varsigma + M_{\gamma,\nu}$ must be a 
\textit{good} fraction of $M$. For this to occur, 
the round bracket in \eqref{eq:zimpok11} must not become too small. 
For the 
solution with $m_\ast \approx m_\mathrm{H}$, the central degenerate core
(or, a black hole)
carries  a mass of about $10^6\,M_\odot$ while rebounding \es cloud, and 
the associated burst of energy, carries almost the entire mass. 
For $m\approx m_\ast$, the mass of the  remanent core becomes of the order of
$M$ itself. That one cannot make a more precise statement for the mass of the rebounding cloud near 
$m_\ast = m$ is due to the order-of-magnitude nature of the 
calculation and may be considered as a drawback (there is no 
prediction from order-of-magnitude estimates) or as a virtue (the rebounding cloud is very sensitive to Elko details and thus may probe Elko physics). Unless $m$ is very finely tuned the round bracket in \eqref{eq:zimpok11} is 
still of the order of unity and thus again the rebounding \es cloud carries a mass of the order of $M$%
\footnote{In other words, for $m_\ast\approx m$ the $M_\mathrm{Ch}$ 
becomes of the order of the initial 
mass of the \es cloud $M$ 
itself, and hence its difference from the latter 
ceases to have a reliable meaning.}.
For $m_\ast \gg m$ the order-of-magnitude estimate is more robust.

In \textit{favour} of the solution $m\approx m_\ast$ one 
is tempted to note that,
apart from the dark matter problem, there are two 
outstanding cases where the fermions of
the Standard Model of particle physics fail to
provide the astrophysical consequences expected of them.
These are pulsar kicks \cite{Kusenko:1997sp,Kusenko:2004mm} and 
supernova explosions \cite{Ahluwalia-Khalilova:2004wf}.
Kusenko \cite{Kusenko:1997sp,Kusenko:2004mm}, 
and Dolgov and Hansen \cite{Dolgov:2000ew}
have argued that a sterile component with
a mass of about $2$\textendash$20\,\,\mbox{keV}/c^2$ provides
a good candidate to explain pulsar kicks and that the
same particle may also be cosmological dark matter. However,
for the argument to work for pulsar kicks it is essential that the
sterile component must carry an intrinsic parity
asymmetry. Such an asymmetry is naturally built in 
the \es particles, $\varsigma$ and $\stackrel{\neg} {\varsigma}$. 
But, as long as one confines 
oneself
to the set
of assumptions we have used, 
the $4.2\;\mbox{keV}/c^2$ identification 
noted here gives rise to too high a value for the galactic core.
It thus
violates the minimal viability criteria. In addition, there seems to be
a more fundamental problem with such an identification: how is one
to add a mass dimension one component (Elko) to a mass dimension 
three-halves field (neutrinos)?
The problem appears to be non-trivial and has no clear answer due to
subtle questions which mixing of local and  
non-local field on the one hand, and  mass
dimension three-halves
and one on the other, raises.

In \textit{favour} of the solution $m\approx m_\mathrm{H}$ 
we have the additional support
from two recent works. 
From observations on $0.511\;\mbox{MeV}/c^2$ 
gamma-ray line seen by the European Space
Agency's INTEGRAL gamma-ray satellite   \cite{Weidenspointner:2004my},
Boehm \textit{et al} 
\cite{Boehm:2003bt,Boehm:2004gt}, followed
by additional observations of Beacom \textit{et al} 
\cite{Beacom:2004pe}, purport
to read a dark  matter particle mass in the range of $1$\textendash$20$ MeV.

The $m_\ast = m_\mathrm{H}$ identification 
meets the minimal viability criteria well. Yet, the strength
of this viability and identification must not be overestimated.
There are several open points regarding these considerations.

\begin{enumerate}
\item[$\bullet$]
Quartic Elko self-interactions
have been neglected. These may contribute in an essential 
way in the energy range considered. Moreover, 
for stability reasons the coupling constant 
in front of this term has to be such that a repulsive
interaction emerges. As a result this may contribute
to the rebound of the collapsing Elko cloud. 

\item[$\bullet$]
Non-locality has been neglected, but because it scales 
essentially with $m^{-1}$ contributions
from it could be relevant here.

\item[$\bullet$]
Non-standard gravitational interactions could be
of importance in this context \textendash~ 
for instance, the `square' of \es field, 
which is a scalar of mass dimension two,
may couple to the Ricci scalar with a dimensionless 
coupling constant, much like the Jordan\textendash Brans\textendash Dicke 
field 
\cite{Jordan:1959eg,Brans:1961sx}
does in scalar tensor theories or quintessence models 
\cite{Wetterich:1988fm,Wang:1998gt,Carroll:1998zi,Zlatev:1998tr}.

\end{enumerate}

Therefore, our simple order-of-magnitude 
considerations 
above cannot be used to determine uniquely the mass of Elko, 
but it appears to be likely that it lies in the range 
of $1\;\mbox{keV}/c^2$ to $20\;\mbox{MeV}/c^2$ 
if \es is to explain dark matter. It is desirable 
to improve these limits further, but to this end one has to address 
the caveats mentioned above in a more detailed study, possibly 
combining the two scenarios with
critical temperatures of $T_H$ and $T_\varsigma$, respectively.

In conclusion,
the gravitationally induced collapse of an \es cloud, if 
$\varsigma$ and $\stackrel{\neg} {\varsigma}$ 
are to serve  as dark matter, must be qualitatively similar to
type-II supernova explosions of stellar objects which leave a degenerate core
of fermionic matter and are accompanied by (a) an expanding
envelope of matter, and (b) an intense electromagnetic and neutrino
radiation carrying several solar masses.  
For an \es cloud, the neutronic core  is replaced by
a massive degenerate core of $\varsigma$ (or, a black hole) 
 \textemdash~ indicated here to be of the order of $10^6$ solar
masses \textemdash~ while the 
gamma-ray and neutrino
radiation may carry a mass of the order of galactic mass and the
burst temperature $T_\ast$ may be
characterized by  two characteristic masses: the Higgs mass and the mass of
$\varsigma$.

Given that we are confronted with a truly unknown cosmic phenomenon, 
it is not beyond reason that the  following sequence of multiple
rebounds is realized. In that event one can no longer reply
on the one-mass scale dominance scenario outlined above. Therefore, no
quantitative analysis can be presented at this stage of our work. 
In this scenario, if one assumes that the Electroweak and GUT fields
do not carry any significant coupling to \es then only one additional 
mass scale comes into the picture, and it is given by $m_\mathrm{P}$. As 
the collapsing \es cloud successively soars to this temperature
one may expect that apart from an explosion induced by 
quartic self-interaction
of \es itself another explosive phase occurs
at the Higgs temperature, and
an unknown quantum-gravity induced effect 
occurs at the Planck temperature. 
Should the latter carry an explosive element, and should
the \es self-interaction and Higgs-mediated explosions
not succeed in causing a significant rebound, then
a one-mass scale scenario predicts \es mass to be
\beq
\hspace*{-210pt}
m = \left( \frac{ m_\mathrm{p}^2\, m_\mathrm{P}}
{\alpha_\mathrm{star}}\right)^{1/3}\,.
\label{eq:zimpok14}
\eeq
With $m_\mathrm{P} = 1.2 \times 10^{19}\;\mbox{GeV}/c^2$, this results in
the \es mass being  $0.5\; \mbox{TeV}/c^2$. In this scenario almost
all of the \es clouds rebounds without leaving a remanent core.

\subsection{Similarities and differences from the  mirror matter
proposal}
\label{Sec:mirrormatter}

We recall that mirror matter, 
which restores parity symmetry by introducing
a parallel universe, is postulated to be endowed with the
following properties.

\begin{enumerate}

\item[$\bullet$]
Particle masses in the mirror world are set degenerate to the  masses of
the Standard Model particles of the world in which we reside.

\item[$\bullet$]
The mirror world carries same gauge group as that of the Standard Model,
with all left-handed fields replaced by right-handed fields
and vice versa.
 
\item[$\bullet$]
The mirror world has the following dominant 
interactions with the SM fields:
\beq
\hspace*{-10pt}
\mathcal{L}(x) = \lambda_\mathrm{M}\;    \phi^\dagger(x)\phi(x)\,
{\phi^{\prime}}^\dagger(x)
\phi^\prime(x) + 
\epsilon_\mathrm{M}\;  F^{\mu\nu}(x)  F^\prime_{\mu\nu}(x)
\eeq 
where $\phi(x)$ is the Standard Model Higgs doublet, and
$F^{\mu\nu}(x)$ is the electromagnetic  field strength tensor
\textemdash~
with primed fields representing their mirror counterparts,
while $\lambda_\mathrm{M}$ and $\epsilon_\mathrm{M}$ are dimensionless coupling 
constants. 

\end{enumerate}
It is a viable and attractive dark matter candidate
\cite{Foot:2004pa,Foot:2004wz,Berezhiani:2003xm}. Over the
years it had been suspected that mirror matter, apart from
being a dark matter candidate,
could also provide a resolution
to the orthopositronium life time puzzle which
had acquired a statistical significance well
above the $5$-$\sigma$ level. Latest measurements by two
independent groups,
however, have discovered a systematic source of errors
and now these latest experiments 
provide a decay rate which agrees well with predictions
of quantum electrodynamics \cite{Asai:2003hs,Vallery:2003iz}.

The \es proposal for dark matter has some noteworthy similarities
and differences from  the mirror matter framework.
The similarity between equations (\ref{eq:Higgs})
and (\ref{eq:zimpok17}), and the  mirror
counterpart just enumerated, is obvious, and carries 
its roots in  mass dimension
one of the \es field. Yet, apart from the assumption
that standard power-counting 
renormalizability arguments provide a good guide,
in case of \es no postulate has to be
made to restrict the interactions with the Standard Model
matter to the specific form.

Mirror symmetry predicts mirror stars, and 
claims  an  indirect support for their existence 
\cite{Blinnikov:1999ky,Foot:1999hm}.
Yet, we suspect these claims for the simple
reason that the evidence must have been overwhelming
already
because mirror stars would have already  been observed as
numerous dark companions of luminous SM-model stellar objects
\textit{unless} one invokes some arguments as to why the
dark matter and luminous matter of galaxies occupy 
overlapping  spatial regions and yet the indicated circumstance
does not seem to arise.

One of advantages of our proposal, which to some
extent is a subjective perception, is that the
\es framework does not have to invoke a parallel universe.
It thus avoids the inevitable formation of mirror stars,
and mirror planets \textemdash~ which are
contained in the framework of unbroken mirror symmetry. Broken
mirror symmetry does not seem to be a   way out as it 
destroys the very motivation upon which the mirror symmetry is based.
Yet, an interesting possibility with parity symmetry being spontaneously 
broken has been suggested by  Berezhiani and Mohapatra 
\cite{Berezhiani:1995yi}.

On the other hand, the mirror\textendash matter framework is local
at a quantum field theoretic level.
The \es framework carries
a well defined non-locality, whose physics motivation have been
discussed at some length in  \cite{Ahluwalia-Khalilova:2004sz}.

These remarks aside, both mirror matter and \es matter
provide most natural candidates for dark matter without
invoking any untested spacetime symmetries. For this
reason, and due to their similarities, it should be
considered as an urgent task to distinguish their 
implications for astroparticle physics, cosmology, and
experiments such as those  envisaged for mirror matter
\cite{Badertscher:2003rk}.
While pursuing these studies,
the following observations should be kept in mind.
 
\begin{enumerate}

\item[$\bullet$]
The fermions of the mirror matter carry mass dimension three-halves.
In contrast, the fermion[s] of the \es matter carry mass 
dimension one.

\item[$\bullet$]
The limited interactions which the \es matter carries
with itself, and with the Standard Model fields, are
dictated by its mass dimension. At present, is not known what effects 
the intrinsic \es non-locality carries. The mirror matter
postulates the absence of a large class of possible interactions
with the Standard Model fields. 

\item[$\bullet$]
Given differing mass dimensionalities of the mirror and \es fermions,
it is more natural that the suspected sterile component of neutrinos
\textit{may} be a mirror neutrino. Such a suggestion has been
made in \cite{Foot:1995pa,Berezhiani:1995yi}. 
The introduction of an \es fermion as
a sterile component has not yet been investigated, due to
subtle questions which non-locality and its mass dimension 
raises. 

\item[$\bullet$]
Mirror matter demands  the existence of mirror stars, planets, and
even  meteorites \cite{Foot:2003js}. 
The minimal \es proposal set forth here in this
paper does not suggest any such objects.

\end{enumerate}

\section{Elko particles in a Thirring\textendash Lense gravitational field}
\label{Sec:ThirringLense}

In the above discussion the role of  $\stackrel{\neg} {\varsigma}$  
was entirely overlooked. Here, in this section, we now find that
it may carry significance for the matter\textendash antimatter asymmetry. 
We have not yet developed a full theory based on \es 
in the gravitational background of a rotating gravitational source.
Yet, an important physical implication has already been 
studied \cite{Ahluwalia-Khalilova:2004kv}
in which
{\ecco} single-particle states are considered in
Thirring\textendash Lense gravitational field\footnote{In the 
context of \cite{Ahluwalia-Khalilova:2004kv} the
reader is alerted that the identification of the
described particles with neutrinos and antineutrinos made
there, in view
of the work presented here, is no longer tenable.
The correct identification is with $\varsigma$ and 
$\overline\varsigma$.}.
This example is an important physical case where 
a preferred direction \textit{does} exist and the more general
structure of the spacetime evolution must be invoked. 

The circumstance which allows for a `back-of-the-envelope' 
calculation is the following: the $(1/2,0)$ 
and the $(0,1/2)$  Weyl components, for each of the  
\textit{Elko}, carry opposite helicities. 
Based on the particle  interpretation developed in this paper we 
expect single-particle states to carry the same property. 
Thus if we introduce a self-conjugate state $\vert +,-\rangle_\mathrm{S}$ in the
gravitational environment of a rotating gravitational source, such as 
a neutron star, then each of the 
$(1/2,0)$ and $(0,1/2)$ transforming
components
of the state pick up 
equal and opposite phases (where we now display $\hbar$ explicitly):
\beq
\hspace*{-161pt}&& 
\left({1}/{2},0\right): \quad\exp\left(
-\,\frac{\ri}{\hbar} \times \frac{\hbar}{2} \left\vert \hp\cdot\b\right\vert t    
\right)\label{eq:phasesR} \\
\hspace*{-161pt}&& \left(0,{1}/{2}\right):\quad
\exp\left(
+\,\frac{\ri}{\hbar} \times \frac{\hbar}{2} \left\vert \hp\cdot\b\right\vert t    
\right)\,,\label{eq:phasesL}
\eeq
with $\b$ representing the Thirring\textendash Lense
field of the source star. In the
weak field limit it is given by\footnote{This  
is a valid approximation all the way to a 
neutron star as the dimensionless gravitational 
potential ${G M}/{c^2 R} \approx 0.2$ for a $1.4$ 
solar mass neutron star.}
\beq
\hspace*{-186pt}
\b=\frac{2 G}{c^2}\left(\frac{\Jc - 3 \left(\Jc\cdot\hr\right)}{r^3}
\right)\,.
\eeq
Here $G$ is the Newtonian gravitational constant, 
$c$ is the usual speed of light,
$\Jc$ is the angular momentum of the rotating
gravitational source (denoted by $\mathcal{G}$), and
$r$ specifies the radial
coordinate 
distance of the region where the particle is observed.   
The magnitude of $\Jc$ is given by 
${\mathcal J}\approx \frac{2}{5}M R^2 \omega$,
where $M$ is the mass of $\mathcal{G}$, 
$R$ is its radius, and $\omega$ represents
the associated  angular frequency.
A couple of observations are immediately warranted.

\begin{enumerate}

\item[$\bullet$]
In phases (\ref{eq:phasesR}) and  (\ref{eq:phasesL}), $\hbar$ cancels out.
This happens because the $\hbar$ which appears in a generic quantum
evolution and the $\hbar$ which appears in a particle's spin/helicity 
cancel out.

\item[$\bullet$]
In (\ref{eq:phasesR}) and  (\ref{eq:phasesL}), $t$ refers to the 
ensemble-averaged time as measured by a stationary clock
situated in the gravitational environment at $\r$. 
The qualification, ensemble-averaged, is necessary for
generalized flavour-oscillation clocks 
\cite{Ahluwalia:1997hc,Chryssomalakos:2002bm}.
\end{enumerate}
The spinor associated with 
$\vert \pm,\mp\rangle_\mathrm{S/A}$ are 
$\lambda^\mathrm{S/A}_{\{\pm,\mp\}}(\p)$. 
Using phases 
which the coupling of these states with the $\b$ field induces, we find
the following oscillations\footnote{The calculation has been 
made under the assumption that we prepared the
test particle in such a way that $\hp=\hr$, and that its evolution is
studied in the polar region.}:
\beq
\hspace*{-220pt}
\vert \pm, \mp\rangle_\mathrm{S} \leftrightharpoons 
\vert \pm, \mp\rangle_\mathrm{A} \,,
\eeq  
with oscillation probability $\sin^2(\widetilde{w}_\mathrm{osc.} t)$, where
\beq
\hspace*{-215pt}
\widetilde{\omega}_\mathrm{osc.} =
\frac{4}{5}\left(\frac{G M}{c^2 R}\right)\omega\,. \label{eq:osw}
\eeq
Since in the presence of a preferred direction, 
the study of the wave equation for \es 
associates particle nature to
self-conjugate sector, and provides an antiparticle 
interpretation for
the anti-self-conjugate sector of the \textit{Elko}, 
we are led to conclude that
rotations in gravitational environments induce 
$\varsigma \leftrightharpoons
\stackrel{\neg} {\varsigma}$ oscillations, where 
$\varsigma$ and 
$\stackrel{\neg} {\varsigma}$ represent particles and antiparticles
associated with the neutral $\eta(x)$ field.
A sea composed entirely of $\varsigma$ will in time develop
a $\stackrel{\neg}{\varsigma}$ component,
thus inducing an \es matter\textendash antimatter asymmetry.
Since
we identify the 
$\varsigma$ and $\stackrel{\neg} {\varsigma}$
with the dark matter, these oscillations may have an important role
to play in cosmology, although we hasten to add that the observed 
baryonic matter\textendash antimatter asymmetry is unlikely to be a consequence of this because, after all, \es interacts only very weakly with the matter content of the Standard Model.

\section{A critique and concluding remarks  }
\label{Sec:cc}

In this section we combine the customary concluding remarks
with  additional details of our work.
This section, with exception of sections \ref{Sec:summary0} and
 \ref{Sec:summary}, 
is not a summary of our
work but is  intended as an integral  
part of our exposition. Its purpose is to bring to reader's attention
additional structure which the \textit{Elko}-based quantum field
theory carries, and in addition to point out that 
locality is most likely not a stable feature of quantum field
theories at the Planck scale. 
However, the level of rigour 
now changes and it mirrors
the previous section, giving us the 
luxury of some speculative remarks.

\subsection{Elko as a generalization of Wigner\textendash Weinberg classes}
 \label{Sec:Wigner-Weinberg}

That our effort had a chance of providing us with a theoretically
viable and phenomenologically novel theory seemed assured by 
Wigner's work on the extended Poincar\'e group \cite{Wigner:1962bww}.
In that work a general result is reached that the
Poincar\'e group extended with space, time, 
and space\textendash time reflections
allows every representation
space to support four type of quantum field theories. Each of these theories
differs in the underlying C, P, and T properties\footnote{The 
origin of these observations in fact lies in an unpublished work
of  Bargmann, Wightman, and Wigner.
In \cite{Wigner:1962bww}
Wigner states `Much of the material which follows was taken from
a rather old but unpublished manuscript by 
V. Bargmann, 
A. S. Wightman,
and myself'.}. For the particles of the Standard Model,  
if $\mathcal{C}$ and $\mathcal{P}$  generically represent 
charge conjugation and parity operators, respectively, then
\beq
\hspace*{-39pt}
\mbox{\sc Standard Model:\hspace{36pt}} 
\begin{cases} 
\left\{\mathcal{P},\mathcal{C}\right\} = 0,
\quad    &\mbox{for  fermions:
leptons, quarks} \cr
 \left[\mathcal{P},\mathcal{C}\right] = 0,  \quad   
&\mbox{for bosons: 
gauge/Higgs particles}\,.
\end{cases}
\eeq
In \cite{Wigner:1962bww}, Wigner  
argues that the above class does not exhaust all possibilities
allowed by  Poincar\'e symmetries. In fact, he showed 
that the following additional structure is allowed: 
\beq
\hspace*{-34pt}
\mbox{\sc Wigner Classes:\hspace{36pt}}
\begin{cases}
 \left[\mathcal{P}, \mathcal{C}\right] = 0,
\quad    &\mbox{for  fermions: the presented construct} \cr
   \left\{\mathcal{P},\mathcal{C}\right\} = 0, \quad   
&\mbox{for bosons:  construct of \cite{Ahluwalia:1993zt}}\,.
\end{cases}
\eeq
Yet, for the sake of avoiding confusion we take note that the 
presented construct is a generalization of Wigner\textendash Weinberg 
context \cite{Weinberg:1995sw}. Specifically, in the context 
of Weinberg's work, which in turn 
is an extension of Wigner's considerations, 
equation (2.C.10) of \cite{Weinberg:1995sw},
\beq
\hspace{-160pt}
\mathcal{P} \, \vert\p,\sigma,n\rangle= \sum_m \wp_{n m}\,  
\vert{- \p,\sigma,m}\rangle
\eeq
is no longer sufficiently general.
In reference to the above equation, we give the following notational
details:
the $\vert\p,\sigma,n\rangle$ represent one-particle states, $\p$ is the
momentum vector, $\mathcal{P}$ is the parity operator and carries properties
defined on p 76 of \cite{Weinberg:1995sw}, $\sigma$ represents
one of the $(2j+1)$ spin projections associated with $J_3$, and $n,m$ 
define a degeneracy, and the $\wp_{n m}$ are superposition
coefficients.   In relevance 
to the theory which we present here,
while the degeneracy index may be interpreted as
associated with self-conjugate  and anti-self-conjugate states,
the counterpart of the $\sigma$, i.e., 
the dual-helicity index $\alpha$, introduces additional
structure and allows for the result obtained in
section \ref{Sec:CPT}. This
observation casts the \es formalism into a further generalization
of the Wigner\textendash Weinberg framework, and while remaining in agreement
with Wigner we provide a counter-example to Weinberg's
result contained in equation  (2.C.12) of  \cite{Weinberg:1995sw}.
The counter-example contained in the result
(\ref{eq:cptsq})
 is not to be interpreted as a contradiction, but
as a generalization. 

\subsection{On Lee\textendash Wick non-locality, 
and Snyder\textendash Yang\textendash Mendes algebra}

Lee and Wick have argued that the non-standard  
Wigner classes
must carry  an element of non-locality \cite{Lee:1966tdw}. 
The \textit{Elko}-based
quantum field $\eta(x)$ exhibits a non-locality \eqref{eq:nonlocres}
which is consistent with this expectation.
Yet, we cannot fully identify the non-local element which we discovered with
that of Lee and Wick as our formalism is a generalization 
of the Wigner framework for which Lee\textendash Wick considerations apply.

What, however, can be claimed is the following: locality cannot be
considered a robust feature of Poincar\'e covariant quantum field
theories. This is apparent from Lee\textendash Wick paper as 
well as from the explicit
construct which we present.
The question then arises: why does this circumstance exist?
The answer, we think, lies in the important reminder and observations 
of Mendes that \cite{VilelaMendes:1994zg,VilelaMendes:1999xv}
\begin{enumerate}

\item[(a)]
The Galilean relativity  and classical mechanics are unstable
algebraic structures, i.e., in the mathematical sense they are not
rigid.

\item[(b)]
Their deformation towards stability results in Poincar\'e 
and Heisenberg algebras of special relativity and quantum 
mechanics.
Each of the indicated algebras is stable by itself.

\item[(c)]
The algebra for relativistic quantum field theory 
is the combined Poincar\'e\textendash Heisenberg algebra, which
is unstable. 

\end{enumerate}

The stabilized Poincar\'e\textendash Heisenberg algebra 
as presented by Mendes 
still respects Lorentz algebra\footnote{For a
detailed mathematical and interpretational 
examination of the stability-based framework of Mendes,
we bring to our reader's attention a  recent preprint by
Chryssomalakos and Okon \cite{Chryssomalakos:2004gk}.},
but calls for non-commutative spacetime and 
non-commutative 
energy-mom\-en\-tum-space\footnote{It should be pointed out that 
the idea that quantization of gravity leads to non-locality
 and/or non-commutative spacetime has various roots; 
see, for example, \cite{Douglas:2001ba,Szabo:2001kg} and references
therein. 
In some simple models \textemdash~ for instance, dilaton gravity in two 
dimensions \textemdash~ these ideas turn out to be 
realized straightforwardly, 
i.e., quantization of gravity yields a non-local theory for the 
remaining matter degrees of freedom 
(for reviews \cite{Grumiller:2002nm,Grumiller:2004yq} may be consulted).
Arguments based on incorporating gravitational effects in
a quantum measurement process also suggest an element of
spacetime non-commutativity and non-locality 
\cite{Ahluwalia:1993dd,Doplicher:1994zv}.}.  
This introduces two stabilizing
       dimensionful constants which may be
       identified with Planck length and
       cosmological constant.
It is worth taking note that in 1947 Yang had already argued that 
lack of translational invariance of Snyder's algebra, also suggested 
earlier in 
the same year, is remedied if spacetime is allowed to carry curvature 
\cite{Yang:1947ud,Snyder:1946qz}, and that the algebra found on
algebra-stability grounds by Mendes is precisely the one that Yang had
arrived at more than five decades ago.

A preliminary examination of 
the Snyder\textendash Yang\textendash Mendes algebra suggests that the usual
expectation on locality commutators/anticommutators cannot hold. We phrase
it as `locality is not a stable feature of quantum field
theories', and suspect that it
will completely lose its conventional meaning in
quantum field theories based on stabilized  
Poincar\'e\textendash Heisenberg algebra. Following Mendes,
that stability of algebra associated with
a physical theory should be considered as an  important physical 
criterion for the viability of a theory has also been emphasized by
 Chryssomalakos \cite{Chryssomalakos:2001nd,Chryssomalakos:2004wc,
Chryssomalakos:2004gk}. 

Yet, just as Dirac theory will suffer modifications in 
any extension which respects the  
stabilized Poincar\'e\textendash Heisenberg algebra,
i.e., the Snyder\textendash Yang\textendash Mendes algebra,
 but still remains a 
useful low-energy model, the same may be expected 
for the \es theory presented here. It may serve as a low-energy model
to explore consequences of non\textendash locality.

\subsection{A hint for non-commutative momentum space}
 
Evaluating the determinant of the operator ${\mathcal O}$ in 
(\ref{meq}) with (\ref{br}), (\ref{bl}) and (\ref{identities}) establishes
\begin{equation}
\hspace*{-70pt}
Det[{\mathcal O}]= \frac{
\left(
m^2+p^2-(2 m +E)^2
\right)^2\;
\left(m^2+p^2-E^2
\right)^2}{
\left({2 m (E + m)}\right)^4 }
\,.
\end{equation}
The momentum\textendash space wave operator, 
${\mathcal O}$, {\em appears\/} 
to support two type of spinors: those associated
with the usual  dispersion relation, 
\beq
\hspace*{-150pt}
E^2=m^2+p^2\,, \qquad \mbox{multiplicity =  2} \label{ds1}
\eeq
and those
associated with
\beq
\hspace*{-80pt}
E=\begin{cases}-\;2 m -\sqrt{m^2+p^2}\,,\qquad \mbox{multiplicity =  1}\cr
         -\;2 m +\sqrt{m^2+p^2}\,,\qquad \mbox{multiplicity  = 1\,.}
\end{cases}
\label{nd}
\eeq
However, it is now to be noted that 
the $\mbox{Det}\left[{\cal O}\right]$ is independent of 
${\cal A}$, and that dispersion relations (\ref{nd})
yield
\begin{equation}
\hspace*{-73pt}  
\label{eq:inconsistent}
  \rb \lb\Big\vert_{E= -\;2 m \pm\sqrt{m^2+p^2}}\;\; = 
\lb \rb\Big\vert_{E= -\;2 m \pm\sqrt{m^2+p^2}}\;\; = - \mathbb{I} \,.
\end{equation}
Obviously, (\ref{eq:inconsistent}) is inconsistent with 
the starting point as (\ref{br}) is the inverse of (\ref{bl}). 
Thus, there are only two options: either the solutions (\ref{nd}) 
are discarded or the relation between $\rb$ and $\lb$ is modified. 
\textit{It is conceivable that for example in the context of 
non-commutative 
momentum space the latter scenario might be realized.}
This possibility is again taken up in appendix \ref{Sec:noncomm}.

An explicit calculation shows that\footnote{To be precise, we have verified 
the result (\ref{eq:rdnc}) only for $j=1$ and $j=\frac{3}{2}$. As such for 
$j>\frac{3}{2}$, equation  (\ref{eq:rdnc}) should be considered a 
conjecture.}
\begin{equation}
\hspace*{-105pt} 
  \kappa^{(j,0)}\kappa^{(0,j)} \Big\vert_{E= -\;2 m \pm\sqrt{m^2+p^2}}\;\; = 
 \kappa^{(j,0)}\kappa^{(0,j)} \Big\vert_{E= -\;2 m \pm\sqrt{m^2+p^2}}\;\;
\ne - \mathbb{I} \,,\label{eq:rdnc}
\end{equation}
\textit{unless} $j=\frac{1}{2}$. It is indicative of the fact that the 
above-mentioned 
non-commutativity is probably  representation-space dependent .

\subsection{Generalization to higher spins}

It may be worth noting that a generalization of the 
$\left({1}/{2},0\right)\oplus \left(0,{1}/{2}\right)$ \es
field to higher  $(j,0)\oplus(0,j)$ spins is perhaps not too difficult
a conceptual exercise. If such an exercise is undertaken it may
reveal that in the absence of a preferred direction 
\textit{all} such higher spin fields will carry mass dimension one. 
This may have important  consequence for the physical understanding of
representation spaces which carry a single-spin $j$ content.    

For other representation spaces, such as spinor-vector 
Rarita\textendash Schwinger 
field,  an \textit{ab initio} analysis along the lines of 
\cite{Kirchbach:2002nu,Kaloshin:2004jh} may 
also be worth considering. In such an 
analysis, the spinor sector would have to be given  an \es structure, and
the resulting propagator will almost certainly carry new physical content. 

The purpose of these remarks is simply to emphasize that the physical
content of various physically relevant representation spaces
is far from complete, and a serious \textit{ab initio} study appears
to carry significant physical promise as to give such a task an 
element of urgency. This becomes even more justified when one realizes 
that various \textit{exotica}, such as non-locality, 
non-commutative 
spacetime, modified dispersion relations, all arise in 
constructs which go far beyond  the point-particle context. As such, the 
suggested programme may be considered as a minimal departure from the 
standard high-energy physics framework with a potential to allow
a systematic study of the indicated  \textit{exotica}.

\subsection{A reference guide to some of the key
equations}
 \label{Sec:summary0}

For a reader interested in the main theoretical expressions we now
provide a brief reference list.
Equations (\ref{eq:zimpok13})\textendash (\ref{eq:chconj})
provide the formal structure of the \textit{Elko}. 
Their dual-helicity
nature follows from the result (\ref{y}).
Like the Dirac dual, there exists a new dual for \textit{Elko}.
It is defined in equation ~(\ref{eq:md}). 
With respect to this dual, the orthonormality and completeness
relations are enumerated in equations (\ref{zd1})\textendash (\ref{z1}).
Wigner's expectation for
commutativity, as opposed to \textit{Diracian} anticommutativity, of the 
$C$ and $P$ operators while acting on \es
is contained in equation (\ref{eq:cpelko}),
while  the square of the $CPT$ operator 
as contained in equation (\ref{eq:cptsq}) shows another difference
from the Dirac construct.
The \es quantum field and its dual are given in equations 
(\ref{eq:4.36ap}) and (\ref{eq:4.36bp}).
The action and the Lagrangian density for
\es are contained in equations (\ref{eq:zimpok16A}) and
(\ref{eq:zimpok16B}). Equation  (\ref{eq:zimpok16C})
gives the canonically conjugate momentum.
The covariant amplitude which determines the \es propagator
in terms of the \es spin sums is found in equation (\ref{eq:6.a}).
The spin sums which finally give the \es propagator 
are presented in equations (\ref{spinsumS}) and (\ref{spinsumA}),
with the resulting covariant amplitude given in 
equation (\ref{eq:noteworthy}).
The propagator, in the absence of a preferred direction, is 
written in equation (\ref{eq:noteworthy4}).
It is to be contrasted with the Dirac propagator
recorded in equation (\ref{eq:dm}). The non-locality
anticommutators are found in equations (\ref{eq:locality}),
(\ref{eq:nonlocres}), and in a remark which
encloses equation (\ref{eq:pipi}). 
An estimate of the sensitivity of the \es non-locality to its
mass can be obtained from (\ref{eq:nonlocren}).
The non-locality as manifest in 
$\mathcal{T}\left(\eta(x)\stackrel{\neg}
{\eta}(y)\eta(z)\stackrel{\neg}{\eta}(w)\right)$ 
is discusssed around equation  \eqref{eq:8.3.14}.
The quartic
self-interaction of \es is to be read at
equation (\ref{eq:quartic}), while equation (\ref{eq:Higgs})
provides Higgs\textendash \es interaction. A remarkably simple 
equation which connects the number of stars
in a typical galaxy with three relevant elementary particle
masses, including that of \textit{Elko}, is seen at
equation (\ref{eq:astar}). It defines the physical viability
for the identification of the \es framework with dark matter.
Equations (\ref{eq:zimpok15}) and
(\ref{eq:zimpok14}) give analytical expressions for possible
\es masses. 
The fractional  \es contribution to the total cosmic 
energy density, $\Omega_\varsigma$,
is given by equation  (\ref{eq:omega_varsigmaD}). 
 Although interactions with gravity were not our main concern,
 a brief study of Elko 
particles in a Thirring\textendash Lense gravitational background
revealed the interesting possibility of oscillations between 
self-  and anti-self-conjugate states, with an oscillation frequency 
given by equation \eqref{eq:osw}.  
A hint for non-commutative momentum space,
and its possible dependence on representation space 
(i.e., spin content of the probing test particle), 
is suggested by the results contained in 
equations (\ref{eq:inconsistent})
and (\ref{eq:rdnc}).

\subsection{Summary}
 \label{Sec:summary}

The unexpected theoretical result of this paper is: a fermionic quantum
field based on dual-helicity eigenspinors\footnote{We 
have called these \es after their German name.} 
 of the spin-$1/2$ 
charge conjugation operator carries mass dimension one, and not
three-halves.
This circumstance forbids a large class of interactions with
gauge and matter fields of the Standard Model, 
while allowing for an interaction with the Higgs field. In addition, owing
to mass dimension one, the introduced field is endowed with a quartic
self-interaction. This suggests a first-principle 
identification of the new field with dark matter.
Thus, regarding the question we asked in the introduction as 
to what constitutes dark matter, we have provided
a new possible answer, namely {\em Elko}. The question on 
dark energy  shall perhaps be the subject of a subsequent paper.
The indicated interaction calls for some very specific properties
for a gravitationally induced collapse of a 
galactic-mass cloud
of the new particles. In particular it asks for
a supernova-like explosion for the collapsing cloud.
A semi-quantitative argument yields three different values for the
mass of the new dark matter candidate. These values are
$3 \;\mbox{keV}$, $1$\textendash $1.2 \;\mbox{MeV}$, and $0.5\;\mbox{TeV}$
as lower bounds.
The first of these values arises if the quartic self-interaction
is held responsible for the explosion, while the second value 
results from Higgs being behind  the phenomena, and
the last value results from the possibility of Planck-scale 
physics. Considerations on the relic abundance 
of \textit{Elko}, and various astrophysical observations, 
suggest a $20\;\mbox{MeV}$  mass for \es particles.

We conclude this long exposition with two remarks. 

\begin{enumerate}
\item[(a)]
The non-locality that appears in \es theory carries
no free parameters and is governed entirely by the 
\es mass and resides in the well established
spacetime symmetries. Therefore, for a theoretical 
physicist it may serve as a fertile playing ground 
for examining phenomenological consequences of 
non-locality.  

\item[(b)] Ordinarily, dark matter is postulated  not to carry
any Standard Model interactions. 
The presented theory does not postulate this absence as an 
input but requires it in the sense made precise above.
\end{enumerate}

\section*{Acknowledgments}

We thank Richard Arnowitt, Abhay Ashtekar, Chryssomalis Chryssomalakos, 
Naresh Dadhich,
Salvatore Esposito, Steen Hansen, Achim Kempf, Yong Liu, Terry Pilling, 
Lewis Ryder, 
Parampreet Singh, George Sudarshan, and Dima Vassilevich
for correspondence, at times extended, and/or discussions. 
We thank a \textit{JCAP} referee for his/her detailed and constructive
review of this
long paper and for her/his suggestions which, in particular, 
led to the addition of sections \ref{Sec:FurtherOnNonLocality}  
and \ref{Sec:Relic}. 
The questions  raised also helped us improve
the presentation of section  \ref{Sec:cem}.

Parts of this paper are based  on  Concluding Remarks 
and an Invited Talk \cite{Ahluwalia:2002nk} presented at `Physics beyond 
the Standard Model: Beyond the Desert 2002 (Oulu, Finland)';
and in that context we thank all the organizers, and in particular Hans
Klapdor-Kleingrothaus for his early encouragement.
DG gratefully acknowledges financial support 
and enjoyable hospitality from 
CIU at the University of Zacatecas in autumn 2003 
while part of this paper was conceived.  
Muchas gracias a Gema Mercado Sanchez por la organizacion de mi estancia en Zacatecas.

DVA-K acknowledges 
CONACyT (Mexico) 
   for funding this research through Project
   32067-E. IUCAA (India) for its hospitality which
   supported part of this work.
DG's work has been supported by 
  the Erwin Schr{\"o}dinger fellowship J-2330-N08 of the Austrian
  Science Foundation (FWF).

\begin{appendix}

\section{Appendix:  Auxiliary details}\label{app:aux}

\def\aux{\mbox{\boldmath$\displaystyle\mathbf{\phi_L^\pm(\0)}$}}

\subsection{The $\aux$}\label{app:b}

Representing the unit vector along $\p$, 
as 
\beq
\hspace*{-115pt}
\hp =\Big(\sin(\theta)\cos(\phi),\,
\sin(\theta)\sin(\phi),\,\cos(\theta)\Big)\,,
\eeq
the $\phi^\pm_\mathrm{L}(\0)$ take the explicit form
\beq
\hspace*{-150pt}&&\phi_\mathrm{L}^+(\0) =
\sqrt{m} \mathrm{e}^{\ri\vartheta_1} 
\left(
\begin{array}{c}
\cos(\theta/2) \mathrm{e}^{-\ri\phi/2}\\
\sin(\theta/2) \mathrm{e}^{\ri\phi/2}
\end{array}
\right)\,,\\
\hspace*{-150pt}&&\phi_\mathrm{L}^-(\0) =
\sqrt{m} \mathrm{e}^{\ri\vartheta_2} 
\left(
\begin{array}{c}
\sin(\theta/2) \mathrm{e}^{-\ri\phi/2}\\
-\cos(\theta/2) \mathrm{e}^{\ri\phi/2}
\end{array}
\right)\,.
\eeq
In this paper we take $\vartheta_1$ and $\vartheta_2$ to be zero. 

For the evaluation of spin sums (cf appendix \ref{app:i}) 
the following identities are useful:
\beq
\hspace*{-150pt}&&\sigma_2\phi_\mathrm{L}^+(\0) =
\sqrt{m} 
\left(
\begin{array}{c}
-\ri\sin(\theta/2) \mathrm{e}^{\ri\phi/2}\\
\ri\cos(\theta/2) \mathrm{e}^{-\ri\phi/2}
\end{array}
\right)\,,\\
\hspace*{-150pt}&&\sigma_2\phi_\mathrm{L}^-(\0) =
\sqrt{m}  
\left(
\begin{array}{c}
\ri\cos(\theta/2) \mathrm{e}^{\ri\phi/2}\\
\ri\sin(\theta/2) \mathrm{e}^{-\ri\phi/2}
\end{array}
\right)\,,
\eeq
as they imply
\begin{eqnarray}
\hspace*{-205pt}&& 
(\phi_\mathrm{L}^\pm(\0))^\dagger \sigma_2 (\phi_\mathrm{L}^\pm
(\0))^\ast = 0\,, \\
\hspace*{-205pt}&& (\phi_\mathrm{L}^\pm(\0))^\dagger \sigma_2 
(\phi_\mathrm{L}
^\mp(\0))^\ast = \pm \ri\,.
\end{eqnarray}
Additional helpful relations are
\begin{eqnarray}
\hspace*{-225pt}&& (\phi_\mathrm{L}^\pm(\0))^\dagger 
(\phi_\mathrm{L}^\pm(\0)) = 1\,, \\
\hspace*{-225pt}&& (\phi_\mathrm{L}^\pm(\0))^\dagger 
(\phi_\mathrm{L}^\mp(\0)) = 0\,.
\end{eqnarray}

\def\aux2{\mbox{\boldmath$\displaystyle\mathbf{\Theta (\phi_\mathrm{L}
^\pm(\0))^\ast}$}}
\subsection{Helicity properties of $\aux2$}
\label{app:a}

Complex conjugating equation (\ref{x})
gives
\beq
\hspace*{-160pt}{\s}^\ast\cdot{{\hp}} 
\;\left[\phi_\mathrm{L}^\pm (\0)\right]^\ast= 
\pm\;\left[\phi_\mathrm{L}^\pm(\0)\right]^\ast\,.  
\eeq
Substituting for $\s^\ast$ from equation (\ref{wigner}) then results in
\beq
\hspace*{-140pt}\Theta 
\s \Theta^{-1}\cdot{\hp} \,\left[\phi_\mathrm{L}^\pm (\0)\right]^\ast
= 
\mp\,\left[\phi_\mathrm{L}^\pm(\0)\right]^\ast\, .
\eeq
But $\Theta^{-1} = -\Theta$. So 
\beq
\hspace*{-137pt}
-\; \Theta \s \Theta
\cdot{\hp} \,\left[\phi_\mathrm{L}^\pm (\0)\right]^\ast= 
\mp\,\left[\phi_\mathrm{L}^\pm(\0)\right]^\ast \,.
\eeq
Or, equivalently,
\beq
\hspace*{-140pt}
\Theta^{-1} \s \Theta
\cdot{\hp} \,\left[\phi_\mathrm{L}^\pm (\0)\right]^\ast= 
\mp\,\left[\phi_\mathrm{L}^\pm(\0)\right]^\ast \,.
\eeq
Finally, left-multiplying  both sides of the preceding 
equation by $\Theta$, and moving $\Theta$ through 
${\hp}$, yields equation(\ref{y}).

\section{Appendix: {\ecco}logy details \label{app:maj}}

\def\aux3{\mbox{\boldmath$\displaystyle\mathbf{\lambda(\p)}$}}
\subsection{Bi-orthonormality relations for 
$\aux3$ spinors}\label{app:c}

On setting $\vartheta_1$ and $\vartheta_2$ to be zero \textemdash~ a fact
that we explicitly note \cite{Ahluwalia:1994pf,Ahluwalia:1996uy} \textemdash~ 
we find the following \textit{ 
bi-orthonormality\/} relations for the self-conjugate 
spinors:
\beq
\hspace*{-35pt} \overline{\lambda}^\mathrm{S}_{\{-,+\}}(\p) 
\lambda^\mathrm{S}_{\{-,+\}} (\p) = 0\,,
\qquad 
\overline{\lambda}^\mathrm{S}_{\{-,+\}}(\p) 
\lambda^\mathrm{S}_{\{+,-\}} (\p) = + 2 \ri m
\,,&& \label{bo1}\\
\hspace*{-35pt} \overline{\lambda}^\mathrm{S}_{\{+,-\}}(\p) 
\lambda^\mathrm{S}_{\{-,+\}} (\p) = - 2 \ri m\,,
\qquad 
\overline{\lambda}^\mathrm{S}_{\{+,-\}}(\p) 
\lambda^\mathrm{S}_{\{+,-\}} (\p) = 0
\,.\label{bo2}&&
\eeq
Their counterpart for anti-self-conjugate spinors reads 
\beq
\hspace*{-35pt} 
\overline{\lambda}^\mathrm{A}_{\{-,+\}}(\p) \lambda^\mathrm{A}_{\{-,+\}} 
(\p) = 0\,,
\qquad 
\overline{\lambda}^\mathrm{A}_{\{-,+\}}(\p) 
\lambda^\mathrm{A}_{\{+,-\}} (\p) = - 2 \ri m
\,,&&\label{bo3}\\
\hspace*{-35pt} 
\overline{\lambda}^\mathrm{A}_{\{+,-\}}(\p) 
\lambda^\mathrm{A}_{\{-,+\}} (\p) = + 2 \ri m\,,
\qquad 
\overline{\lambda}^\mathrm{A}_{\{+,-\}}(\p) \lambda^\mathrm{A}_{\{+,-\}} 
(\p) = 0
\,,\label{bo4}&&
\eeq
while all combinations  of the type 
$\overline{\lambda}^\mathrm{A}(\p) \lambda^\mathrm{S}(\p)$ and
$\overline{\lambda}^\mathrm{S}(\p) \lambda^\mathrm{A}(\p)$ identically vanish.
We take note that the bi\textendash orthogonal norms of the \es
are intrinsically {\em imaginary.}
The associated completeness relation is
\def\da{{\{+,-\}}}
\def\ua{{\{-,+\}}}

\beq
\hspace*{-125pt}&&-\frac{1}{2 \ri m}
{\Bigg(}\left[\lambda^\mathrm{S}_{\{-,+\}}(\p) 
\overline{\lambda}^\mathrm{S}_{\da}(\p)
-\lambda^\mathrm{S}_{\da}(\p) 
\overline{\lambda}^\mathrm{S}_{\{-,+\}}(p)\right]   \nonumber\\ 
\hspace*{-150pt}&&\hspace*{50pt} \quad -
\left[\lambda^\mathrm{A}_{\{-,+\}}(\p) \overline{\lambda}^\mathrm{A}
_{\da}(\p)
-\lambda^\mathrm{A}_{\da} (\p)\overline{\lambda}^\mathrm{A}_{\{-,+\}}(\p)
\right]{\Bigg)}
 = \mathbb{I}\,. \label{lc}
\eeq

\def\aux4{\mbox{\boldmath$\displaystyle\mathbf{\rho(\p)}$}}
\subsection{The $\aux4$ spinors}\label{app:d}

Now, $(1/2,0)\oplus(0,1/2)$ is a four-dimensional representation space.
Therefore, there cannot be more than four independent spinors.
Consistent with this observation, we find that 
the $\rho(\p)$ spinors are related to the $\lambda(\p)$ spinors 
through the following identities:
\beq
\hspace{-70pt}&&\rho^\mathrm{S}_{\da}(\p) = 
+\, \ri \lambda^\mathrm{A}_{\da}(\p)\,,\quad
\rho^\mathrm{S}_{\ua}(\p) = - \,\ri \lambda^\mathrm{A}_{\ua}(\p),\label{id1}\\
\hspace{-70pt}&&\rho^\mathrm{A}_{\da}(\p) = 
- \,\ri \lambda^\mathrm{S}_{\da}(\p)\,,\quad
\rho^\mathrm{A}_{\ua}(\p) = +\, \ri \lambda^\mathrm{S}_{\ua}(\p)\,.\label{id2}
\eeq
Using these identities, one may immediately obtain 
the bi-orthonormality
and completeness relations for the $\rho(\p)$ spinors. 
In the massless limit, $\rho^\mathrm{S}_{\da}(\p)$ and $\rho^\mathrm{A}_
{\da}(\p)$
{\em identically vanish.\/}
A particularly simple orthonormality, as opposed to bi-orthonormality,
relation exists between the 
$\lambda(\p)$ and $\rho(\p)$ spinors:
\beq
\hspace{-75pt}\overline{\lambda}^\mathrm{S}_{\ua}(\p) 
\rho^\mathrm{A}_{\ua}(\p) = -2 m = \overline{\lambda}^\mathrm{A}_{\ua}(\p)
\rho^\mathrm{S}_{\ua} (\p)\\  
\hspace{-75pt}\overline{\lambda}^\mathrm{S}_{\da}(\p) 
\rho^\mathrm{A}_{\da}(\p) = - 2 m = \overline{\lambda}^\mathrm{A}_{\da}(\p)
\rho^\mathrm{S}_{\da} (\p).
\eeq
An associated completeness relation also exists, and it reads
\beq
\hspace*{-125pt}&&-\frac{1}{2  m}
{\Bigg(}\left[\lambda^\mathrm{S}_{\ua}(\p) \overline{\rho}^\mathrm{A}_{\ua}(\p)
+\lambda^\mathrm{S}_{\da}(\p) \overline{\rho}^\mathrm{A}_
{\da}(p)\right] \nonumber \\
\hspace*{-150pt}&&\hspace*{50pt}\quad+
\left[\lambda^\mathrm{A}_{\ua}(\p) \overline{\rho}^\mathrm{S}_{\ua}(\p)
+\lambda^\mathrm{A}_{\da} (\p)\overline{\rho}^\mathrm{S}_{\da}
(\p)\right]{\Bigg)} 
= \mathbb{I}\,.
\label{lrcompleteness}
\eeq
The results of this section are in the spirit of 
\cite{Mc1957,Case1957,Ahluwalia:1994pf,Ahluwalia:1996uy}.

The completeness relation (\ref{lc}) confirms
that a physically complete theory of 
fundamentally neutral particle spinors must incorporate the 
self as well as anti-self-conjugate spinors. However, one has a choice.
One may either work with the set $\{\lambda^\mathrm{S}(\p),
\lambda^\mathrm{A}(\p\})$,
or with the physically and mathematically equivalent set,
  $\{\rho^\mathrm{S}(\p),\rho^\mathrm{A}(\p)\}$. 
One is also free to choose some 
appropriate combinations of neutral particle spinors 
from these two sets.

 \subsection{{\ecco} in the Majorana realization}\label{app:e}

 The $\lambda^\mathrm{S,A}(\p)$ obtained above are in Weyl realization  
 (subscripted by $\mathrm{W}$).
 In  Majorana realization (subscripted by $\mathrm{M}$) 
these spinors are given by:
 \beq
\hspace*{-210pt}
 \lambda^\mathrm{S,A}_\mathrm{M}(\p) 
= {\mathcal S} \,\lambda^\mathrm{S,A}_\mathrm{W}(\p)\,,
 \eeq
 where
 \beq
\hspace*{-160pt}
 {\mathcal S} = \frac{1}{2} \left(
 \begin{array}{cc}
 \mathbb{I} + \ri\Theta &{\;\;} \mathbb{I} - \ri\Theta \\
 -\left(\mathbb{I} - \ri\Theta\right) & {\;\;}\mathbb{I} + \ri\Theta 
 \end{array}\right)\,.
 \eeq 
 Calculations show that the 
 $\lambda^\mathrm{S}_\mathrm{M}(\p)$ are real, 
while the $\lambda^\mathrm{A}_\mathrm{M}(\p)$
 are imaginary.

\subsection{Spin sums}\label{app:i}

The evaluation of the spin sums is straightforward. We perform 
it here explicitly for the self-dual spinors and sketch briefly 
the result for the 
anti-self-dual ones. The definition (\ref{eq:madu1}) 
together with (\ref{id2}), (\ref{lsup}) and (\ref{lsdown}) yields
\begin{equation}
  \label{eq:ss1}
 \hspace*{-90pt} \sum_{\alpha=\{-,+\},\{+,-\}} \lambda^\mathrm{S}_\alpha(\p)
\stackrel{\neg}\lambda_\alpha^\mathrm{S}(\p) = 
\ri\,\frac{(E+m)^2-p^2}{2m(E+m)} \, {\mathcal S}\,,
\end{equation}
with
\begin{equation}
  \label{eq:ss2}
\hspace*{-80pt}
  {\mathcal S} := \left(\lambda^\mathrm{S}_{\{-,+\}}(\0)
\bar{\lambda}^\mathrm{S}_{\{+,-\}}(\0)-
\lambda^\mathrm{S}_{\{+,-\}}(\0)\bar{\lambda}^\mathrm{S}_{\{-,+\}}(\0)
\right)\,.
\end{equation}
Note that (\ref{eq:ss2}) contains only `ordinary' Dirac bars.
By virtue of the identities
\begin{equation}
  \label{eq:ss3}
\hspace*{-120pt}
  \phi^+_\mathrm{L}(\0)(\phi_\mathrm{L}^-(\0))^\dagger-
\phi^-_\mathrm{L}(\0)(\phi_\mathrm{L}^+(\0))^\dagger = -\ri 
m{\mathcal A}^\mathrm{S}
\end{equation}
and
\begin{equation}
  \label{eq:ss4}
\hspace*{-235pt}
  \sigma_2({\mathcal A}^\mathrm{S})^\ast\sigma_2=-{\mathcal A}^\mathrm{S}
\end{equation}
one obtains
\begin{equation}
  \label{eq:ss5}
\hspace*{-215pt}
  {\mathcal S} = -\ri m \left(
\begin{array}{cc}
\mathbb{I}    & \mathcal{A}^\mathrm{S} \\
\mathcal{A}^\mathrm{S} & \mathbb{I}
\end{array}\right)\,.
\end{equation}
In conjunction with the dispersion relation (\ref{ds1}),
the result (\ref{eq:spinsum}) is finally produced.

The definition (\ref{eq:madu2}) together with (\ref{id1}), (\ref{laup}) and (\ref{ladown}) establishes the same result as in (\ref{eq:ss1}) and (\ref{eq:ss2}) with superscript $S$ replaced by $A$. Consequently, the whole 
expression acquires an overall sign and ${\mathcal A}^\mathrm{S}$ has 
to be replaced by ${\mathcal A}^\mathrm{A}$. Having inserted 
${\mathcal A}^\mathrm{A}=-{\mathcal A}^\mathrm{S}$ the result is displayed 
in (\ref{eq:spinsum2}).

In a similar fashion the following spin sums may be evaluated:
\begin{equation}
  \label{eq:ss6}
\hspace*{-95pt}
  \sum_{\alpha=\{-,+\},\{+,-\}} \lambda^\mathrm{S}_\alpha(\p)
\left(\lambda_\alpha^\mathrm{S}(\p)\right)^\dagger = 
(E-p)\left(\mathbb{I}+\mawave\right)\,, 
\end{equation}
and 
\begin{equation}
  \label{eq:ss7}
\hspace*{-95pt}
  \sum_{\alpha=\{-,+\},\{+,-\}} \lambda^\mathrm{A}_\alpha(\p)
\left(\lambda_\alpha^\mathrm{A}(\p)\right)^\dagger = 
(E+p)\left(\mathbb{I}-\mawave\right)\,, 
\end{equation}
with $\mawave$ as defined in (\ref{eq:gammaphi}).

The `twisted' spin sums relevant to non-locality turns out as
\def\st{{\sin(\theta)}}
\def\ct{{\cos(\theta)}}
\beq
\hspace*{-65pt}&&  \sum_\beta\left[
\lambda^\mathrm{S}_{\beta}(\p) 
\left(\lambda^\mathrm{A}_{\beta}(\p) \right)^\mathrm{T}
+
\lambda^\mathrm{S}_{\beta}(-\p) 
\left(\lambda^\mathrm{A}_{\beta}(-\p) \right)^\mathrm{T}\right]\nonumber\\
\hspace*{-65pt}&&\hspace{55pt} =
2 \left(
\begin{array}{cccc}
e^{-\, \ri\, \phi} \,p \,\ct & p\, \st & 0 & - \ri\, E \\
p\, \st & - \mathrm{e}^{+\,\ri\, \phi}\, p\, \ct & \ri\, E & 0      \\
0 &  -\, \ri \,E &  - \mathrm{e}^{-\, \ri\, \phi} p\, \ct & - p \, \st \\ 
 \ri \,E & 0 & - p\, \st & \mathrm{e}^{+\,\ri\, \phi}\, p\, \ct
\end{array}
\right)\,, \label{eq:ss17}
\eeq
and
\beq
\hspace*{-16pt}&&\sum_\beta\left[
\left(\stackrel{\neg}\lambda^\mathrm{S}_{\beta}(\p)\right)^\dagger 
\stackrel{\neg}\lambda^\mathrm{A}_{\beta}(\p) 
+
\left(\stackrel{\neg}\lambda^\mathrm{S}_{\beta}(-\p)\right)^\dagger 
\stackrel{\neg}\lambda^\mathrm{A}_{\beta}(-\p) 
\right]\nonumber\\
\hspace*{-16pt}&&\hspace{55pt}
=2\left(
\begin{array}{cccc}
 \sqrt{p^2+m^2} & 0 & \ri\, p \sin(\theta) & 
-\, \ri\, \mathrm{e}^{-\,\ri\,\phi}\, p\, \cos(\theta) \\
0 &  \sqrt{p^2+m^2} &  - \, \ri\,
 \mathrm{e}^{+\,\ri\,\phi}\, p\, \cos(\theta) 
&- \, \ri\, p\, \sin(\theta)\\
 \ri\, p\, \sin(\theta) &  - \, \ri\, \mathrm{e}^{-\,\ri\,\phi}\, p \,
\cos(\theta) & -   \sqrt{p^2+m^2} & 0\\
 - \, \ri\, \mathrm{e}^{+\,\ri\,\phi} \,p 
\,\cos(\theta) & - \, \ri\, p\, \sin(\theta) 
& 0 & -  \sqrt{p^2+m^2}
\end{array}
\right)\,.
\nonumber\\
\hspace*{-16pt}\label{eq:ssnl}
\eeq
Finally, the identities
\beq
\hspace*{-193pt}&& \left(\lambda^\mathrm{S/A}_{\{-,+\}}(\p)\right)^\dagger
\lambda^\mathrm{S/A}_{\{+,-\}}(\p) = 0\,,\label{emptwist}\\
\hspace*{-193pt}&& \left(\lambda^\mathrm{S/A}_{\{+,-\}}(\p)\right)^\dagger
\lambda^\mathrm{S/A}_{\{-,+\}}(\p) = 0\,,\label{epptwist}
\eeq
and
\beq
\hspace*{-158pt}&& \left(\lambda^\mathrm{S/A}_{\{-,+\}}(\p)\right)^\dagger
\lambda^\mathrm{S/A}_{\{-,+\}}(\p) = 2 (E-p)\,,\label{emp}\\
\hspace*{-158pt}&& \left(\lambda^\mathrm{S/A}_{\{+,-\}}(\p)\right)^\dagger
\lambda^\mathrm{S/A}_{\{+,-\}}(\p) = 2 (E+p)\,,\label{epp}
\eeq
may be useful in various contexts.

\def\aux5{\mbox{\boldmath$\displaystyle\mathbf{\{\eta,\eta\}}$}}
\subsection{Distributional part of $\aux5$}\label{app:int}

An integral such as that in  (\ref{eq:int2}) can be evaluated 
by methods described in \cite{Gelfand:1964} in the context 
of the Fourier transformation of $\theta(x)x^\lambda$. The general result
\begin{equation}
  \label{eq:gelfand}
\hspace*{-50pt}
  \lim_{\epsilon\to 0}\int
_0^\infty x^\lambda 
\mathrm{e}^{\ri(k+\ri\epsilon)x}\,\mathrm{d}x = 
\ri \mathrm{e}^{\ri\lambda\pi/2}\Gamma(\lambda+1)
\lim_{\epsilon\to 0}\left(k+\ri\epsilon\right)^{-\lambda-1}
\end{equation}
with $\lambda=1$ can be applied to obtain
\begin{eqnarray}
\hspace*{-40pt}&&  \int
_0^\infty x\sin{kx}\,\mathrm{d}x := 
\lim_{\epsilon\to 0}\frac{1}{2 \ri} 
\int
_0^\infty x \left(\mathrm{e}^{\ri(k+\ri
\epsilon)x}-\mathrm{e}^{-\ri(k-\ri
\epsilon)x}\right) = \lim_{\epsilon\to 0} 
\frac{1}{2\ri}\left(\frac{1}{k^2-\ri\epsilon}-
\frac{1}{k^2+\ri\epsilon}\right) \nonumber \\
\hspace*{-40pt}&&  = \lim_{\epsilon\to 0} 
\frac{1}{2\ri}\left(P\left(\frac{1}{k^2}\right)+\ri\pi\delta(k^2)-
P\left(\frac{1}{k^2}\right)+\ri\pi\delta(k^2)\right)  = 
\pi\delta(k^2)\,.
  \label{eq:apint1}
\end{eqnarray}
The symbol $P$ denotes the principal value. In our case 
the quantity $k$ is nothing but the radius $r$. An alternative 
representation of (\ref{eq:apint1}) follows from
\begin{eqnarray}
\hspace*{-20pt}&&  \int
_0^\infty x\sin{kx}\,\mathrm{d}x 
:= -\frac{\mathrm{d}}{\mathrm{d}k}\lim_{\epsilon\to 0}\frac{1}{2} 
\int
_0^\infty \left(\mathrm{e}^{\ri(k+\ri\epsilon)x}-
\mathrm{e}^{-\ri(k-\ri\epsilon)x}\right) = 
\frac{\mathrm{d}}{\mathrm{d}k}\lim_{\epsilon\to 0} \frac{\ri}{2}
\left(\frac{1}{k-\ri\epsilon}-
\frac{1}{k+\ri\epsilon}\right) \nonumber \\
\hspace*{-20pt}&&  = \frac{\mathrm{d}}{\mathrm{d}k}\lim_{\epsilon\to 0} 
\frac{\ri}{2}\left(P\left(\frac{1}{k}\right)+\ri\pi\delta(k)-
P\left(\frac{1}{k}\right)+\ri\pi\delta(k)\right)  = 
-\pi\delta^\prime(k)\,.
  \label{eq:apint2}
\end{eqnarray}

\def\auxX{\mbox{\boldmath$\displaystyle\mathbf{\phi }$}}
\def\aux7{\mbox{\boldmath$\displaystyle\mathbf{\mathcal{O}}$}}

\subsection{On the $\auxX$ dependence of  $\aux7$ for {\ecco} 
and non-standard dispersion relations}
\label{Sec:noncomm}

The matrix $\mawave$ depends on a direction $\n$ which is 
orthogonal to the direction of propagation, $\hp$, but it is 
independent from $|\p|$ and $p_0$, in contrast to the Dirac case. 
A different way of writing the projection operators is
\begin{equation}
  \label{eq:proj}
\hspace*{-207pt}
  \frac12 (\mathbb{I}\pm\mawave)=\frac{
\mathbb{I}\pm\gamma^5\gamma_\mu n^\mu}{2}\,,
\end{equation}
with $n_\mu=(0,\n)$. We emphasize that $\n=(1/\sin{\theta})
\mathrm{d}\hp/\mathrm{d}\phi$ 
is not independent from $\p$ as it depends on $\phi$.

In the light of this angular dependence we consider the possibility of a boost 
operator different from the standard one, implying in general also a 
non-standard dispersion relation. To this end let us replace (\ref{br}) 
and (\ref{bl}) by
\begin{equation}
\hspace*{-150pt}
  \rb = \exp{A}\,,\quad \lb = \exp{B}\,,\label{nc1}
\end{equation}
with some as yet unspecified operators $A,B$. The exponential 
representation has been chosen in order to make invertibility manifest, 
but for specific cases other representations might be more useful. 
All identities derived in section \ref{sec:spinsumconnection} still 
hold. As the operator ${\mathcal O}$ has block form with mutually 
commuting non-singular entries, its determinant is given by 
\begin{equation}
  \label{eq:detO}
\hspace*{-181pt}
  \det{{\mathcal O}}=\det{\left(\mathbb{I}-\matD\matD^{-1}\right)}=0\,,
\end{equation}
with $\matD$ as defined in (\ref{eq:D}). Thus, the determinant of ${\mathcal O}$ vanishes without implying further restrictions. 

Regarding the multiplicity of the dispersion relations it should be 
noted that the matrix ${\mathcal O}$ in (\ref{meq}) maximally has half 
rank because the lower block linearly depends from the upper one. 
If $\matD\propto\mathbb{I}$ the rank of ${\mathcal O}$ is 1, else it is 2. 
The first case is trivial and may arise only for very special choices of 
$A,B$ and ${\mathcal A}$. Thus, generically there will be either one 
dispersion relation with multiplicity 2 as in (\ref{ds1}) or two 
dispersion relations with multiplicity 1 as in (\ref{nd}). Clearly, 
the explicit form of the dispersion relations will depend on the choice 
of the boost operator, but {\em not} on the matrix ${\mathcal A}$. For 
the standard choice (\ref{br}), (\ref{bl}) only the standard dispersion 
relation (\ref{ds1}) appears.

Finally, the question will be addressed to what extent \es particles 
may probe non-commutativity of energy\textendash momentum space or 
deformations of the Lorentz group different from the way Dirac particles 
do. Such a difference, if any, can be traced back to the behavior of the 
matrix ${\mathcal A}$ encoding the $CPT$ properties which is proportional 
to the unit matrix only for Dirac particles. The dispersion relation 
is independent from it, but the spin sum operator ${\mathcal O}$ is 
sensitive to it. Consequently, \es particles may probe aspects of 
non-standard dispersion relations in a way different from Dirac particles. 

For the sake of concreteness we suppose 
$A=A_\mu\sigma^\mu$ and $B=B_\mu\sigma^\mu$ 
with $\sigma^\mu=(\mathbb{I},\s)$. It is useful 
to introduce an adapted orthogonal Dreibein $\hp, 
\n=(1/\sin{\theta})d\hp/d\phi, \l:=\n\times\hp=\mathrm{d}\hp/
\mathrm{d}\theta$ and to decompose $\s$ with respect to it.
Note that in the Dirac case ${\mathcal A}$ 
trivially commutes with all these projections, 
while for \es we obtain
\begin{equation}
  \label{eq:commutators}
\hspace*{-90pt}
  [{\mathcal A},\sigma^0]=0\,,\qquad\{{\mathcal A},\s\cdot\p\}=0\,,
\qquad[{\mathcal A},\s\cdot\n]=0\,,\qquad\{{\mathcal A},\l\cdot\p\}=0\,.
\end{equation}
Therefore, for Dirac particles $\matD = 
\exp{(A_\mu\sigma^\mu)}\exp{(B_\mu\sigma^\mu)}\mathcal{A}$, 
while for \es particles $\matD = 
\exp{(A_\mu\sigma^\mu)}\exp{(\tilde{B}_\mu\sigma^\mu)}\mathcal{A}$, 
where $\tilde{B}$ can be derived from $B$ by virtue of 
(\ref{eq:commutators}). Because $B$ typically has a non-vanishing 
$\hp$ component, $\tilde{B} \neq B$ in general. 
A possibility for non-standard boost operators which 
can be considered as `natural' in the context of 
\es particles has been addressed in 
section \ref{Sec:cc}.

\subsection{Some other anticommutators in the 
context of  non-locality discussion}
\label{Sec:other}

Here we collect the anticommutators 
$\{\eta^\dagger(\x,t),\eta^\dagger(\x^\prime,t)\}$ and $\{\stackrel{\neg}{\eta}(\x,t),\stackrel{\neg}{\eta}(\x^\prime,t)\}$. The former just follows from (\ref{eq:nonlocres}) by Hermitean conjugation and thus 
provides
\begin{equation}
\hspace*{-85pt}
\Big\langle\hspace{11pt}\Big\vert
\left\{\eta^\dagger(\x,t),\eta^\dagger(\x^\prime,t)\right\}
 \Big\vert\hspace{11pt}\Big\rangle = 
 - \frac{1+mr}{4\,m\,\pi r^3}e^{-mr} \gamma^0\gamma^1 - \frac{1}{2\,m\,\pi 
r^2}\delta(r^2)\gamma^2\gamma^5\,,\label{eq:nonlocher}
\end{equation}
because all $\gamma$ matrices are Hermitean and 
anticommute with each other. It should be noted that (\ref{eq:nonlocher}) 
is just the negative of (\ref{eq:nonlocres}).
By virtue of the identities
\begin{equation}
  \label{eq:antiid}
\hspace*{-80pt}
  \left\{\stackrel{\neg}{\eta}(\x,t),\stackrel{\neg}{\eta}(\x^\prime,t)
\right\}\sim\gamma^0\left\{\eta^\dagger(\x,t),\eta^\dagger(\x^\prime,t)
\right\}\gamma^0\sim-\gamma^0\left\{\eta(\x,t),\eta(\x^\prime,t)\right\}
\gamma^0\,,
\end{equation}
where $\sim$ means equivalence of corresponding vacuum expectation values,
one obtains after trivial rearrangements of $\gamma$ matrices, 
\begin{equation}
\hspace*{-95pt}
\Big\langle\hspace{11pt}\Big\vert
\left\{\stackrel{\neg}{\eta}(\x,t),\stackrel{\neg}{\eta}(\x^\prime,t)\right\}
 \Big\vert\hspace{11pt}\Big\rangle = \frac{1+mr}{4\,m\,\pi r^3} 
e^{-mr} \gamma^0\gamma^1 - \frac{1}{2\,m\,\pi r^2}
\delta(r^2)\gamma^2\gamma^5\,. \label{eq:nonlocdua}
\end{equation}
This result exhibits the same behavior of the distributional part as (\ref{eq:nonlocher}) and the same behavior of the remaining 
part as (\ref{eq:nonlocres}).

\vspace{16pt}

\noindent\small
\textit{Note added in proof.\/} While this paper was being proof read,
\cite{daRocha:2005ti} appeared, which places \textit{Elko}
as Lounesto's class 5 spinors. It further emphasizes its differences
and similarities with the Majorana spinors. The authors also 
confirm our result contained in equation (\ref{eq:cpelko}).

\end{appendix}


\begin{thebibliography}{10%
0}

\small

\bibitem{Nicastro:2004fb}
{F. Nicastro {\it et al.}}, {\it The mass of the missing baryons in the x-ray
  forest of the warm-hot intergalactic medium},  {\em Nature} (2005) 495--498,
  [\href{http://xxx.lanl.gov/abs/astro-ph/0412378}{{\tt astro-ph/0412378}}].

\bibitem{Zwicky:1933gu}
F.~Zwicky, {\it Spectral displacement of extra galactic nebulae},  {\em Helv.
  Phys. Acta} {\bf 6} (1933) 110.

\bibitem{Einasto:2004rf}
J.~Einasto, {\it {Dark matter: early considerations}},
  \href{http://xxx.lanl.gov/abs/astro-ph/0401341}{{\tt astro-ph/0401341}}.

\bibitem{Rees:2004ta}
M.~J. Rees, {\it {Dark matter: introduction}},  {\em Phil. Trans. Roy. Soc.
  Lond.} {\bf 361} (2003) 2427.

\bibitem{Olive:2003iq}
K.~A. Olive, {\it {TASI} lectures on dark matter},
  \href{http://xxx.lanl.gov/abs/astro-ph/0301505}{{\tt astro-ph/0301505}}.

\bibitem{Wigner:1962bww}
E.~P. Wigner, {\it Unitary representations of the inhomogeneous {L}orentz group
  including reflections},  in {\em Group theoretical concepts and methods in
  elementary particle physics, Lectures of the Istanbul summer school of
  theoretical physics (1962)} (F.~G\"ursey, ed.), pp.~37--80.
\newblock Gordon and Breach, New York, 1964.

\bibitem{Lee:1966tdw}
T.~D. Lee and G.~C. Wick, {\it {Space inversion, time reversal, and other
  discrete symmetries in local field theories}},  {\em Phys. Rev.} {\bf 148}
  (1966) 1385--1404.

\bibitem{Weinberg:1995sw}
S.~Weinberg, {\em {The Quantum theory of fields. Vol. 1: Foundations}}.
\newblock Cambridge University Press, 1995.

\bibitem{Ahluwalia:2000pj}
D.~V. Ahluwalia and M.~Kirchbach, {\it {(1/2,1/2) representation space: an ab
  initio construct}},  {\em Mod. Phys. Lett.} {\bf A16} (2001) 1377--1384,
  [\href{http://xxx.lanl.gov/abs/hep-th/0101009}{{\tt hep-th/0101009}}].

\bibitem{Ruegg:2003ps}
H.~Ruegg and M.~Ruiz-Altaba, {\it {The Stueckelberg field}},
  \href{http://xxx.lanl.gov/abs/hep-th/0304245}{{\tt hep-th/0304245}}.

\bibitem{Reuter:2004nx}
M.~Reuter and H.~Weyer, {\it Quantum gravity at astrophysical distances?},
  \href{http://xxx.lanl.gov/abs/hep-th/0410119}{{\tt hep-th/0410119}}.

\bibitem{Milgrom:1983pn}
M.~Milgrom, {\it A modification of the {N}ewtonian dynamics: implications for
  galaxies},  {\em Astrophys. J.} {\bf 270} (1983) 371--389.

\bibitem{Sanders:2002pf}
R.~H. Sanders and S.~S. McGaugh, {\it Modified {Newtonian} dynamics as an
  alternative to dark matter},  {\em Ann. Rev. Astron. Astrophys.} {\bf 40}
  (2002) 263--317, [\href{http://xxx.lanl.gov/abs/astro-ph/0204521}{{\tt
  astro-ph/0204521}}].

\bibitem{Ahluwalia:2002eh}
D.~V. Ahluwalia, N.~Dadhich, and M.~Kirchbach, {\it On the spin of
  gravitational bosons},  {\em Int. J. Mod. Phys.} {\bf D11} (2002) 1621--1634,
  [\href{http://xxx.lanl.gov/abs/gr-qc/0205072}{{\tt gr-qc/0205072}}].

\bibitem{Kiselev:2004vy}
V.~V. Kiselev, {\it Vector field and rotational curves in dark galactic halos},
   {\em Class. Quant. Grav.} {\bf 22} (2005) 541--558,
  [\href{http://xxx.lanl.gov/abs/gr-qc/0404042}{{\tt gr-qc/0404042}}].

\bibitem{Bekenstein:2004ne}
J.~D. Bekenstein, {\it Relativistic gravitation theory for the {MOND}
  paradigm},  {\em Phys. Rev.} {\bf D70} (2004) 083509,
  [\href{http://xxx.lanl.gov/abs/astro-ph/0403694}{{\tt astro-ph/0403694}}]. J.
  D. Bekenstein, Phys. Rev. D. {\bf 71} (2005) 069901 (erratum).

\bibitem{Merrifield:2004tr}
M.~R. Merrifield, {\it Dark matter on galactic scales (or the lack thereof)},
  \href{http://xxx.lanl.gov/abs/astro-ph/0412059}{{\tt astro-ph/0412059}}.

\bibitem{Visser:2004bf}
M.~Visser, {\it Cosmography: cosmology without the {E}instein equations},
  \href{http://xxx.lanl.gov/abs/gr-qc/0411131}{{\tt gr-qc/0411131}}.

\bibitem{Wigner:1939cj}
E.~P. Wigner, {\it On unitary representations of the inhomogeneous {L}orentz
  group},  {\em Annals Math.} {\bf 40} (1939) 149--204,
  [\href{http://xxx.lanl.gov/abs/{Reprinted in: Nucl. Phys. Proc. Suppl. {\bf
  6} (1989) 9}}{{\tt {Reprinted in: Nucl. Phys. Proc. Suppl. {\bf 6} (1989)
  9}}}].

\bibitem{Ahluwalia:1993zt}
D.~V. Ahluwalia, M.~B. Johnson, and T.~Goldman, {\it {A
  Bargmann-Wightman-Wigner type quantum field theory}},  {\em Phys. Lett.} {\bf
  B316} (1993) 102--108, [\href{http://xxx.lanl.gov/abs/hep-ph/9304243}{{\tt
  hep-ph/9304243}}].

\bibitem{Ahluwalia-Khalilova:2004sz}
D.~V. Ahluwalia-Khalilova and D.~Grumiller, {\it {Dark matter: a spin one half
  fermion field with mass dimension one?}},
  \href{http://xxx.lanl.gov/abs/hep-th/0410192}{{\tt hep-th/0410192}}.

\bibitem{Douglas:2001ba}
M.~R. Douglas and N.~A. Nekrasov, {\it {Noncommutative field theory}},  {\em
  Rev. Mod. Phys.} {\bf 73} (2001) 977--1029,
  [\href{http://xxx.lanl.gov/abs/hep-th/0106048}{{\tt hep-th/0106048}}].

\bibitem{Szabo:2001kg}
R.~J. Szabo, {\it {Quantum field theory on noncommutative spaces}},  {\em Phys.
  Rept.} {\bf 378} (2003) 207--299,
  [\href{http://xxx.lanl.gov/abs/hep-th/0109162}{{\tt hep-th/0109162}}].

\bibitem{Athanassopoulos:1996jb}
{C. Athanassopoulos {\textit{et al.}} [LSND Collaboration]}, {\it Evidence for
  $\bar\nu_\mu \to \bar\nu_e$ oscillation from the {LSND} experiment at the
  {Los Alamos Meson Physics Facility}},  {\em Phys. Rev. Lett.} {\bf 77} (1996)
  3082--3085, [\href{http://xxx.lanl.gov/abs/nucl-ex/9605003}{{\tt
  nucl-ex/9605003}}].

\bibitem{Athanassopoulos:1997pv}
{C. Athanassopoulos {\textit{et al.}} [LSND Collaboration]}, {\it Evidence for
  $\nu_\mu \to \nu_e$ neutrino oscillations from {LSND}},  {\em Phys. Rev.
  Lett.} {\bf 81} (1998) 1774--1777,
  [\href{http://xxx.lanl.gov/abs/nucl-ex/9709006}{{\tt nucl-ex/9709006}}].

\bibitem{Ahluwalia:1998xb}
D.~V. Ahluwalia, {\it {Reconciling super-{K}amiokande, {LSND}, and Homestake
  neutrino oscillation data}},  {\em Mod. Phys. Lett. A} {\bf 13} (1998)
  2249--2264, [\href{http://xxx.lanl.gov/abs/hep-ph/9807267}{{\tt
  hep-ph/9807267}}].

\bibitem{Murayama:2000hm}
H.~Murayama and T.~Yanagida, {\it {LSND, SN1987A, and CPT violation}},  {\em
  Phys. Lett. B} {\bf 520} (2001) 263--268,
  [\href{http://xxx.lanl.gov/abs/hep-ph/0010178}{{\tt hep-ph/0010178}}].

\bibitem{Raaf:2004hx}
{J. L. Raaf {\textit{et al.}} [BooNE Collaboration]}, {\it {MiniBooNE status}},
   2004.
\newblock \href{http://xxx.lanl.gov/abs/hep-ex/0408008}{{\tt hep-ex/0408008}}.

\bibitem{Majorana:1937vz}
E.~Majorana, {\it Theory of the symmetry of electrons and positrons},  {\em
  Nuovo Cim.} {\bf 14} (1937) 171--184.

\bibitem{Marshak:1961ecg}
R.~E. Marshak and E.~C.~G. Sudarshan, {\em Introduction to elementary particle
  physics}.
\newblock John Wiley, 1961.

\bibitem{Mohapatra:1991ng}
R.~N. Mohapatra and P.~B. Pal, {\it Massive neutrinos in physics and
  astrophysics},  {\em World Sci. Lect. Notes Phys.} {\bf 41} (1991) 1--318.
  see, Sec. 4.2.

\bibitem{LHR1996}
L.~H. Ryder, {\em Quantum Field Theory}.
\newblock Cambridge University Press, Cambridge, 1996.
\newblock see ch. 2.

\bibitem{Klapdor-Kleingrothaus:hans2001}
H.~V. Klapdor-Kleingrothaus, ed., {\em Sixty years of Double Beta Decay}.
\newblock World Scientific, Singapore, 2001.

\bibitem{Klapdor-Kleingrothaus:2001ke}
H.~V. Klapdor-Kleingrothaus, A.~Dietz, H.~L. Harney, and I.~V. Krivosheina,
  {\it Evidence for neutrinoless double beta decay},  {\em Mod. Phys. Lett.}
  {\bf A16} (2001) 2409--2420,
  [\href{http://xxx.lanl.gov/abs/hep-ph/0201231}{{\tt hep-ph/0201231}}].

\bibitem{Klapdor-Kleingrothaus:2002md}
H.~V. Klapdor-Kleingrothaus, A.~Dietz, and I.~V. Krivosheina, {\it Status of
  evidence for neutrinoless double beta decay},  {\em Found. Phys.} {\bf 32}
  (2002) 1181--1223, [\href{http://xxx.lanl.gov/abs/hep-ph/0302248}{{\tt
  hep-ph/0302248}}]. H. V. Klapdor-Kleingrothaus, A. Dietz, I. V. Krivosheina,
  Found. Phys. {\bf 33} (2003) 679-684 (erratum).

\bibitem{Klapdor-Kleingrothaus:2004gs}
H.~V. Klapdor-Kleingrothaus, A.~Dietz, I.~V. Krivosheina, C.~Dorr, and
  C.~Tomei, {\it Support of evidence for neutrinoless double beta decay},  {\em
  Phys. Lett.} {\bf B578} (2004) 54--62,
  [\href{http://xxx.lanl.gov/abs/hep-ph/0312171}{{\tt hep-ph/0312171}}].

\bibitem{Klapdor-Kleingrothaus:2004ge}
H.~V. Klapdor-Kleingrothaus, A.~Dietz, I.~V. Krivosheina, and O.~Chkvorets,
  {\it {Data acquisition and analysis of the Ge-76 double beta experiment in
  Gran Sasso 1990-2003}},  {\em Nucl. Instrum. Meth.} {\bf A522} (2004)
  371--406, [\href{http://xxx.lanl.gov/abs/hep-ph/0403018}{{\tt
  hep-ph/0403018}}].

\bibitem{Case1957}
K.~M. Case, {\it {Reformulation of Majorana theory of the neutrino}},  {\em
  Phys. Rev.} {\bf 107} (1957) 307--316.

\bibitem{Mc1957}
J.~A. McLennan, Jr., {\it Parity nonconservation and the theory of neutrino},
  {\em Phys. Rev.} {\bf 106} (1957) 821--822.

\bibitem{Ahluwalia:1993cz}
D.~V. Ahluwalia, T.~Goldman, and M.~B. Johnson, {\it {Majorana-like $(j,0)
  \oplus (0,j)$ representation spaces: construction and physical
  interpretation}},  {\em Mod. Phys. Lett. A} {\bf 9} (1994) 439--450,
  [\href{http://xxx.lanl.gov/abs/hep-th/9307118}{{\tt hep-th/9307118}}].

\bibitem{Ahluwalia:1996uy}
D.~V. Ahluwalia, {\it {Theory of neutral particles: McLennan-Case construct for
  neutrino, its generalization, and a fundamentally new wave equation}},  {\em
  Int. J. Mod. Phys.} {\bf A11} (1996) 1855--1874,
  [\href{http://xxx.lanl.gov/abs/hep-th/9409134}{{\tt hep-th/9409134}}].

\bibitem{Ahluwalia-Khalilova:2003jt}
D.~V. Ahluwalia-Khalilova, {\it Extended set of {M}ajorana spinors, a new
  dispersion relation, and a preferred frame},
  \href{http://xxx.lanl.gov/abs/hep-ph/0305336}{{\tt hep-ph/0305336}}.

\bibitem{Bernabei:2003su}
{R. Bernabei \textit{et al.}}, {\it Results from {DAMA/NaI} and perspectives
  for {DAMA/LIBRA}},  in {\em {Proceedings of the Fourth Tegernsee
  International Conference on Particle Physics Beyond the Standard Model,
  BEYOND 2003, Germany}} (H.~V. Klapdor-Kleingrothaus, ed.), pp.~541--560,
  June, 2003.

\bibitem{Nath:1997qm}
P.~Nath and R.~Arnowitt, {\it Non-universal soft {SUSY} breaking and dark
  matter},  {\em Phys. Rev.} {\bf D56} (1997) 2820--2832,
  [\href{http://xxx.lanl.gov/abs/hep-ph/9701301}{{\tt hep-ph/9701301}}].

\bibitem{Dimopoulos:1996gy}
S.~Dimopoulos, G.~F. Giudice, and A.~Pomarol, {\it Dark matter in theories of
  gauge-mediated supersymmetry breaking},  {\em Phys. Lett.} {\bf B389} (1996)
  37--42, [\href{http://xxx.lanl.gov/abs/hep-ph/9607225}{{\tt
  hep-ph/9607225}}].

\bibitem{Chattopadhyay:2001vx}
U.~Chattopadhyay and P.~Nath, {\it Upper limits on sparticle masses from g-2
  and the possibility for discovery of susy at colliders and in dark matter
  searches},  {\em Phys. Rev. Lett.} {\bf 86} (2001) 5854--5857,
  [\href{http://xxx.lanl.gov/abs/hep-ph/0102157}{{\tt hep-ph/0102157}}].

\bibitem{Bottino:2002ry}
A.~Bottino, N.~Fornengo, and S.~Scopel, {\it Light relic neutralinos},  {\em
  Phys. Rev.} {\bf D67} (2003) 063519,
  [\href{http://xxx.lanl.gov/abs/hep-ph/0212379}{{\tt hep-ph/0212379}}].

\bibitem{Feng:2000gh}
J.~L. Feng, K.~T. Matchev, and F.~Wilczek, {\it Neutralino dark matter in focus
  point supersymmetry},  {\em Phys. Lett.} {\bf B482} (2000) 388--399,
  [\href{http://xxx.lanl.gov/abs/hep-ph/0004043}{{\tt hep-ph/0004043}}].

\bibitem{Baer:1997ai}
H.~Baer and M.~Brhlik, {\it Neutralino dark matter in minimal supergravity:
  direct detection vs. collider searches},  {\em Phys. Rev.} {\bf D57} (1998)
  567--577, [\href{http://xxx.lanl.gov/abs/hep-ph/9706509}{{\tt
  hep-ph/9706509}}].

\bibitem{Bertone:2004pz}
G.~Bertone, D.~Hooper, and J.~Silk, {\it Particle dark matter: evidence,
  candidates and constraints},  {\em Phys. Rept.} {\bf 405} (2005) 279--390,
  [\href{http://xxx.lanl.gov/abs/hep-ph/0404175}{{\tt hep-ph/0404175}}].

\bibitem{Nigam:1956}
B.~P. Nigam and L.~L. Foldy, {\it {Representation of Charge Conjugation for
  Dirac Fields}},  {\em Phys. Rev.} {\bf 102} (1956) 1410--1412. Typographical
  error: In the right hand side of Eq. (37) of this reference $U_s(\theta)$
  should be corrected to read $U_s(0)$. DVA-K thanks B.~P.~Nigam for a private
  communication on the subject.

\bibitem{Rajaraman:1982is}
R.~Rajaraman, {\em Solitons and Instantons: An introduction to solitons and
  instantons in quantum field theory}.
\newblock North-Holland, Amsterdam, 1982.

\bibitem{Mavromatos:2004sz}
N.~E. Mavromatos, {\it {CPT violation and decoherence in quantum gravity}},
  \href{http://xxx.lanl.gov/abs/gr-qc/0407005}{{\tt gr-qc/0407005}}.

\bibitem{Datta:2003dg}
A.~Datta, R.~Gandhi, P.~Mehta, and S.~U. Sankar, {\it {Atmospheric neutrinos as
  a probe of CPT and Lorentz violation}},
  \href{http://xxx.lanl.gov/abs/hep-ph/0312027}{{\tt hep-ph/0312027}}.

\bibitem{Ahluwalia:1993dd}
D.~V. Ahluwalia, {\it Quantum measurements, gravitation, and locality},  {\em
  Phys. Lett. B} {\bf 339} (1994) 301--303,
  [\href{http://xxx.lanl.gov/abs/gr-qc/9308007}{{\tt gr-qc/9308007}}].

\bibitem{VilelaMendes:1994zg}
R.~Vilela~Mendes, {\it Deformations, stable theories and fundamental
  constants},  {\em J. Phys.} {\bf A27} (1994) 8091--8104.

\bibitem{Doplicher:1994zv}
S.~Doplicher, K.~Fredenhagen, and J.~E. Roberts, {\it Space-time quantization
  induced by classical gravity},  {\em Phys. Lett. B} {\bf 331} (1994) 39--44,
  [\href{http://xxx.lanl.gov/abs/DESY 94-065}{{\tt DESY 94-065}}].

\bibitem{Ahluwalia:2000iw}
D.~V. Ahluwalia, {\it Wave-particle duality at the planck scale: Freezing of
  neutrino oscillations},  {\em Phys. Lett. A} {\bf 275} (2000) 31--35,
  [\href{http://xxx.lanl.gov/abs/gr-qc/0002005}{{\tt gr-qc/0002005}}].

\bibitem{Ahluwalia:1997lhr}
D.~V. Ahluwalia, {\it {Book review of Quantum Field Theory by L. H. Ryder}},
  {\em Found. Phys.} {\bf 28} (1998) 527--529,
  [\href{http://xxx.lanl.gov/abs/physics/9702005}{{\tt physics/9702005}}].

\bibitem{Gaioli:1995ra}
F.~H. Gaioli and E.~T. Garcia~Alvarez, {\it {Some remarks about intrinsic
  parity in Ryder's derivation of the Dirac equation}},  {\em Am. J. Phys.}
  {\bf 63} (1995) 177--178, [\href{http://xxx.lanl.gov/abs/hep-th/9807211}{{\tt
  hep-th/9807211}}]. See also Sec. 5.5 of \cite{Weinberg:1995sw}. It provides
  another important perspective on the matter. Note, in particular, the
  relation to parity covariance.

\bibitem{Dirac:1933pam}
P.~A.~M. Dirac, ``Theory of electrons and positrons.''
\newblock Nobel Lecture 1933, available, e.g., at:
  http://www.nobel.se/physics/laureates/1933/dirac-lecture.pdf.

\bibitem{Schweber:1994sss}
S.~S. Schweber, {\em {QED and the men who made it: Dyson, Feynman, Schwinger,
  and Tomonaga}}.
\newblock Princeton University Press, Princeton, 1994.
\newblock See Secction 1.6 and 1.7.

\bibitem{Lamb:1947npp}
W.~E. Lamb, Jr. and R.~C. Retherford, {\it Fine structure of the hydrogen atom
  by a microwave method},  {\em Phys. Rev.} {\bf 72} (1947) 241--243.

\bibitem{Bethe1947lsp}
H.~A. Bethe, {\it The electromagnetic shift of energy levels},  {\em Phys.
  Rev.} {\bf 72} (1947) 339--341.

\bibitem{Anderson:1933mb}
C.~D. Anderson, {\it The positive electron},  {\em Phys. Rev.} {\bf 43} (1933)
  491--494.

\bibitem{Blackett:1933pms}
P.~M.~S. Blackett and G.~P.~S. Occhialini, {\it Some photographs of the tracks
  of pentrating radition},  {\em Proc. Roy. Soc. London A} {\bf 139} (1933)
  699--720.

\bibitem{Kirchbach:2002nu}
M.~Kirchbach and D.~V. Ahluwalia, {\it Spacetime structure of massive
  gravitino},  {\em Phys. Lett.} {\bf B529} (2002) 124--131,
  [\href{http://xxx.lanl.gov/abs/hep-th/0202164}{{\tt hep-th/0202164}}].

\bibitem{Kaloshin:2004jh}
A.~E. Kaloshin and V.~P. Lomov, {\it The {R}arita\textendash {S}chwinger field:
  dressing procedure and spin-parity content},
  \href{http://xxx.lanl.gov/abs/hep-ph/0409052}{{\tt hep-ph/0409052}}.

\bibitem{PR1989}
P.~Ramond, {\em {Field theory: A modern primer}}.
\newblock Addison-Wesley Publishing Co., California, 1989.
\newblock see sec. 1.4.

\bibitem{Stueckelberg:1941th}
{E. C. G. St\"uckelberg}, {\it {La mecanique du point materiel en theorie de
  relativite et en theorie des quanta}},  {\em Helv. Phys. Acta} {\bf 15}
  (1942) 23--37.

\bibitem{Feynman:1949hz}
R.~P. Feynman, {\it The theory of positrons},  {\em Phys. Rev.} {\bf 76} (1949)
  749--759.

\bibitem{Ahluwalia:2001md}
D.~V. Ahluwalia and M.~Kirchbach, {\it Primordial space-time foam as an origin
  of cosmological matter-antimatter asymmetry},  {\em Int. J. Mod. Phys.} {\bf
  D10} (2001) 811--824, [\href{http://xxx.lanl.gov/abs/astro-ph/0107246}{{\tt
  astro-ph/0107246}}].

\bibitem{Weinberg:1972gcb}
S.~Weinberg, {\em Gravitation and cosmology}.
\newblock John Wiley \& sons, New York, 1972.

\bibitem{Streater:1989vi}
R.~F. Streater and A.~S. Wightman, {\em {PCT, Spin and Statistics, and All
  That}}.
\newblock {Addison-Wesley (Advanced book classics), Redwood City}, 1989.

\bibitem{Peskin:1995mep}
M.~E. Peskin and D.~V. Schroeder, {\em An {I}ntroduction to {Q}uantum {F}ield
  {T}heory}.
\newblock Addison-Wesley, 1995.

\bibitem{Sudarshan:2004ecg}
{E. C. G. Sudarshan}.
\newblock private communication (2004).

\bibitem{Ryder:1986mc}
L.~H. Ryder, {\em Elementary {P}articles and {S}ymmetries}.
\newblock Gordon and Breach, New York, 1986.
\newblock p. 32.

\bibitem{Velo:1969bt}
G.~Velo and D.~Zwanziger, {\it {Propagation and quantization of
  Rarita-Schwinger waves in an external electromagnetic potential}},  {\em
  Phys. Rev.} {\bf 186} (1969) 1337--1341.

\bibitem{Ahluwalia:xa}
D.~V. Ahluwalia and M.~Sawicki, {\it {Front form pinors in the
  {W}einberg-{S}oper formalism and generalized {M}elosh transformations for any
  spin}},  {\em Phys. Rev. D} {\bf 47} (1993) 5161--5168,
  [\href{http://xxx.lanl.gov/abs/nucl-th/9603019}{{\tt nucl-th/9603019}}].

\bibitem{Dvoeglazov:1995kn}
V.~V. Dvoeglazov, {\it {Neutral particles in light of the Majorana-Ahluwalia
  ideas}},  {\em Int. J. Theor. Phys.} {\bf 34} (1995) 2467--2490,
  [\href{http://xxx.lanl.gov/abs/hep-th/9504158}{{\tt hep-th/9504158}}].

\bibitem{Kirchbach:2002mkn}
{M. Kirchbach}.
\newblock private communication (2002).

\bibitem{Kirchbach:2004qn}
M.~Kirchbach, C.~Compean, and L.~Noriega, {\it {Beta decays with momentum space
  {M}ajorana spinors}},  {\em Eur. Phys. J. A} {\bf 22} (2004) 149--155,
  [\href{http://xxx.lanl.gov/abs/hep-ph/0411316}{{\tt hep-ph/0411316}}].

\bibitem{Kirchbach:2003mk3}
M.~Kirchbach, {\it Single $\beta$ and double $0\nu\beta\beta$ decays with
  majorana spinors},  in {\em {Proceedings of the Fourth Tegernsee
  International Conference on Particle Physics Beyond the Standard Model,
  BEYOND 2003, Castle Ringberg, 9-14 June 2003}} (H.~V. Klapdor-Kleingrothaus,
  ed.), pp.~429--441, Springer, 2003.

\bibitem{Hladik:spb}
J.~Hladik, {\em {Spinors in Physics}}.
\newblock Springer, New York, 1999.
\newblock see pp. 167-168.

\bibitem{Hatfield:bri1992}
B.~Hatfield, {\em Quantum Field Theory of Point Particles and Strings}.
\newblock Addison-Wesley Pub. Co., California, 1992.
\newblock see Sec. 4.3.

\bibitem{Ralston:2003pf}
J.~P. Ralston and P.~Jain, {\it The virgo alignment puzzle in propagation of
  radiation on cosmological scales},  {\em Int. J. Mod. Phys. D (in press)}
  (2004) [\href{http://xxx.lanl.gov/abs/astro-ph/0311430}{{\tt
  astro-ph/0311430}}].

\bibitem{Zee:ant2003}
A.~Zee, {\em Quantum Field Theory in a Nutshell}.
\newblock Princeton University Press, Princeton, 2003.
\newblock see pp. 107-108.

\bibitem{Nachtmann:1989onz}
O.~Nachtmann, {\em Elementary {P}article {P}hysics}.
\newblock Springer-Verlag, Berlin, 1989.
\newblock see, Sec. 4.5.3.

\bibitem{Weinberg:1964ev}
S.~Weinberg, {\it Feynman rules for any spin. {II}. {M}assless particles},
  {\em Phys. Rev.} {\bf 134} (1964) B882--B896.

\bibitem{Liu:2004yong}
{Y. Liu}.
\newblock private communication (2004).

\bibitem{Accioly:2004bm}
A.~Accioly and R.~Paszko, {\it Photon mass and gravitational deflection},  {\em
  Phys. Rev.} {\bf D69} (2004) 107501.

\bibitem{Luo:2003rz}
{J. Luo, L.-C. Tu, Z.-K. Hu, and E.-J. Luan}, {\it New experimental limit on
  the photon rest mass with a rotating torsion balance},  {\em Phys. Rev.
  Lett.} {\bf 90} (2003) 081801.

\bibitem{Meszaros:2001ig}
P.~Meszaros, {\it Gamma-ray bursts: afterglow implications, progenitor clues
  and prospects},  {\em Science} {\bf 291} (2001) 79--84,
  [\href{http://xxx.lanl.gov/abs/astro-ph/0102255}{{\tt astro-ph/0102255}}].

\bibitem{Boehm:2003bt}
C.~Boehm, D.~Hooper, J.~Silk, and M.~Casse, {\it {MeV} dark matter: has it been
  detected?},  {\em Phys. Rev. Lett.} {\bf 92} (2004) 101301,
  [\href{http://xxx.lanl.gov/abs/astro-ph/0309686}{{\tt astro-ph/0309686}}].

\bibitem{Boehm:2004gt}
C.~Boehm and Y.~Ascasibar, {\it More evidence in favour of light dark matter
  particles?},  {\em Phys. Rev.} {\bf D70} (2004) 115013,
  [\href{http://xxx.lanl.gov/abs/hep-ph/0408213}{{\tt hep-ph/0408213}}].

\bibitem{Beacom:2004pe}
J.~F. Beacom, N.~F. Bell, and G.~Bertone, {\it Gamma-ray constraint on galactic
  positron production by {MeV} dark matter},  {\em Phys. Rev. Lett.} {\bf 94}
  (2005) 171301, [\href{http://xxx.lanl.gov/abs/astro-ph/0409403}{{\tt
  astro-ph/0409403}}].

\bibitem{Casse:2004gw}
M.~Casse, P.~Fayet, S.~Schanne, B.~Cordier, and J.~Paul, {\it Integral and
  light dark matter},  \href{http://xxx.lanl.gov/abs/astro-ph/0404490}{{\tt
  astro-ph/0404490}}.

\bibitem{Fayet:2004kz}
P.~Fayet, {\it Light dark matter},
  \href{http://xxx.lanl.gov/abs/hep-ph/0408357}{{\tt hep-ph/0408357}}.

\bibitem{Bergstroem:2004Goobar}
{Bergstr$\ddot{o}$m, L. and Goobar, A.}, {\em Cosmology and {P}article
  {A}strophysics}.
\newblock Springer-Verlag and Praxis, 2004.

\bibitem{Dodelson:2003sco}
{Dodelson, S.}, {\em Modern cosmology}.
\newblock Academic Press, 2003.

\bibitem{Kolb:1994Tur}
E.~W. Kolb and M.~Turner, {\em The early universe}.
\newblock Westview Press, 1994.

\bibitem{Sievers:2002tq}
{J. L. Sievers \textit{et al}}, {\it {Cosmological parameters from cosmic
  background imager observations and comparisons with BOOMERANG, DASI, and
  MAXIMA}},  {\em Astrophys. J.} {\bf 591} (2003) 599--622,
  [\href{http://xxx.lanl.gov/abs/astro-ph/0205387}{{\tt astro-ph/0205387}}].

\bibitem{Ahluwalia-Khalilova:2004ab}
D.~V. Ahluwalia-Khalilova and D.~Grumiller, {\it Spin half fermions with mass
  dimension one: {T}heory, phenomenology, and dark matter},
  \href{http://xxx.lanl.gov/abs/hep-th/0412080}{{\tt hep-th/0412080}}.

\bibitem{Vanderveld:2005tq}
R.~A. Vanderveld and I.~Wasserman, {\it Spherical gravitational collapse of
  annihilating dark matter and the minimum mass of {CDM} black holes},
  \href{http://xxx.lanl.gov/abs/astro-ph/0505497}{{\tt astro-ph/0505497}}.

\bibitem{Kusenko:1997sp}
A.~Kusenko and G.~Segre, {\it Neutral current induced neutrino oscillations in
  a supernova},  {\em Phys. Lett.} {\bf B396} (1997) 197--200,
  [\href{http://xxx.lanl.gov/abs/hep-ph/9701311}{{\tt hep-ph/9701311}}].

\bibitem{Kusenko:2004mm}
A.~Kusenko, {\it Pulsar kicks from neutrino oscillations},
  \href{http://xxx.lanl.gov/abs/astro-ph/0409521}{{\tt astro-ph/0409521}}.

\bibitem{Ahluwalia-Khalilova:2004wf}
D.~V. Ahluwalia-Khalilova, {\it Neutrino oscillations and supernovae},  {\em
  Gen. Rel. Grav.} {\bf 36} (2004) 2183--2187,
  [\href{http://xxx.lanl.gov/abs/astro-ph/0404055}{{\tt astro-ph/0404055}}].

\bibitem{Dolgov:2000ew}
A.~D. Dolgov and S.~H. Hansen, {\it Massive sterile neutrinos as warm dark
  matter},  {\em Astropart. Phys.} {\bf 16} (2002) 339--344,
  [\href{http://xxx.lanl.gov/abs/hep-ph/0009083}{{\tt hep-ph/0009083}}].

\bibitem{Weidenspointner:2004my}
{G. Weidenspointner \textit{et al.}}, {\it {SPI} observations of positron
  annihilation radiation from the 4th galactic quadrant: sky distribution},
  \href{http://xxx.lanl.gov/abs/astro-ph/0406178}{{\tt astro-ph/0406178}}.

\bibitem{Jordan:1959eg}
P.~Jordan, {\it The present state of {D}irac's cosmological hypothesis},  {\em
  Z. Phys.} {\bf 157} (1959) 112--121.

\bibitem{Brans:1961sx}
C.~Brans and R.~H. Dicke, {\it Mach's principle and a relativistic theory of
  gravitation},  {\em Phys. Rev.} {\bf 124} (1961) 925--935.

\bibitem{Wetterich:1988fm}
C.~Wetterich, {\it Cosmology and the fate of dilatation symmetry},  {\em Nucl.
  Phys.} {\bf B302} (1988) 668.

\bibitem{Wang:1998gt}
L.-M. Wang and P.~J. Steinhardt, {\it Cluster abundance constraints on
  quintessence models},  {\em Astrophys. J.} {\bf 508} (1998) 483--490,
  [\href{http://xxx.lanl.gov/abs/astro-ph/9804015}{{\tt astro-ph/9804015}}].

\bibitem{Carroll:1998zi}
S.~M. Carroll, {\it Quintessence and the rest of the world},  {\em Phys. Rev.
  Lett.} {\bf 81} (1998) 3067--3070,
  [\href{http://xxx.lanl.gov/abs/astro-ph/9806099}{{\tt astro-ph/9806099}}].

\bibitem{Zlatev:1998tr}
I.~Zlatev, L.-M. Wang, and P.~J. Steinhardt, {\it Quintessence, cosmic
  coincidence, and the cosmological constant},  {\em Phys. Rev. Lett.} {\bf 82}
  (1999) 896--899, [\href{http://xxx.lanl.gov/abs/astro-ph/9807002}{{\tt
  astro-ph/9807002}}].

\bibitem{Foot:2004pa}
R.~Foot, {\it Mirror matter-type dark matter},
  \href{http://xxx.lanl.gov/abs/astro-ph/0407623}{{\tt astro-ph/0407623}}.

\bibitem{Foot:2004wz}
R.~Foot and R.~R. Volkas, {\it Spheroidal galactic halos and mirror dark
  matter},  \href{http://xxx.lanl.gov/abs/astro-ph/0407522}{{\tt
  astro-ph/0407522}}.

\bibitem{Berezhiani:2003xm}
Z.~Berezhiani, {\it Mirror world and its cosmological consequences},  {\em Int.
  J. Mod. Phys.} {\bf A19} (2004) 3775--3806,
  [\href{http://xxx.lanl.gov/abs/hep-ph/0312335}{{\tt hep-ph/0312335}}].

\bibitem{Asai:2003hs}
S.~Asai, O.~Jinnouchi, and T.~Kobayashi, {\it Solution of orthopositronium
  lifetime puzzle},  {\em Int. J. Mod. Phys.} {\bf A19} (2004) 3927--3938,
  [\href{http://xxx.lanl.gov/abs/hep-ex/0308031}{{\tt hep-ex/0308031}}].

\bibitem{Vallery:2003iz}
R.~S. Vallery, P.~W. Zitzewitz, and D.~W. Gidley, {\it Resolution of the
  orthopositronium lifetime puzzle},  {\em Phys. Rev. Lett.} {\bf 90} (2003)
  203402.

\bibitem{Blinnikov:1999ky}
S.~Blinnikov, {\it Gamma-ray bursts produced by mirror stars},
  \href{http://xxx.lanl.gov/abs/astro-ph/9902305}{{\tt astro-ph/9902305}}.

\bibitem{Foot:1999hm}
R.~Foot, {\it Have mirror stars been observed?},  {\em Phys. Lett.} {\bf B452}
  (1999) 83--86, [\href{http://xxx.lanl.gov/abs/astro-ph/9902065}{{\tt
  astro-ph/9902065}}].

\bibitem{Berezhiani:1995yi}
Z.~G. Berezhiani and R.~N. Mohapatra, {\it Reconciling present neutrino
  puzzles: Sterile neutrinos as mirror neutrinos},  {\em Phys. Rev.} {\bf D52}
  (1995) 6607--6611, [\href{http://xxx.lanl.gov/abs/hep-ph/9505385}{{\tt
  hep-ph/9505385}}].

\bibitem{Badertscher:2003rk}
{A. Badertscher, \textit{et al.}}, {\it An apparatus to search for mirror dark
  matter via the invisible decay of orthopositronium in vacuum},  {\em Int. J.
  Mod. Phys.} {\bf A19} (2004) 3833--3848,
  [\href{http://xxx.lanl.gov/abs/hep-ex/0311031}{{\tt hep-ex/0311031}}]. The
  information as presented here is taken from the ar{X}ived preprint. The
  published version carries a single author, {S. N. Gninenko}, and an
  abbreviated title. {T}he {J}ournal-ref. for both versions is \textit{same}.

\bibitem{Foot:1995pa}
R.~Foot and R.~R. Volkas, {\it Neutrino physics and the mirror world: how exact
  parity symmetry explains the solar neutrino deficit, the atmospheric neutrino
  anomaly and the {LSND} experiment},  {\em Phys. Rev.} {\bf D52} (1995)
  6595--6606, [\href{http://xxx.lanl.gov/abs/hep-ph/9505359}{{\tt
  hep-ph/9505359}}].

\bibitem{Foot:2003js}
R.~Foot and S.~Mitra, {\it Have mirror micrometeorites been detected?},  {\em
  Phys. Rev.} {\bf D68} (2003) 071901,
  [\href{http://xxx.lanl.gov/abs/hep-ph/0306228}{{\tt hep-ph/0306228}}].

\bibitem{Ahluwalia-Khalilova:2004kv}
D.~V. Ahluwalia-Khalilova, {\it Charge conjugation and {L}ense-{T}hirring
  effect: a new asymmetry},  {\em Int. J. Mod. Phys.} {\bf D13} (2004)
  2361--2367, [\href{http://xxx.lanl.gov/abs/gr-qc/0405112}{{\tt
  gr-qc/0405112}}].

\bibitem{Ahluwalia:1997hc}
D.~V. Ahluwalia, {\it {On a new non-geometric element in gravity}},  {\em Gen.
  Rel. Grav.} {\bf 29} (1997) 1491--1501,
  [\href{http://xxx.lanl.gov/abs/gr-qc/9705050}{{\tt gr-qc/9705050}}].

\bibitem{Chryssomalakos:2002bm}
C.~Chryssomalakos and D.~Sudarsky, {\it {On the geometrical character of
  gravitation}},  {\em Gen. Rel. Grav.} {\bf 35} (2003) 605--617,
  [\href{http://xxx.lanl.gov/abs/gr-qc/0206030}{{\tt gr-qc/0206030}}].

\bibitem{VilelaMendes:1999xv}
R.~Vilela~Mendes, {\it {Geometry, stochastic calculus and quantum fields in a
  non-commutative space-time}},  {\em {J. Math. Phys.}} {\bf {41}} (2000)
  156--186, [\href{http://xxx.lanl.gov/abs/math-ph/9907001}{{\tt
  math-ph/9907001}}].

\bibitem{Chryssomalakos:2004gk}
C.~Chryssomalakos and E.~Okon, {\it Generalized quantum relativistic
  kinematics: a stability point of view},  {\em Int. J. Mod. Phys.} {\bf D13}
  (2004) 2003--2034, [\href{http://xxx.lanl.gov/abs/hep-th/0410212}{{\tt
  hep-th/0410212}}].

\bibitem{Grumiller:2002nm}
D.~Grumiller, W.~Kummer, and D.~V. Vassilevich, {\it {Dilaton gravity in two
  dimensions}},  {\em Phys. Rept.} {\bf 369} (2002) 327--429,
  [\href{http://xxx.lanl.gov/abs/hep-th/0204253}{{\tt hep-th/0204253}}].

\bibitem{Grumiller:2004yq}
D.~Grumiller, {\it Virtual black holes and the {S}-matrix},  {\em Int. J. Mod.
  Phys.} {\bf D13} (2004) 1973--2002,
  [\href{http://xxx.lanl.gov/abs/hep-th/0409231}{{\tt hep-th/0409231}}].

\bibitem{Yang:1947ud}
C.~N. Yang, {\it On quantized space-time},  {\em Phys. Rev.} {\bf 72} (1947)
  874.

\bibitem{Snyder:1946qz}
H.~S. Snyder, {\it Quantized space-time},  {\em Phys. Rev.} {\bf 71} (1947)
  38--41.

\bibitem{Chryssomalakos:2001nd}
{C. Chryssomalakos}, {\it {Stability of Lie superalgebras and branes}},  {\em
  {Mod. Phys. Lett. A}} {\bf {16}} (2001) 197--209,
  [\href{http://xxx.lanl.gov/abs/hep-th/0102134}{{\tt hep-th/0102134}}].

\bibitem{Chryssomalakos:2004wc}
C.~Chryssomalakos and E.~Okon, {\it Linear form of 3-scale relativity algebra
  and the relevance of stability},  {\em Int. J. Mod. Phys. D} {\bf 13} (2004)
  1817--1822, [\href{http://xxx.lanl.gov/abs/hep-th/0407080}{{\tt
  hep-th/0407080}}].

\bibitem{Ahluwalia:2002nk}
D.~V. Ahluwalia, {\it {Evidence for Majorana neutrinos: Dawn of a new era in
  spacetime structure}},  in {\em Beyond the Desert 2002} (H.~V.
  Klapdor-Klengrothaus, ed.), pp.~143--160.
\newblock Institute of Physics Publishing, Bristol, 2003.
\newblock \href{http://xxx.lanl.gov/abs/hep-ph/0212222}{{\tt hep-ph/0212222}}.

\bibitem{Ahluwalia:1994pf}
D.~V. Ahluwalia, T.~Goldman, and M.~B. Johnson, {\it {(j,0) + (0,j)
  representation space: Majorana - like construct}},  {\em Acta Phys. Polon.}
  {\bf B25} (1994) 1267--1278,
  [\href{http://xxx.lanl.gov/abs/hep-th/9312091}{{\tt hep-th/9312091}}].

\bibitem{Gelfand:1964}
I.~Gelfand and G.~Shilov, {\em Generalized Functions -- Properties and
  Operations}, vol.~1.
\newblock Academic Press, 1964.

\bibitem{daRocha:2005ti}
R.~da~Rocha and W.~A. Rodrigues~Jr, {\it Where are {Elko} spinor fields in
  {Lounesto} spinor field classification?},
  \href{http://xxx.lanl.gov/abs/math-ph/0506075}{{\tt math-ph/0506075}}.

\end{thebibliography}

\providecommand{\href}[2]{#2}\begingroup\raggedright\endgroup

\end{document}